\documentclass[pdftex,twocolumn,epjc3_preprint,runningheads]{svjour3}
\pdfoutput=1

\usepackage[T1]{fontenc}
\usepackage{lmodern}
\usepackage{calc}
\usepackage{graphicx}
\usepackage{booktabs}
\usepackage{textcomp}
\usepackage{xspace}
\usepackage{relsize}
\usepackage{amssymb}
\usepackage{amsmath}
\usepackage{listings}
\usepackage{microtype}
\usepackage{multirow}
\usepackage{tabularx}
\usepackage{array}
\usepackage{placeins}
\usepackage{cuted}
\usepackage{soul} 
\usepackage{fixltx2e}
\usepackage[numbers,sort&compress]{natbib}
\usepackage[labelfont=bf,font=small]{caption}
\usepackage[skip=-2pt]{subcaption}
\usepackage[colorlinks,citecolor=blue,urlcolor=blue,linkcolor=blue,breaklinks=breakall]{hyperref}
\usepackage{breakurl}
\usepackage[dvipsnames]{xcolor}
\usepackage[clockwise,figuresright]{rotating}
\usepackage{siunitx}
\usepackage{tikz}

\usepackage{etoolbox}
\AfterEndEnvironment{strip}{\leavevmode}

\allowdisplaybreaks

\newcolumntype{L}{>{\raggedright\let\newline\\\arraybackslash\hspace{0pt}}X}
\newcolumntype{R}{>{\raggedleft\let\newline\\\arraybackslash\hspace{0pt}}X}
\newcolumntype{C}{>{\centering\let\newline\\\arraybackslash\hspace{0pt}}X}

\newcommand{\imperial}{Department of Physics, Imperial College London, Blackett Laboratory, Prince Consort Road, London SW7 2AZ, UK}
\newcommand{\nordita}{NORDITA, Roslagstullsbacken 23, SE-10691 Stockholm, Sweden}
\newcommand{\oslo}{Department of Physics, University of Oslo, N-0316 Oslo, Norway}
\newcommand{\adelaide}{Department of Physics, University of Adelaide, Adelaide, SA 5005, Australia}

\newcommand{\monash}{School of Physics and Astronomy, Monash University, Melbourne, VIC 3800, Australia}
\newcommand{\coepp}{Australian Research Council Centre of Excellence for Particle Physics at the Tera-scale}
\newcommand{\okc}{Oskar Klein Centre for Cosmoparticle Physics, AlbaNova University Centre, SE-10691 Stockholm, Sweden}
\newcommand{\su}{Department of Physics, Stockholm University, SE-10691 Stockholm, Sweden}

\newcommand{\annecy}{LAPTh, Universit\'e de Savoie, CNRS, 9 chemin de Bellevue B.P.110, F-74941 Annecy-le-Vieux, France}
\newcommand{\harvard}{Department of Physics, Harvard University, Cambridge, MA 02138, USA}
\newcommand{\grappa}{GRAPPA, Institute of Physics, University of Amsterdam, Science Park 904, 1098 XH Amsterdam, Netherlands}

\newcommand{\gambitacknos    }{We warmly thank the Casa Matem\'aticas Oaxaca, affiliated with the Banff International Research Station, for hospitality whilst part of this work was completed, and the staff at Cyfronet, for their always helpful supercomputing support.  \GB has been supported by STFC (UK; ST/K00414X/1, ST/P000762/1), the Royal Society (UK; UF110191), Glasgow University (UK; Leadership Fellowship), the Research Council of Norway (FRIPRO 230546/F20), NOTUR (Norway; NN9284K), the Knut and Alice Wallenberg Foundation (Sweden; Wallenberg Academy Fellowship), the Swedish Research Council (621-2014-5772), the Australian Research Council (CE110001004, FT130100018, FT140100244, FT160100274), The University of Sydney (Australia; IRCA-G162448), PLGrid Infrastructure (Poland), Polish National Science Center (Sonata UMO-2015/17/D/ST2/03532), the Swiss National Science Foundation (PP00P2-144674), the European Commission Horizon 2020 Marie Sk\l{}odowska-Curie actions (H2020-MSCA-RISE-2015-691164), the ERA-CAN+ Twinning Program (EU \& Canada), the Netherlands Organisation for Scientific Research (NWO-Vidi 680-47-532), the National Science Foundation (USA; DGE-1339067), the FRQNT (Qu\'ebec) and NSERC/The Canadian Tri-Agencies Research Councils (BPDF-424460-2012).}

\makeatletter

\newcommand{\preprintnumber}[1]{\gdef\@preprintnumber{\begin{flushright}{#1}\end{flushright}}}

\g@addto@macro\bfseries{\boldmath}
\makeatother

\newcommand{\subparagraph}{} 
\usepackage{titlesec}
\titleformat*{\paragraph}{\bfseries}

\journalname{Eur. Phys. J. C}
\smartqed
\let\underscore\_
\renewcommand{\_}{\discretionary{\underscore}{}{\underscore}}

\makeatletter
\let\orgdescriptionlabel\descriptionlabel
\renewcommand*{\descriptionlabel}[1]{%
  \let\orglabel\label
  \let\label\@gobble
  \phantomsection
  \protected@edef\@currentlabel{#1}%
  \let\label\orglabel
  \orgdescriptionlabel{#1}%
}
\makeatother

\newcommand\postnewlinemarker{\hbox{\ensuremath{\hookrightarrow}}}
\newcommand\cpp[1]{{\lstinline!#1!}}  
\newcommand\cpppragma[1]{{\CPPcommentstyle#1}}
\newcommand\yaml[1]{{\lstset{style=yaml}\lstinline!#1!\lstset{style=cpp}}}
\newcommand\yamlvalue[1]{{\YAMLvaluestyle\ttfamily#1}}
\newcommand\term[1]{{\lstset{style=terminal}\lstinline!#1!\lstset{style=cpp}}}
\newcommand\fortran[1]{{\lstset{style=fortran}\lstinline!#1!\lstset{style=cpp}}}
\newcommand\py[1]{{\lstset{style=python}\lstinline!#1!\lstset{style=cpp}}}
\newcommand\customtilde{{\raisebox{0.2ex}{\scalebox{0.6}{\boldmath$\sim$}}}}
\newcommand\mathematica[1]{{\lstset{style=Mathematica}\lstinline!#1!\lstset{style=cpp}}}

\lstnewenvironment{lstlistingyaml}{\lstset{style=yaml}}{\lstset{style=cpp}}
\lstnewenvironment{lstlistingterm}{\lstset{style=terminal}}{\lstset{style=cpp}}
\lstnewenvironment{lstlistingfortran}{\lstset{style=fortran}}{\lstset{style=cpp}}
\lstnewenvironment{lstcpp}{\lstset{style=cpp}}{\lstset{style=cpp}}
\lstnewenvironment{lstcppalt}{\lstset{style=cppalt}}{\lstset{style=cpp}}
\lstnewenvironment{lstcppnum}{\lstset{style=cppnum}}{\lstset{style=cpp}}
\lstnewenvironment{lstyaml}{\lstset{style=yaml}}{\lstset{style=cpp}}
\lstnewenvironment{lstterm}{\lstset{style=terminal}}{\lstset{style=cpp}}
\lstnewenvironment{lsttermalt}{\lstset{style=terminalalt}}{\lstset{style=cpp}}
\lstnewenvironment{lsttext}{\lstset{style=text}}{\lstset{style=cpp}}
\lstnewenvironment{lstfortran}{\lstset{style=fortran}}{\lstset{style=cpp}}
\lstnewenvironment{lstpy}{\lstset{style=python}}{\lstset{style=cpp}}
\lstnewenvironment{lstmathematica}{\lstset{style=mathematica}}{\lstset{style=cpp}}

\newcommand{\tmpname}{}
\newcommand{\tmplistingname}{}
\makeatletter
\newif\ifATOlabelname
\lst@Key{labelname}{Listing}{\def\ATOlabelname{#1}\global\ATOlabelnametrue}
\makeatother
\lstnewenvironment{lstcpplabel}[1][]{
  \lstset{style=cpp,#1} 
  \ifATOlabelname
    \renewcommand{\tmpname}{\lstlistingname}
    \renewcommand{\tmplistingname}{\lstlistlistingname}
    \renewcommand{\lstlistingname}{\ATOlabelname}
    \renewcommand{\lstlistlistingname}{List of \lstlistingname s}
  \fi
}{
  \renewcommand{\lstlistingname}{\tmpname}
  \renewcommand{\lstlistlistingname}{\tmplistingname}
  \lstset{style=cpp}
}
\definecolor{solarized@base03}{HTML}{002B36}
\definecolor{solarized@base02}{HTML}{073642}
\definecolor{solarized@base01}{HTML}{586e75}
\definecolor{solarized@base00}{HTML}{657b83}
\definecolor{solarized@base0}{HTML}{839496}
\definecolor{solarized@base1}{HTML}{93a1a1}
\definecolor{solarized@base2}{HTML}{EEE8D5}
\definecolor{solarized@base3}{HTML}{FDF6E3}
\definecolor{solarized@yellow}{HTML}{B58900}
\definecolor{solarized@orange}{HTML}{CB4B16}
\definecolor{solarized@red}{HTML}{DC322F}
\definecolor{solarized@magenta}{HTML}{D33682}
\definecolor{solarized@violet}{HTML}{6C71C4}
\definecolor{solarized@blue}{HTML}{268BD2}
\definecolor{solarized@cyan}{HTML}{2AA198}
\definecolor{solarized@green}{HTML}{859900}
\definecolor{darkred}{HTML}{550003}
\definecolor{darkgreen}{HTML}{00AA00}

\newcommand\YAMLstringstyle{\footnotesize\color{solarized@green}\mdseries}
\newcommand\YAMLkeystyle{\footnotesize\color{solarized@blue}\ttfamily}
\newcommand\YAMLvaluestyle{\footnotesize\color{blue}\mdseries}
\newcommand\ProcessThreeDashes{\llap{\color{cyan}\mdseries-{-}-}}

\newcommand\CPPidentifierstyle{\color{solarized@blue}\footnotesize\ttfamily}
\newcommand\CPPcommentstyle{\color{solarized@violet}\footnotesize\ttfamily}
\newcommand\CPPdirectivestyle{\color{solarized@magenta}\footnotesize\ttfamily}
\newcommand\termplainstyle{\footnotesize\ttfamily}

\newcommand\processLongMacroDelimiter
{%
\CPPdirectivestyle%
\#define%
}

\lstdefinestyle{cpp}
{
  language=C++,
  basicstyle=\footnotesize\ttfamily,
  basewidth={0.53em,0.44em}, 
  numbers=none,
  tabsize=2,
  breaklines=true,
  escapeinside={@}{@},
  showstringspaces=false,
  numberstyle=\tiny\color{solarized@base01},
  keywordstyle=\color{solarized@orange},
  stringstyle=\color{solarized@red}\ttfamily,
  identifierstyle=\color{solarized@blue},
  commentstyle=\CPPcommentstyle,
  directivestyle=\CPPdirectivestyle,
  emphstyle=\color{solarized@green},
  frame=single,
  rulecolor=\color{solarized@base2},
  rulesepcolor=\color{solarized@base2},
  literate={~} {\customtilde}1,
  moredelim=*[directive]\ \ \#,
  moredelim=*[directive]\ \ \ \ \#
}

\lstdefinestyle{cppalt}
{
  language=C++,
  basicstyle=\footnotesize\ttfamily,
  basewidth={0.53em,0.44em}, 
  numbers=none,
  tabsize=2,
  breaklines=true,
  escapeinside={*@}{@*},
  showstringspaces=false,
  numberstyle=\tiny\color{solarized@base01},
  keywordstyle=\color{solarized@orange},
  stringstyle=\color{solarized@red}\ttfamily,
  identifierstyle=\color{solarized@blue},
  commentstyle=\CPPcommentstyle,
  directivestyle=\CPPdirectivestyle,
  emphstyle=\color{solarized@green},
  frame=single,
  rulecolor=\color{solarized@base2},
  rulesepcolor=\color{solarized@base2},
  literate={~}{\customtilde}1,
  moredelim=**[is][\processLongMacroDelimiter]{BeginLongMacro}{EndLongMacro} 
}

\lstdefinestyle{cppnum}
{
  language=C++,
  basicstyle=\footnotesize\ttfamily,
  basewidth={0.53em,0.44em}, 
  numbers=none,
  tabsize=2,
  breaklines=true,
  escapeinside={@}{@},
  numberstyle=\tiny\color{solarized@base01},
  showstringspaces=false,
  numberstyle=\tiny\color{solarized@base01},
  keywordstyle=\color{solarized@orange},
  stringstyle=\color{solarized@red}\ttfamily,
  identifierstyle=\color{solarized@blue},
  commentstyle=\CPPcommentstyle,
  directivestyle=\CPPdirectivestyle,
  emphstyle=\color{solarized@green},
  frame=single,
  rulecolor=\color{solarized@base2},
  rulesepcolor=\color{solarized@base2},
  literate={~} {\customtilde}1,
  moredelim=*[directive]\ \ \#,
  moredelim=*[directive]\ \ \ \ \#
}

\lstdefinestyle{python}
{
  language=Python,
  basicstyle=\footnotesize\ttfamily,
  basewidth={0.53em,0.44em},
  numbers=none,
  tabsize=2,
  breaklines=true,
  escapeinside={@}{@},
  showstringspaces=false,
  numberstyle=\tiny\color{solarized@base01},
  keywordstyle=\color{blue},
  stringstyle=\color{orange}\ttfamily,
  identifierstyle=\color{darkred},
  commentstyle=\color{purple},
  emphstyle=\color{green},
  frame=single,
  rulecolor=\color{solarized@base2},
  rulesepcolor=\color{solarized@base2},
  literate = {~}{\customtilde}1
             {\ as\ }{{\color{blue}\ as\ \color{black}}}3
}

\lstdefinestyle{fortran}
{
  language=Fortran,
  basicstyle=\footnotesize\ttfamily,
  basewidth={0.53em,0.44em},
  numbers=none,
  tabsize=2,
  breaklines=true,
  escapeinside={@}{@},
  showstringspaces=false,
  numberstyle=\tiny\color{solarized@base01},
  keywordstyle=\color{blue},
  stringstyle=\color{orange}\ttfamily,
  identifierstyle=\color{Periwinkle},
  commentstyle=\color{purple},
  emphstyle=\color{green},
  morekeywords={and, or, true, false},
  frame=single,
  rulecolor=\color{solarized@base2},
  rulesepcolor=\color{solarized@base2},
  literate={~}{\customtilde}1
}

\lstdefinestyle{terminal}
{
  language=bash,
  basicstyle=\termplainstyle,
  numbers=none,
  tabsize=2,
  breaklines=true,
  escapeinside={@}{@},
  frame=single,
  showstringspaces=false,
  numberstyle=\tiny\color{solarized@base01},
  keywordstyle=\color{solarized@orange},
  stringstyle=\color{solarized@red}\ttfamily,
  identifierstyle=\color{black},
  commentstyle=\color{solarized@violet},
  emphstyle=\color{solarized@green},
  frame=single,
  rulecolor=\color{solarized@base2},
  rulesepcolor=\color{solarized@base2},
  morekeywords={gambit, cmake, make, mkdir},
  deletekeywords={test},
  literate = {\ gambit}{{\ }{\color{black}}gambit}7
             {/gambit}{{/}{\color{black}}gambit}6
             {gambit/}{{\color{black}}gambit{/}}6
             {/include}{{/}{\color{black}}include}8
             {cmake/}{{\color{black}}cmake/}6
             {.cmake}{{.}{\color{black}}cmake}6
             {~}{\customtilde}1
}

\lstdefinestyle{terminalalt}
{
  language=bash,
  basicstyle=\footnotesize\ttfamily,
  numbers=none,
  tabsize=2,
  breaklines=true,
  escapeinside={*@}{@*},
  frame=single,
  showstringspaces=false,
  numberstyle=\tiny\color{solarized@base01},
  keywordstyle=\color{solarized@orange},
  stringstyle=\color{solarized@red}\ttfamily,
  identifierstyle=\color{black},
  commentstyle=\color{solarized@violet},
  emphstyle=\color{solarized@green},
  frame=single,
  rulecolor=\color{solarized@base2},
  rulesepcolor=\color{solarized@base2},
  morekeywords={gambit, cmake, make, mkdir},
  deletekeywords={test},
  literate = {\ gambit}{{\ }{\color{black}}gambit}7
             {/gambit}{{/}{\color{black}}gambit}6
             {gambit/}{{\color{black}}gambit{/}}6
             {/include}{{/}{\color{black}}include}8
             {cmake/}{{\color{black}}cmake/}6
             {.cmake}{{.}{\color{black}}cmake}6
             {~}{\customtilde}1
}

\lstdefinestyle{text}
{
  language={},
  basicstyle=\footnotesize\ttfamily,
  identifierstyle=\color{black},
  numbers=none,
  tabsize=2,
  breaklines=true,
  escapeinside={*@}{@*},
  showstringspaces=false,
  frame=single,
  rulecolor=\color{solarized@base2},
  rulesepcolor=\color{solarized@base2},
  literate={~}{\customtilde}1
}

\lstdefinestyle{yaml}
{
  language=bash,
  escapeinside={@}{@},
  keywords={true,false,null},
  otherkeywords={},
  keywordstyle=\color{solarized@base0}\bfseries,
  basicstyle=\footnotesize\color{black}\ttfamily,
  identifierstyle=\YAMLkeystyle,
  sensitive=false,
  commentstyle=\color{solarized@orange}\ttfamily,
  morecomment=[l]{\#},
  morecomment=[s]{/*}{*/},
  stringstyle=\YAMLstringstyle\ttfamily,
  moredelim=**[s][\YAMLkeystyle]{,}{:},   
  moredelim=**[l][\YAMLvaluestyle]{:},    
  morestring=[b]',
  morestring=[b]",
  literate =    {---}{{\ProcessThreeDashes}}3
                {>}{{\textcolor{solarized@red}\textgreater}}1
                {|}{{\textcolor{solarized@red}\textbar}}1
                {\ -\ }{{\mdseries\color{black}\ -\ \negmedspace}}3
                {\}}{{{\color{black} \}}}}1
                {\{}{{{\color{black} \{}}}1
                {[}{{{\color{black} [}}}1
                {]}{{{\color{black} ]}}}1
                {~}{\customtilde}1,
  breakindent=0pt,
  breakatwhitespace,
  columns=fullflexible
}

\lstdefinestyle{mathematica}
{
  language={Mathematica},
  basicstyle=\footnotesize\ttfamily,
  basewidth={0.53em,0.44em},
  numbers=none,
  tabsize=2,
  breaklines=true,
  escapeinside={@}{@},
  numberstyle=\tiny\color{black},
  showstringspaces=false,
  numberstyle=\tiny\color{solarized@base01},
  keywordstyle=\color{solarized@orange},
  stringstyle=\color{solarized@red}\ttfamily,
  identifierstyle=\color{solarized@orange}\ttfamily,
  commentstyle=\color{solarized@gray}\ttfamily,
  directivestyle=\color{solarized@orange}\ttfamily,
  emphstyle=\color{solarized@green},
  frame=single,
  rulecolor=\color{solarized@base2},
  rulesepcolor=\color{solarized@base2},
  literate={~} {\customtilde}1,
  moredelim=*[directive]\ \ \#,
  moredelim=*[directive]\ \ \ \ \#,
  mathescape=true
}

\newcommand{\cross}[1]{\ref{#1}}
\newcommand{\doublecross}[2]{\hyperref[#2]{\textbf{#1}}}
\newcommand{\doublecrosssf}[2]{\hyperref[#2]{\textbf{\textsf{#1}}}}
\newcommand{\gitem}[1]{\item[\textbf{#1}\label{#1}]}
\newcommand{\gsfitem}[1]{\item[\textbf{\textsf{#1}}\label{#1}]}

\newcommand{\startglossary}{\section{Glossary}\label{glossary}Here we explain some terms that have specific technical definitions in \GB.\begin{description}}
\newcommand{\finishglossary}{\end{description}}

\newcommand{\bcode}{\begin{lstlisting}}
\newcommand{\ecode}{\end{lstlisting}}
\newcommand{\metavarf}[1]{\textit{\color{darkgreen}\footnotesize\textrm{#1}}}
\newcommand{\metavars}[1]{\textit{\color{darkgreen}\scriptsize\textrm{#1}}}
\newcommand{\metavar}{\metavarf}

\newcommand{\sss}{\scriptscriptstyle}
\newcommand{\ms}{m_{\sss S}}
\newcommand{\lhs}{\lambda_{h\sss S}}
\newcommand{\ls}{\lambda_{\sss S}}

\newcommand{\DR}{$\overline{DR}$\xspace}

\newcommand{\MSbar}{$\MSBar$\xspace}
\newcommand{\MSBar}{\overline{MS}}

\newcommand{\gambit}{\textsf{GAMBIT}\xspace}
\newcommand{\gambitversion}{1.0.0}
\newcommand{\gambitVer}{\gambit \textsf{\gambitversion}\xspace}
\newcommand{\darkbit}{\textsf{DarkBit}\xspace}
\newcommand{\colliderbit}{\textsf{ColliderBit}\xspace}
\newcommand{\flavbit}{\textsf{FlavBit}\xspace}
\newcommand{\specbit}{\textsf{SpecBit}\xspace}
\newcommand{\decaybit}{\textsf{DecayBit}\xspace}
\newcommand{\precisionbit}{\textsf{PrecisionBit}\xspace}
\newcommand{\scannerbit}{\textsf{ScannerBit}\xspace}

\newcommand{\GB}{\gambit}

\newcommand{\higgsbounds}{\textsf{HiggsBounds}\xspace}

\newcommand{\higgssignals}{\textsf{HiggsSignals}\xspace}

\newcommand{\threebit}{\textsf{3-BIT-HIT}\xspace}

\newcommand{\feynhiggs}{\textsf{FeynHiggs}\xspace}
\newcommand{\FH}{\feynhiggs}

\newcommand\flexiblesusy{\FlexibleSUSY}
\newcommand\FlexibleSUSY{\textsf{FlexibleSUSY}\xspace}
\newcommand\FlexibleEFTHiggs{\textsf{FlexibleEFTHiggs}\xspace}
\newcommand\SOFTSUSY{\textsf{SOFTSUSY}\xspace}
\newcommand\SUSPECT{\textsf{SuSpect}\xspace}
\newcommand\NMSSMCalc{\textsf{NMSSMCALC}\xspace}
\newcommand\NMSSMTools{\textsf{NMSSMTools}\xspace}
\newcommand\NMSPEC{\textsf{NMSPEC}\xspace}
\newcommand\NMHDECAY{\textsf{NMHDECAY}\xspace}
\newcommand\HDECAY{\textsf{HDECAY}\xspace}
\newcommand\prophecy{\textsf{PROPHECY4F}\xspace}
\newcommand\SDECAY{\textsf{SDECAY}\xspace}
\newcommand\SUSYHIT{\textsf{SUSY-HIT}\xspace}
\newcommand\susyhd{\textsf{SUSYHD}\xspace}
\newcommand\HSSUSY{\textsf{HSSUSY}\xspace}
\newcommand\susyhit{\SUSYHIT}
\newcommand\gmtwocalc{\textsf{GM2Calc}\xspace}
\newcommand\SARAH{\textsf{SARAH}\xspace}
\newcommand\SPheno{\textsf{SPheno}\xspace}
\newcommand\superiso{\textsf{SuperIso}\xspace}
\newcommand\SFOLD{\textsf{SFOLD}\xspace}
\newcommand\HFOLD{\textsf{HFOLD}\xspace}
\newcommand\FeynHiggs{\textsf{FeynHiggs}\xspace}
\newcommand\Mathematica{\textsf{Mathematica}\xspace}

\newcommand\xx{\raisebox{0.2ex}{\smaller ++}\xspace}
\newcommand\Cpp{\textsf{C\xx}\xspace}

\newcommand\Fortran{\textsf{Fortran}\xspace}
\newcommand\YAML{\textsf{YAML}\xspace}

\newcommand\beq{\begin{equation}}
\newcommand\eeq{\end{equation}}

\renewcommand{\url}[1]{\href{#1}{#1}}

\usepackage{enumitem} 
\usepackage{titlesec}
\usepackage{slashed}
\usepackage{pbox}
\newcommand{\ctr}{\centering}
\titleformat{\section}
   {\Large\bfseries} 
   {\normalfont\Large\bfseries \thesection} 
   {1em}
   {}
\titleformat*{\subsection}{\normalsize\bfseries} 
\titleformat*{\subsubsection}{\normalsize} 
\titleformat{\paragraph} 
{\normalsize\itshape}{\theparagraph}{1em}{}
\titlespacing*{\paragraph}
{0pt}{3.25ex plus 1ex minus .2ex}{1.5ex plus .2ex}
\titleformat{\subparagraph}[runin]{\bfseries}{}{}{}[]
\titlespacing{\subparagraph}{0pt}{0pt}{5pt}

\makeatletter
\renewcommand\tableofcontents{%
    \subsection*{\contentsname}
    \@starttoc{toc}%
    \addtocontents{toc}{\begingroup\protect\small}%
    \AtEndDocument{\addtocontents{toc}{\endgroup}}%
    }
\makeatother

\renewcommand{\cross}[1]{\hyperref[#1]{\textbf{#1}}}

\newcommand{\capability}{\doublecross{cap}{capability}\-\doublecross{ability}{capability}}
\newcommand{\capabilities}{\doublecross{cap}{capability}\-\doublecross{abilities}{capability}}
\newcommand{\dependency}{\doublecross{dep}{dependency}\-\doublecross{en}{dependency}\-\doublecross{dency}{dependency}}
\newcommand{\dependencies}{\doublecross{dep}{dependency}\-\doublecross{en}{dependency}\-\doublecross{dencies}{dependency}}
\newcommand{\backend}{\doublecross{back}{backend}\-\doublecross{end}{backend}}
\newcommand{\backends}{\doublecross{back}{backend}\-\doublecross{ends}{backend}}
\newcommand{\modulefunction}{\doublecross{module}{module function} \doublecross{function}{module function}}
\newcommand{\modulefunctions}{\doublecross{module}{module function} \doublecross{functions}{module function}}
\newcommand{\dependencyresolver}{\doublecross{dep}{dependency resolver}\-\doublecross{en}{dependency resolver}\-\doublecross{dency}{dependency resolver} \doublecross{re}{dependency resolver}\-\doublecross{solver}{dependency resolver}}
\newcommand{\dependencytree}{\doublecross{dep}{dependency tree}\-\doublecross{en}{dependency tree}\-\doublecross{dency}{dependency tree} \doublecross{tree}{dependency tree}}

\preprintnumber{NORDITA 2017-078, CoEPP-MN-17-7}

\title{SpecBit, DecayBit and PrecisionBit: GAMBIT modules for computing mass spectra, particle decay rates and precision observables}

\author{The GAMBIT Models Workgroup:
Peter Athron\thanksref{inst:a,inst:b, e1} \and
Csaba Bal\'azs\thanksref{inst:a,inst:b} \and
Lars A.\ Dal\thanksref{inst:c} \and
Joakim Edsj\"o\thanksref{inst:d,inst:e} \and
Ben Farmer\thanksref{inst:d,inst:e, e2} \and
Tom\'as E.\ Gonzalo\thanksref{inst:c} \and
Anders Kvellestad\thanksref{inst:f} \and
James McKay\thanksref{inst:g} \and
Antje Putze\thanksref{inst:h} \and
Chris Rogan\thanksref{inst:i} \and
Pat Scott\thanksref{inst:g, e3} \and
Christoph Weniger\thanksref{inst:j} \and
Martin White\thanksref{inst:k,inst:b}}

\institute{%
  \monash\label{inst:a} \and
  \coepp\label{inst:b} \and
  \oslo\label{inst:c} \and
  \okc\label{inst:d} \and
  \su\label{inst:e} \and
  \nordita\label{inst:f} \and
  \imperial\label{inst:g} \and
  \annecy\label{inst:h} \and
  \harvard\label{inst:i} \and
  \grappa\label{inst:j} \and
  \adelaide\label{inst:k}
}

\titlerunning{SpecBit, DecayBit and PrecisionBit}
\authorrunning{GAMBIT Models Workgroup}

\thankstext{e1}{peter.athron@coepp.org.au}
\thankstext{e2}{benjamin.farmer@fysik.su.se}
\thankstext{e3}{p.scott@imperial.ac.uk}

\date{Received: 16 March 2017 / Accepted: 16 November 2017}

\begin{document}

\maketitle

\begin{abstract}
We present the \GB modules \specbit, \decaybit and \precisionbit.
Together they provide a new framework for linking publicly available
spectrum generators, decay codes and other precision observable
calculations in a physically and statistically consistent manner.
This allows users to automatically run various combinations of
existing codes as if they are a single package. The modular
design allows software packages fulfilling the same
role to be exchanged freely at runtime, with the results presented in
a common format that can be easily passed to downstream dark
matter, collider and flavour codes.  These modules constitute an
essential part of the broader \GB framework, a major new software
package for performing global fits.  In this paper we present the
observable calculations, data, and likelihood functions implemented in
the three modules, as well as the conventions and assumptions used in
interfacing them with external codes.  We also present \threebit, a
command-line utility for computing mass spectra, couplings, decays and
precision observables in the MSSM, which shows how the three modules
can be easily used independently of \GB.

\end{abstract}

\tableofcontents

\section{Introduction}

Run II of the Large Hadron Collider (LHC) is engaged in a wide-ranging
search for evidence of physics Beyond the Standard Model (BSM).  Such
models typically have large parameter spaces, so understanding
their phenomenology requires detailed calculations using computer
programs.  Phenomenological studies therefore often involve a large
chain of public codes that must be linked together.

This set of codes includes spectrum generators (to determine the
masses of new particles), decay
calculators (to obtain decay widths), and
packages capable of predicting low-energy precision
observables, such as the anomalous magnetic moment of the muon.  These
codes need to be linked together in such a way that information from
the spectrum generator can be passed to the other calculators, and
their outputs can in turn be used in other programs.

For the Minimal Supersymmetric Standard Model
(MSSM) and the Next-to-minimal Supersymmetric Standard Model (NMSSM), there are the SLHA \cite{Skands:2003cj} and
SLHA2 \cite{Allanach:2008qq} conventions, which simplify matters somewhat.  However,
even in these cases it can be far more convenient to have
the programs automatically linked, as can be testified by the
popularity of packages that incorporate individual codes, such as
\SUSYHIT \cite{Djouadi:2006bz} for the MSSM and
\NMSSMTools \cite{Ellwanger:2004xm,Ellwanger:2005dv,Ellwanger:2006rn}
for the NMSSM.  On the other hand, keeping
individual codes distinct in a modular framework allows for easier isolation of the origin of
differences in results, and for the ability to mix and match different
codes according to preference or the specific advantages of one tool or
another.  Indeed, the need to compare several
software packages for cross-checks, catching bugs and revealing
uncertainties in each calculation has been demonstrated on numerous
occasions (e.g.~Refs.\ \cite{Allanach:2003jw,Allanach:2004rh,Staub:2015aea,Drechsel:2016htw}).

Here we present a framework that provides the best of both worlds,
designed to work with all models including both
non-supersymmetric models and exotic SUSY models beyond the MSSM and
NMSSM.  The framework consists of three packages: \specbit, for handing
renormalisation group running and calculation of mass spectra, \decaybit, for computing branching
ratios and widths, and \precisionbit, for calculating other precision observables.  All three packages are interfaced in a common way
within \gambitVer \cite{gambit}, allowing specific functions and external software to be exchanged at runtime, whilst
being run from exactly the same input parameters under the same physical conventions. The user interface
is designed to be as simple and general as possible, so that information can be extracted and inserted in an intuitive way.

\GB (the Global and Modular Beyond-SM Inference Tool) is a multi-purpose physics tool for performing parameter scans and global fits in BSM models using either frequentist or Bayesian statistics.  Each of the three packages we describe here is a \cross{module} within the broader \GB framework.  Here we \doublecross{highlight}{glossary} terms that may be considered \GB jargon, and provide a simple summary of their meanings in the \doublecross{glossary}{glossary} (Appendix \ref{glossary}).  Each module collects a series of \modulefunctions{} based around a common theme.  Each module function typically calculates a single observable or likelihood component for use in a fit, or performs some calculation that is needed by another module function in order to eventually arrive at the value of an observable or likelihood.  Module functions that require the results of other module functions can declare \dependencies{} on the physical or mathematical quantities that they require.  \GB defines a series of theoretical models available for analysis, and various rules that relate the different module functions to each other, to models, and to functions that are available for modules to call from external physics codes, known as \backends{}.  At runtime, the \GB \dependencyresolver{} identifies the module and backend functions required to compute the observables and likelihoods requested by a user, and arranges them into a \dependencytree{}.  It then uses methods from graph theory to `solve' the tree, and determine the order in which the functions must be called so that all dependencies and model consistency requirements are guaranteed to be satisfied. \GB's sampling module \scannerbit \cite{ScannerBit} runs the user's choice of sampling algorithm on the solved dependency tree, and saves the resulting parameter samples, derived observables and likelihoods in an output database for subsequent statistical analysis and plotting.

\specbit, \decaybit and \precisionbit are used to compute and return spectra, decay widths and precision observables, run associated \backend{} codes, interpret the results thus obtained, and provide them to other \doublecross{modules}{module} as required for subsequent calculations. Other \GB modules calculate dark
matter observables (\darkbit \cite{DarkBit}), high-energy collider
signatures (\colliderbit \cite{ColliderBit}) and quantities from flavour physics (\flavbit \cite{FlavBit}).
An extended description of the structure, features and abilities of \GB can be found in Ref.\ \cite{gambit}.

\specbit, \decaybit and \precisionbit are designed so that a user may easily add new models and
interfaces to new backends. Models that have already been implemented
in \GB are automatically supported.  The first release features various incarnations of the MSSM and the scalar singlet dark matter model, as these
are the models on which the first \GB scans \cite{CMSSM,MSSM,SSDM} have been performed.

The three modules have been initially set up to exploit several external codes\footnote{When any of these external codes are used as part of \gambit or the \specbit, \decaybit and \precisionbit modules, the references for that code listed here should be cited along with this manual. A full list of references associated with all external codes currently utilised as \GB backends can be found in the file \term{README.md} in the main directory of the \GB source code.}:
\begin{itemize}
\item \FlexibleSUSY \cite{Athron:2014yba} and \SPheno \cite{Porod:2003um,Porod:2011nf} for spectrum generation, performing calculations in the dimensional reduction ($\overline{DR}$) scheme.
\item \FeynHiggs \cite{Heinemeyer:1998yj,Heinemeyer:1998np,Degrassi:2002fi,Frank:2006yh,Hahn:2013ria,Bahl:2016brp} for additional Higgs and $W$ mass calculations in a mixed
$\overline{DR}/$on-shell (OS) scheme.
\item \HDECAY\footnote{We interface with \SDECAY and \HDECAY via
\SUSYHIT \cite{Djouadi:2006bz}.\addtocounter{footnote}{-1}}\cite{Djouadi:1997yw,Spira:1997dg,Butterworth:2010ym} and \FeynHiggs \cite{Heinemeyer:1998yj,Heinemeyer:1998np,Degrassi:2002fi,Frank:2006yh,Hahn:2013ria} for Higgs decays.
\item \SDECAY \footnotemark \cite{Muhlleitner:2003vg} for sparticle decays
\item \FeynHiggs, \superiso \cite{Mahmoudi:2007vz,Mahmoudi:2008tp,Mahmoudi:2009zz} and \gmtwocalc \cite{gm2calc} for calculation of additional precision observables.
\end{itemize}

The models and backends supported by \specbit, \decaybit and \precisionbit will be continually updated, and
are expected to grow rapidly. There are immediate plans to add interfaces to the spectrum generators
\SOFTSUSY \cite{Allanach:2001kg,Allanach:2009bv,Allanach:2011de,Allanach:2014nba} and
\SUSPECT \cite{Djouadi:2002ze} for the MSSM, next-to-minimal \SOFTSUSY \cite{Allanach:2013kza},
\NMSSMCalc \cite{Ender:2011qh,Graf:2012hh,Baglio:2013iia,King:2015oxa}
and \NMSPEC \cite{Ellwanger:2006rn} for the NMSSM, and many additional models via \FlexibleSUSY \cite{Athron:2014yba} and
\SARAH\ / \SPheno \cite{Staub:2008uz,Staub:2010jh,Staub:2012pb,Staub:2013tta,Goodsell:2014bna,Porod:2011nf}.
For decays in the MSSM, \HFOLD \cite{Frisch:2010gw} and
\SFOLD \cite{Hlucha:2011yk} could also be added. We are also adding \susyhd \cite{arXiv:1504.05200} as a first example for a forthcoming \Mathematica\footnote{\href{http://www.wolfram.com/mathematica/}{http://www.wolfram.com/mathematica}} interface and we will also add similar pure and hybrid effective field theory calculations of \HSSUSY \cite{Bagnaschi:2015pwa,Athron:2016fuq} and \FlexibleEFTHiggs \cite{Athron:2016fuq} from \FlexibleSUSY.  NMSSM decays will be obtainable from
\NMSSMCalc \cite{Ender:2011qh,Graf:2012hh,Baglio:2013iia,King:2015oxa} and \NMHDECAY \cite{Ellwanger:2004xm,Ellwanger:2005dv}.  Decays in other
models can be added from \SARAH\ / \SPheno and the
upcoming \textsf{FlexibleDecay} extension of \FlexibleSUSY.

In Sec.~\ref{Sec:SpecBit} we describe the module \specbit.  This
includes how \specbit manages the running of spectrum generators, how
to access and use the information it extracts from them, and how it
can be used to analyse vacuum stability.  Next we provide a detailed
description of the module
\decaybit (Sec.~\ref{Sec:DecayBit}), including the decay data it contains, decay calculations it performs internally, and its use of backends.
In Sec.~\ref{Sec:PrecBit} we detail \precisionbit, including its likelihood functions and interfaces to external precision codes.  Sec.\ \ref{Sec:threebit} gives some examples of each of the modules in action within \GB, and presents \threebit, a simple utility that performs a similar function to \SUSYHIT, and serves as a basic example of the standalone use of \specbit, \decaybit and \precisionbit outside of the \GB framework.  We give a brief summary in Sec.\ \ref{sec:summary}. In  Appendix~\ref{Sec:PhysBack} we review the physics of the models discussed in this paper, and the conventions that we adopt for them.  We provide details of the interface with \SPheno in Appendix \ref{app:SPheno_backend}, explicit documentation of some of the classes involved in \specbit in Appendices \ref{app:subspec_contents}--\ref{app:gettermaps}, instructions for adding new spectrum generators to \specbit in Appendix \ref{sec:adding_new_models}, and an explicit example of how to build spectrum classes for new models in Appendix \ref{app:wrapper_worked_examples}.  In Appendix \ref{glossary}, we give the glossary of \GB terms highlighted at various points in this paper.

\specbit, \decaybit and \precisionbit are released under the standard 3-clause BSD license\footnote{\href{http://opensource.org/licenses/BSD-3-Clause}{http://opensource.org/licenses/BSD-3-Clause}.  Note that \textsf{fjcore} \cite{Cacciari:2011ma} and some outputs of \flexiblesusy \cite{Athron:2014yba} (incorporating routines from \SOFTSUSY \cite{Allanach:2001kg}) are also shipped with \GB \textsf{1.0}.  These code snippets are distributed under the GNU General Public License (GPL; \href{http://opensource.org/licenses/GPL-3.0}{http://opensource.org/licenses/GPL-3.0}), with the special exception, granted to \GB by the authors, that they do not require the rest of \GB to inherit the GPL.}, and can be downloaded from \href{http://gambit.hepforge.org}{gambit.hepforge.org}.

\section{SpecBit}
\label{Sec:SpecBit}
\begin{figure}[t]
\centering
\includegraphics[width=\columnwidth]{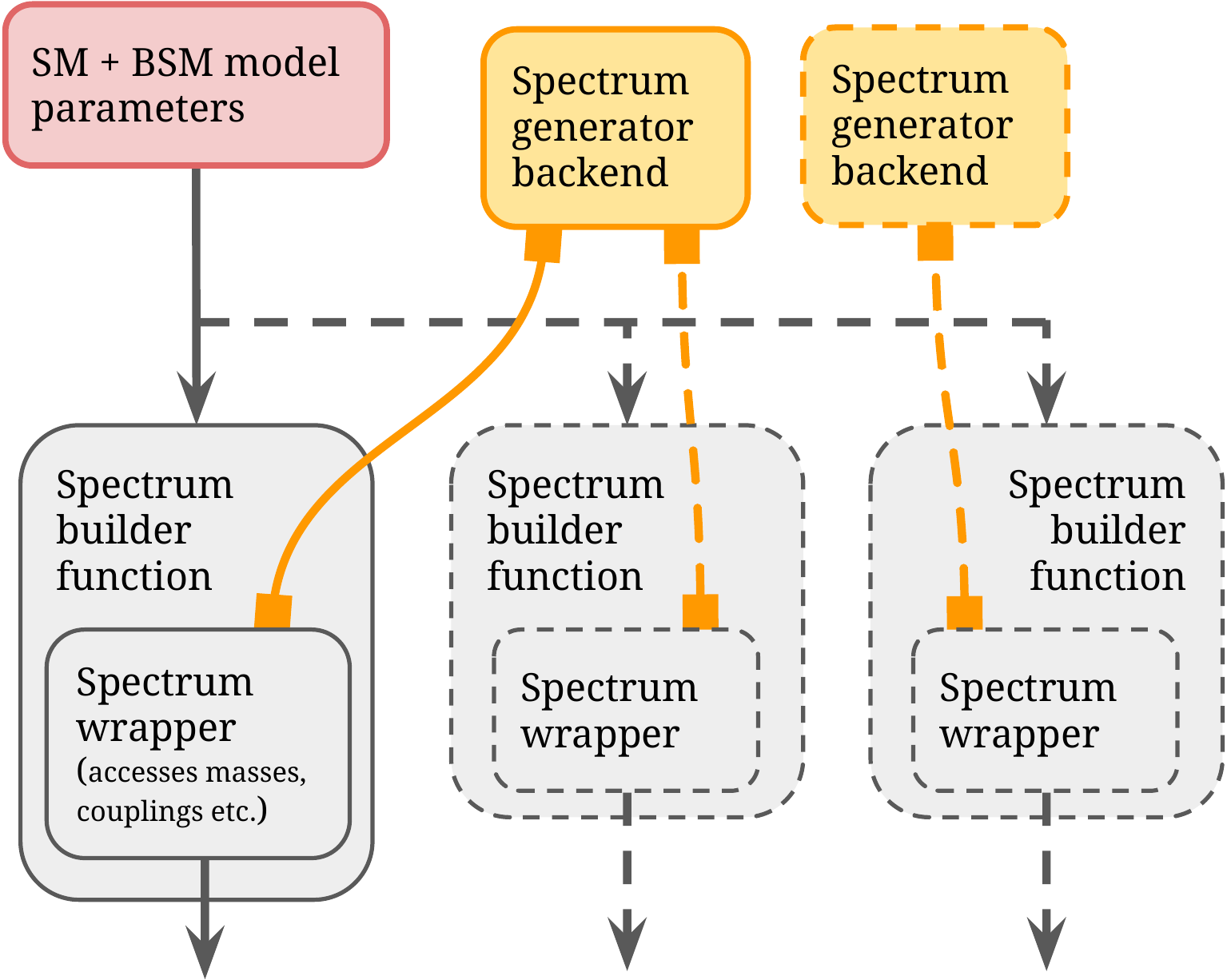}
\caption{Schematic representation of the structure of \specbit. Model parameters {\color[HTML]{D64655}(red box)} are obtained from the \GB Core and fed to \specbit module functions {\color[HTML]{7A7A7A}(grey boxes)}. These module functions run spectrum generator backend codes {\color[HTML]{FA9800}(yellow boxes)} and embed the results within \cpp{Spectrum} wrapper objects {\color[HTML]{7A7A7A}(inner grey boxes)}, which carry the spectrum information out to other \GB functions that require it. In some cases the backend code remains connected to the wrapper object, allowing RGEs to be called to translate couplings and mass parameters to other scales. In this diagram solid outlines indicate ``active'' elements for a hypothetical scan, while dashed outlines indicate inactive elements. The ``active'' elements are those that are activated by the central \GB dependency resolution system, or manually run in a standalone code. The ``inactive'' elements represent alternate calculation pathways not required in a scan, but can be switched on instead if the user chooses.
\label{fig:specbit_overall}
}
\end{figure}

All information in \GB about the spectrum of particle masses and their
couplings comes from module functions in \specbit.  This includes the
pole masses and mixings of all physical states in the model, scheme
dependent parameters, such as those defined
in \DR \cite{Siegel:1979wq,Capper:1979ns,Jack:1994rk}
or \MSbar \cite{Bardeen:1978yd} schemes\footnote{Note that for a given
model, if a user wants to support backend codes that require input
parameters in a different scheme to the one implemented in \specbit,
then the user must add a function that converts between the two
schemes. Consistency can then be ensured by specifying a
scheme-specific spectrum as a \dependency{}.  Currently there are no
conversions between different schemes in \specbit, because all
backends and native calculations can accept \DR and \MSbar parameters
as inputs.}, Higgs couplings and basic SM inputs (such as the top quark pole
mass).  In BSM models where some of this required information is a
calculable prediction of the model, \specbit will obtain it by calling
a spectrum generator, taking inputs as model parameters.  In cases
where the information is not a prediction, but is already specified
directly by the free parameters of the model (e.g.\ if a pole mass is
defined as a model parameter), \specbit will simply take this
information from the model parameters and store it in a similar way to
the information extracted from the spectrum generator. Spectrum data
is then stored inside \cpp{Spectrum} wrapper objects, which transport
the data to other \GB module functions that request it. A general
overview of this process is shown in Figure \ref{fig:specbit_overall}.

Because some calculations require the values of running parameters at a
particular scale, \specbit also allows the \DR or \MSbar
parameters to be run to a scale chosen locally in any module function, using a relevant spectrum generator backend.

\subsection{Supported models and spectrum generators}

The first release of \specbit is extendable to any model, but has built-in support for the scalar singlet dark matter model, and the MSSM.  It also provides a low-energy spectrum object containing SM information.

\subsubsection{Standard model spectrum}
\GB is designed to fit models of BSM physics.  There is thus no full spectrum generator implemented for the SM. However, BSM spectrum generators typically rely on SM inputs as low energy boundary conditions.  To store this information, \specbit contains a \cpp{QedQcd} spectrum object, which was originally part of \SOFTSUSY and is also used in \FlexibleSUSY.  This is the source of some of the SM data that can be extracted from \specbit\footnote{The SM data extractable from \specbit are shown in Table \ref{tab:SM_contents}.}  and details of the relevant \capabilities{} are given in Sec.~\ref{sec:SM_options} and Table \ref{tab:specbitsmcap}.  More details on this \cpp{QedQcd} object can be found in the \SOFTSUSY manual \cite{Allanach:2001kg}.

\subsubsection{Spectrum generators for the MSSM}
\label{Sec:MSSMspecgens}
MSSM mass spectra are typically obtained by finding solutions to the
renormalisation group equations (RGEs) that simultaneously satisfy
boundary conditions (BCs) at high and low scales, before using self
energies to calculate the pole masses.  The low-scale BC matches the
spectrum to observed SM data, and the high-scale BC places constraints
on the soft SUSY-breaking masses.

The first version of \specbit comes with interfaces to two different
spectrum generators: \FlexibleSUSY and \SPheno.  It also contains an
interface to \FH, which can be used to obtain Higgs and sparticle pole masses;
details of our interface to \FH are given in Sec.\ \ref{decays:MSSM}.
\FlexibleSUSY creates a spectrum generator for a given model, defined in a \SARAH input
file, and uses a \FlexibleSUSY model file to specify the BCs for that
model. \SPheno runs in different modes according to the input BCs.

The model is defined by both a set of input parameters that specify
$\overline{DR}$ parameters of the MSSM, and an additional scale at
which the soft SUSY-breaking $\overline{DR}$ parameters are defined.  Note that when there are
constraints relating the parameters to each other, varying this scale
is not equivalent to a reparametrisation, as there will be different
mass splittings that cannot be reproduced by any point in parameter space
when the constraints are applied at a different scale.

\specbit currently implements several spectrum generators
using \FlexibleSUSY \textsf{1.5.1} (with \SARAH\ \textsf{4.9.1}).  In
the spectrum generator \cpp{MSSMatMGUT}, the soft parameters are
defined at the scale where the gauge couplings unify, which is
determined iteratively. In \cpp{MSSMatQ}, the soft parameters are
instead defined at a user-specified scale $Q$.  The \cpp{MSSMatMGUT}
and \cpp{MSSMatQ} spectrum generators support
the \textsf{MSSM63atMGUT} and \textsf{MSSM63atQ} models
respectively. These models represent the most general formulation of
the $CP$-conserving MSSM with the couplings given in
Eq.\ \ref{trilinear_C_couplings} set to
zero (see Sec.\ \ref{sec:MSSMLagrangian} for the full
MSSM Lagrangian). \textsf{MSSM63atMGUT} is defined at
the scale where the gauge couplings unify, and \textsf{MSSM63atQ} is defined at a user-specified
scale $Q$.  These are currently the most general SUSY models
of the \GB model hierarchy (which is described in Sec.\ 5 of
Ref.\ \cite{gambit}).

More constraining BCs create lower-dimensional subspaces of these more
general parameter spaces, so a vast number of possible subspaces
exist.  \specbit uses the \GB model hierarchy to relate these subspaces
to the \textsf{MSSM63atMGUT} or
\textsf{MSSM63atQ} models, and can use the \cpp{MSSMatMGUT} or
\cpp{MSSMatQ} spectrum generators to determine their mass spectra. It
is also possible to directly implement the boundary condition in
a \FlexibleSUSY spectrum generator.  As an example of this, we have
also implemented a specific \cpp{CMSSM} spectrum generator.
This was mostly introduced as a basic check of the model
hierarchy, but remains in \GB for convenience.

Technical details of how
the \FlexibleSUSY spectrum generators are implemented are given in
documentation shipped with the code
(see \term{doc/Adding_FlexibleSUSY_Models.txt}), though an
illustrative example is given in Appendix \ref{app:FS_MSSM_wrapper},
and details required to use them in scans are given in
sections \ref{sec:MSSM_options} and \ref{sec:FS_options}. In the
additional documentation, we also demonstrate how to add
an MSSM variant where soft parameters are fixed at the SUSY scale (\cpp{lowMSSM} in the \FlexibleSUSY naming scheme),
where the SUSY scale is defined as the geometric mean of the $\overline{DR}$
stop masses and is determined iteratively. We will include this variant natively in
the next version of \specbit.

In the case of \SPheno, the running mode is determined by the initialisation of its internal variables. In contrast to \FlexibleSUSY, the available modes are triggered according to the model being scanned (\textsf{CMSSM}, or one of the \textsf{MSSMatMGUT} or \textsf{MSSMatQ} models), not by the usage of different functions. Other running modes available in the out-of-the-box version of \SPheno, such as the NMSSM, GMSB, AMSB, etc., are not covered in the backend version of the software used in \GB but, as mentioned before, there are immediate plans to include these in future releases. The specific details of how to use \SPheno as a spectrum generator in \GB are provided in Secs.\ \ref{sec:MSSM_options} and \ref{sec:SPheno_options}.  A comprehensive description of the backend system can be found in the main \GB paper \cite{gambit}, and in Appendix \ref{app:SPheno_backend} for the specific case of \SPheno.

For each of these spectrum generators the EWSB conditions are used to
fix $|\mu|$ and $b$, so that the Higgs vacuum expectation values (VEVs) are fixed by the measured $Z$
boson mass and the input ratio of the two VEVs
($\tan \beta= \frac{v_u}{ v_d}$).

The pole masses and mixings calculated by the spectrum generator are
stored internally in the \GB\ \cpp{Spectrum} wrapper in a format that
follows SLHA2 conventions \cite{Allanach:2008qq} allowing full family
mixing for sfermions.

\subsubsection{Spectrum generators for the scalar singlet dark matter model} \label{sec:scalar_singlet}

The scalar singlet mass spectrum is available in two forms. One is a
via a simple container object that is set up without any spectrum generation,
in other words the relevant pole masses and coupling, which govern the new physics effects at the TeV scale,
are input directly from scanner-generated parameters. The other option is a spectrum object that interfaces with \FlexibleSUSY to
calculate the spectrum with full RGE running capabilities.

The simple container spectrum is the most efficient
option for a range of calculations in \GB that re-
quire only the masses and coupling at a fixed scale. In
this case, the only new model parameters required for
other module functions are $m_S$ and $\lambda_{hs}$. In the container object
these parameters are stored directly from the \GB model parameters.

If radiative corrections and/or renormalisation group running of model
parameters is necessary, then
a fully-calculable spectrum is required.  For this purpose, we offer a
spectrum object that uses \FlexibleSUSY and the input parameters to calculate pole
masses and couplings.  This spectrum object also has the capacity to evolve
parameters between different scales using RGEs.  This spectrum object interprets the input model
parameters, in particular the masses of the scalar singlet ($m_S$) and
the Higgs boson ($m_h$), as \MSbar quantities, while all other SM
masses are taken according to SLHA2
conventions \cite{Allanach:2008qq}.  These running parameters are
defined at the scale $m_Z$.  The EWSB conditions are then imposed to
calculate the value of the quartic Higgs coupling $\lambda$ (Eq.~\ref{Eq:SM_Lagrangian}), and pole masses are calculated at a user-specified
scale $Q$.  The scale $Q$ may be varied as a nuisance parameter
in the \GB model \yaml{SM_Higgs_running} (see Ref.\ \cite{gambit} for more
details), or left as the top mass by default.  Optimally, this scale
should be set to minimise logarithmic contributions to the RGE $\beta$ functions with the largest
coefficient.

\subsection{User interface and options}
\label{user_interface}

\subsubsection{General settings}
\label{sec:general_settings}
In this section we describe how to use \specbit to run available
spectrum generators via a \GB input \YAML file.\footnote{\YAML is the language that \GB input files are written in; see \href{http://www.yaml.org/}{www.yaml.org}.} At the
most rudimentary level, spectrum information can simply be
written to disk for one model point in an SLHA2-like format,
using \textsf{SLHAea}.\footnote{\textsf{SLHAea} is a \Cpp class allowing internal representation of SLHA files; see \href{http://fthomas.github.io/slhaea}{fthomas.github.io/slhaea}.} At a
more advanced level, the information can be written to disk for every
point in a \GB scan, via the
\GB printer system (for details, see Sec.\ 9 of the \GB Core paper and manual \cite{gambit}), and analysed using external software.
More advanced usage, such as accessing spectrum information at
the \Cpp level (in a \GB module function, for example), is described
in Sec.\ \ref{sec:APIs}.  For details on \YAML usage of the \capabilities{} and associated options described here, please see the \GB manual \cite{gambit} or the \specbit example files in the \term{yaml_files} directory of the \GB source tree. A \term{README} file can be found in this directory, which explains the example files further.

\subsubsection{Standard model} \label{sec:SM_options}

The \capabilities{} available in \specbit related to the SM are given in Table \ref{tab:specbitsmcap}.  The \capability{} \cpp{SM_spectrum} is provided by the function
\begin{itemize}[topsep=3pt]
\item[] \cpp{get_SMINPUTS}.
\end{itemize}
This creates an object containing the low-energy SM parameters obtained directly from the model parameters, which must contain the parameters in the \textsf{StandardModel\_SLHA2} \GB model \cite{gambit}.  This can then be used to build BSM spectrum objects that require this low-energy data.

By default, the \cpp{get_SMINPUTS} function populates its $W$ mass with the observed value of 80.385\,GeV \cite{PDB}.  Users who prefer the $W$ mass returned to respect the tree-level relationship with $m_Z$ and $\sin^2\theta_W$ can choose for it to be calculated at tree level instead, by setting the option \yaml{enforce_tree_level_MW = true} for this function.

The module function
\begin{itemize}[topsep=3pt]
\item[] \cpp{get_QedQcd_spectrum}
\end{itemize}
has a \dependency{} on \cpp{SM_spectrum}, and provides the \capability{} \cpp{qedqcd_subspectrum}.  This  function creates an effective \cpp{QedQcd} object. The returned \cpp{QedQcd} object contains the low energy data, where all running quantities from the \cpp{SMinputs} \dependency{} are now given at the scale $m_Z$.  This is then wrapped along with the original \cpp{SMinputs} into a \cpp{SubSpectrum} object.

Finally, to provide the \capability{} \cpp{SM_spectrum}, the module function
\begin{itemize}[topsep=3pt]
\item[] \cpp{get_SM_spectrum}
\end{itemize}
wraps the \cpp{qedqcd_subspectrum} object into a full spectrum object along with a simple container for the Higgs pole mass and vacuum expectation value, providing a complete SM spectrum.

The \capability{} \cpp{Higgs_Couplings} is described in Sec.\ \ref{sec:higgs_couplings}.
\subsubsection{Scalar singlet dark matter} \label{sec:singletDM_options}

The \capabilities{} available in \specbit that either produce or depend
on the scalar singlet spectrum are given in
Table \ref{tab:specbitsingletdmcap}.

\specbit has a \capability{} \cpp{SingletDM_spectrum} to provide essential details about the scalar singlet spectrum.   This \capability{} is provided by the module functions
\begin{itemize}[topsep=3pt]
\item[] \cpp{get_SingletDM_spectrum_simple} and
\item[] \cpp{get_SingletDM_spectrum_FS},
\end{itemize}
both of which return
 a \cpp{Spectrum} object. The former returns the simple container spectrum, as described in
 Sec.~\ref{sec:parameter_box_wrapper}, and the latter returns a more
 sophisticated spectrum, computed with \FlexibleSUSY, which can run parameters
 as described in Sec.~\ref{sec:subspectrum_structure}.  All options described in
 Sec.~\ref{sec:FS_options} can be given to this function.

If an error is encountered during initial spectrum generation, a \FlexibleSUSY error is passed to \GB. The handling of such an error is controlled by the option \yaml{invalid_point_fatal}; if this option is set to \cpp{true}, invalid spectra will trigger a \specbit error.  It is a common consideration to check for
perturbativity of the dimensionless couplings up to a specific high energy scale \cite{Belanger2013a,Alanne2014,Khan2014}. We provide the ability to do this with the option \yaml{check_perturb}.  The maximum scale to run the couplings up to is set with the option \yaml{check_high_scale}.  If the couplings are found to be non-perturbative, then an invalid point exception is raised as if the initial spectrum calculation encountered an error.  The options, input types and default values for the \cpp{get_SingletDM_spectrum_FS} function are given in the list below.
\begin{itemize}
\item \yaml{check_perturb}: takes a \cpp{bool} to demand that the spectrum be run to \yaml{check_high_scale}. Default \cpp{false}.
\item \yaml{check_high_scale}: takes a \cpp{double} to define the scale that the couplings are run to (given in units of GeV) after spectrum generation. Default $M_{\text{Pl}}=1.22\times 10^{19}$\,GeV.
\item \yaml{invalid_point_fatal}: kill the whole scan if a \FlexibleSUSY error is encountered during spectrum generation. Default \cpp{false}.
\item \FlexibleSUSY options: see Sec.~\ref{sec:FS_options}.
\end{itemize}

The module function
\begin{itemize}[topsep=3pt]
\item[] \cpp{get_SingletDM_spectrum_as_map}
\end{itemize}
also provides
\capability{} \cpp{SingletDM_spectrum}, but as a \Cpp \cpp{map} (aliased as \cpp{map_str_dbl}).
For this function to run, the \cpp{SingletDM_spectrum} needs to be provided as a
\cpp{Spectrum} object. In other words, this function just translates between the \cpp{Spectrum} type
and a \Cpp map. The main use of this function is to print the contents of
the spectrum to an output stream for each data point during a scan (as the \cpp{map_str_dbl} type is
``printable'' but the \cpp{Spectrum} type isn't), as shown in Fig. \ref{fig:specbit_typical}.

There are also some additional \capabilities{} related to vacuum
stability in the scalar singlet model: \cpp{vacuum_stability}, \cpp{check_perturb_min_lambda}, \cpp{VS_likelihood}
and \cpp{expected_lifetime}. These are discussed in detail in
the advanced usage example in Sec.~\ref{sec:vs_code}.

\subsubsection{MSSM} \label{sec:MSSM_options}
\begin{figure}[t]
\centering
\includegraphics[width=\columnwidth]{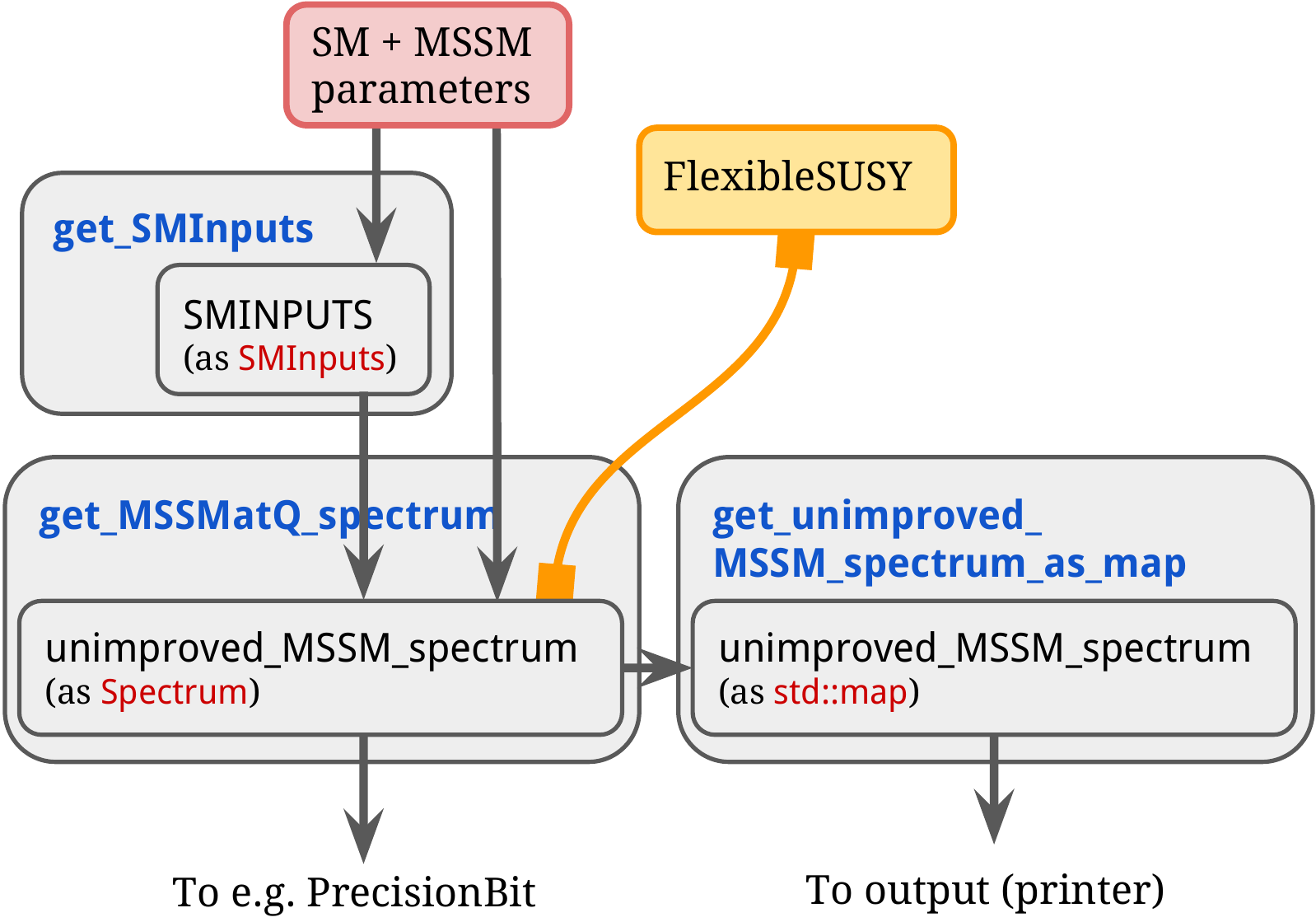}
\caption{A typical flow of information through \specbit, using MSSM module functions as examples. First, model parameters {\color[HTML]{D64655}(red box)} are obtained from the \GB Core. Some of these are used to construct Standard Model input information for the spectrum generation (\cpp{SMINPUTS}, {\color[HTML]{7A7A7A}upper grey box}), while MSSM parameters are passed directly to the spectrum builder function (\cpp{get_MSSMatQ_spectrum}, {\color[HTML]{7A7A7A}lower left grey box}). This function constructs a \cpp{Spectrum} object, which wraps information obtained from \FlexibleSUSY{} {\color[HTML]{FA9800}(yellow box)}, and passes it on to other parts of \GB. In addition, the \cpp{Spectrum} object is translated into a \Cpp map {\color[HTML]{7A7A7A}(lower right grey box)}, which can be parsed by the \GB printer system and written to disk.
\label{fig:specbit_typical}
}
\end{figure}

The \capabilities{} available in \specbit relevant for the MSSM model are given in Table \ref{tab:specbitmssmcap}.

\specbit has a number of module functions that provide the \capability{} \cpp{unimproved_MSSM_spectrum}. We will start by examining three of them:
\begin{itemize}[topsep=3pt]
\item[] \cpp{get_MSSMatMGUT_spectrum},
\item[] \cpp{get_MSSMatQ_spectrum}, and
\item[] \cpp{get_CMSSM_spectrum}.
\end{itemize}
These module functions provide the \cpp{unimproved_MSSM_spectrum} \capability{} in the form of a \cpp{Spectrum} object by calling one of the \FlexibleSUSY spectrum generators described in Sec.~\ref{Sec:MSSMspecgens}.  The module functions all depend on \cpp{SMINPUTS} provided in the form of an \cpp{SMInputs} structure. These spectra are of the most sophisticated type, providing full two-loop RGE running, as described in Sec.~\ref{sec:subspectrum_structure}. Options can be supplied to \FlexibleSUSY via the input \YAML file, and are discussed in Sec.~\ref{sec:FS_options}.

Additionally the module function
\begin{itemize}[topsep=3pt]
\item[] \cpp{get_MSSM_spectrum_SPheno}
\end{itemize}
also provides the \capability{} \cpp{unimproved_MSSM_spectrum}, but in this case obtains the information from a different spectrum generator: \SPheno. As with the module functions that call \FlexibleSUSY, this module function depends on an \cpp{SMInputs} object and provides a \cpp{Spectrum} object, selecting the specific model and running mode of the backended \SPheno at runtime. In contrast to the \FlexibleSUSY case, however, this \cpp{Spectrum} object does not provide RGE running. Is is simply a static spectrum equivalent to the contents of an SLHAea object (\cpp{SLHAstruct}), which is a simple container of SLHA-like information (described in more detail in Sec. \ref{sec:subspectrum_structure} under {\em SLHAea output}). The options available for the backend version of \SPheno used by this module function are in Sec.\ \ref{sec:SPheno_options}.

The module function
\begin{itemize}[topsep=3pt]
\item[] \cpp{get_MSSM_spectrum_from_SLHAfile}
\end{itemize}
also provides this \cpp{unimproved_MSSM_spectrum} \capability{} in the form a \cpp{Spectrum} object.  However, instead of using a spectrum generator to do this it uses an SLHA file.  The \cpp{Spectrum} object in this case does not provide the ability for RGE evolution, but is rather a simple container spectrum. This function takes the options
\begin{itemize}
\item \yaml{filenames}: path to SLHA file to be read in. Can be a list of many file names, in which case each file will be read in sequentially.
\item \yaml{cycles}: number of loops over the \yaml{filenames} list to allow (default \yaml{-1} indicates no limit). When the limit is hit an error will be raised, stopping the scan.
\end{itemize}
This function is mainly for debugging purposes, because one usually generates spectra on the fly during a scan.

Two further \specbit module functions,
\begin{itemize}[topsep=3pt]
\item[] \cpp{get_unimproved_MSSM_spectrum_as_SLHAea} and
\item[] \cpp{get_unimproved_MSSM_spectrum_as_map}
\end{itemize}
are available for translating a \cpp{Spectrum} object to either an SLHAea object (\cpp{SLHAstruct})
or to a \Cpp\ \cpp{map} (as in the analogous scalar singlet DM function), respectively.  These still represent the
spectrum, just with a different type, so the module functions have
both a \dependency{} and a \capability{} of \cpp{unimproved_MSSM_spectrum},
but with different types. See Fig. \ref{fig:specbit_typical} for an example use case of this `translation' behaviour. We have found each format useful for different tasks. For example, some module functions in other parts of \GB request spectrum information in \cpp{SLHAstruct} format, rather than as full \cpp{Spectrum} objects, because these functions work closely with \backend{} codes that were originally designed to work via SLHA files. So keeping the format directly parallel to SLHA was convenient for them. On the other hand, the \cpp{std::map} format is useful because at present neither \cpp{Spectrum} objects nor \cpp{SLHAstruct} objects can be directly written to disk by the \GB printer system \cite{gambit}, so converting to \cpp{std::map} is necessary as an intermediate step for this purpose.

The module functions
\begin{itemize}[topsep=3pt]
\item[] \cpp{get_MSSM_spectrum_as_SLHAea} and
\item[] \cpp{get_MSSM_spectrum_as_map}
\end{itemize}
also translate a \cpp{Spectrum}
object into an \textsf{SLHAea} structure or a \cpp{std::map}.  However in these cases
the \capability{} and \dependency{} is \cpp{MSSM_spectrum}, a \capability{}
that can also be provided by the module functions from \precisionbit,
as shown in module function Table \ref{tab:precisionbit:BE} and
discussed in Sec.~\ref{precisionspectrum}.

In the MSSM it is often useful to be able to identify which Higgs boson
is most similar to the SM Higgs boson.  The module function
\begin{itemize}[topsep=3pt]
\item[] \cpp{most_SMlike_Higgs_MSSM}
\end{itemize}
provides this information by returning a PDG code representing the
$CP$-even Higgs with couplings that are closest to those of the SM Higgs
boson.  We deem the lightest $CP$-even state to be the most SM-like if $\sin(\beta - \alpha)
> \cos(\beta - \alpha)$, where $\alpha$ is the mixing angle defined in
Eq.\ \ref{Eq:alphadef}; otherwise, this function returns the PDG code of the heavier $CP$-even Higgs. This function has
\capability{} \cpp{SMlike_Higgs_PDG_code}, and a \dependency{} on the \cpp{MSSM_spectrum}.

It is also possible to extract the low-energy SM spectrum from the
MSSM spectrum.  The \capability{} \cpp{SM_subspectrum} is
provided by the module function
\begin{itemize}[topsep=3pt]
\item[] \cpp{get_SM_SubSpectrum_from_MSSM_Spectrum},
\end{itemize} which has a \dependency{} on the \cpp{MSSM_spectrum}.

Finally, there are three module functions that require backend functions from \FeynHiggs.  These module functions,
\begin{itemize}[topsep=3pt]
\item[] \cpp{FH_MSSMMasses},
\item[] \cpp{FH_HiggsMasses} and
\item[] \cpp{FH_Couplings},
\end{itemize}
respectively provide MSSM masses with \capability{} \cpp{FH_MSSMMasses}, Higgs masses with capability \cpp{prec_HiggsMasses}, and Higgs couplings with capability \cpp{FH_Couplings_output}.

\subsubsection{FlexibleSUSY options}\label{sec:FS_options}

Any spectrum generator interfaced to \GB will inevitably have its own set of options to control precision and methods of calculation.  Here we briefly summarise the \FlexibleSUSY options, which can be set via \YAML options. For a more detailed explanation of these options, see the \FlexibleSUSY manual \cite{Athron:2014yba}.
\begin{itemize}
\item \yaml{precision_goal}: relative error for the calculation of the $\overline{DR}$ spectrum. Default $1\times 10^{-4}$.
\item \yaml{max_iterations}: maximum number of iterations for the two-scale algorithm. Default $-10\log_{10}$(\yaml{precision_goal}).
\item \yaml{calculate_sm_masses}: compute SM pole masses during spectrum calculation. Default \yaml{false}.
\item \yaml{pole_mass_loop_order}: number of loops in self energies for pole mass calculation. Default 2.
\item \yaml{ewsb_loop_order}: number of loops in radiative corrections to the EWSB conditions in Eq.\ 33 of Ref.\ \cite{Athron:2014yba}. Default 2.
\item \yaml{beta_loop_order}: loop order for RGEs. Default 2.
\item \yaml{threshold_corrections_loop_order}: loop order for threshold corrections as defined in Eqs.\ 6--13 of Ref.\ \cite{Athron:2014yba}. Default 2.
\item Two-loop Higgs pole mass corrections are controlled for specific terms with a boolean input, these are by default all set to \yaml{true}.  The available terms are:
\begin{itemize}
\item \yaml{use_higgs_2loop_at_as}: $\mathcal{O}(\alpha_t \alpha_s)$,
\item \yaml{use_higgs_2loop_ab_as}: $\mathcal{O}(\alpha_b \alpha_s)$,
\item \yaml{use_higgs_2loop_at_at}: $\mathcal{O}(\alpha_t^2+\alpha_t\alpha_b+\alpha_b^2)$,
\item \yaml{use_higgs_2loop_atau_atau}: $\mathcal{O}(\alpha_{\tau}^2)$,
\end{itemize}
\end{itemize}
where for $x=t,b,\tau$, $\alpha_x=y_x^2/(4\pi)$ and $y_x$ is the corresponding Yukawa coupling, and $\alpha_s$ is the strong coupling constant.

Note that the included spectrum generators created by \FlexibleSUSY
perform a fixed-order calculation of the Higgs mass and we do not
currently support additional \FlexibleSUSY codes\footnote{Such options
will be added at a later date, however in the intervening period, an
advanced user may also add these codes themselves, following the
instructions given in this manual.} \HSSUSY and \FlexibleEFTHiggs that
resum large logarithms.  However we refer the reader to
Ref.\ \cite{Athron:2016fuq} for a discussion of how the large
logarithms fortuitously cancel in the fixed order \FlexibleSUSY
calculation, such that the uncertainty is much lower at higher scales
in the MSSM than one would naively expect.

If the \FlexibleSUSY spectrum generator encounters an error during calculation, this may be passed on as a \specbit error, resulting in the termination of a scan, or as an invalid point exception, which will result in the scan point being given an invalid likelihood (an extremely small value defined by \yaml{model_invalid_for_lnlike_below} in the \YAML file). See Ref.\ \cite{gambit} for more details on exception handling.  The management of these exceptions within \GB is controlled with the option \yaml{invalid_point_fatal} discussed in Sec.~\ref{sec:singletDM_options}.

\subsubsection{\SPheno options}\label{sec:SPheno_options}

As with \FlexibleSUSY, there are a number of options available for \SPheno to control certain aspects of the calculation. In \GB only a limited set of all the available options for \SPheno is allowed, because many features or models permitted by \SPheno are not yet covered in \GB. A detailed explanation of all the options available for \SPheno can be found in the manual~\cite{Porod:2011nf}, under the SLHA block \textsf{SPhenoInput}.

The following options can be provided via the \YAML options section in the input \YAML file, and they will be internally assigned to the corresponding variables in \SPheno.

\begin{itemize}
 \item \yaml{SPA_convention}: use the SPA convention~\cite{AguilarSaavedra:2005pw}, effectively setting the parameter output scale to 1\,TeV. Default \yaml{false}.
 \item \yaml{GUTScale}: fixed value for the GUT scale. Default 0.0, which causes the GUT scale to be obtained at runtime.
 \item \yaml{StrictUnification}: force strict unification of the gauge couplings $g_1 = g_2 = g_3$. Default \yaml{false}
 \item \yaml{delta_mass}: relative precision in the calculation of the masses. Default $10^{-6}$.
 \item \yaml{n_run}: maximum number of iterations for the mass calculation. Default 40.
 \item \yaml{TwoLoopRGE}: whether to use two-loop RGEs. Default \yaml{true}.
 \item \yaml{Alpha}: value of the fine structure constant $\alpha(0)$. Default $\tfrac{1}{137}$.
 \item \yaml{gamZ}: value of the decay width of the $Z$ gauge boson. Default $2.49$\,GeV.
 \item \yaml{gamW}: value of the decay width of the $W$ gauge boson. Default $2.04$\,GeV.
 \item \yaml{Use_bsstep_instead_of_rkqs}: use the \yaml{bsstep} algorithm instead of \yaml{rkqs}. Default \yaml{false}.
 \item \yaml{Use_rzextr_instead_of_pzextr}: use the \yaml{rzextr} algorithm instead of \yaml{pzextr}. Default \yaml{false}.
\end{itemize}

Note that the MSSM \SPheno spectrum generator supported currently is
the one immediately available after downloading \SPheno, rather than
including auto-genetated code from \SARAH. It includes only the
fixed-order Higgs mass calculation, so the hybrid EFT / fixed order
calculation based on \FlexibleEFTHiggs \cite{Athron:2016fuq} which has
been developed very recently \cite{Staub:2017jnp} is not currently
available in \specbit.

Errors prompted by \SPheno are collected by the internal integer variable \yaml{kont}, which takes different values according to the source of the error. The specific error messages corresponding to values of \yaml{kont} can be seen in Appendix C of the manual~\cite{Porod:2011nf}. For any of these values, the backend convenience function \cpp{run_SPheno} raises an \cpp{invalid_point} exception, as described in \cite{gambit}.

\subsubsection{Mass cut options}

For any module function that constructs a \cpp{Spectrum} object, it is possible to specify options that enforce user-defined relationships between particle pole masses: \yaml{mass_cut} and \yaml{mass_cut_ratio}. Any spectrum that does not pass the specified cuts is declared invalid (which may invalidate the entire parameter point, depending on whether the likelihood ultimately depends on the spectrum or not \cite{gambit}). While these cuts are not, in general, physically required, we have found this feature useful for certain specialised scans, for example explorations of co-annihilation regions in the MSSM.

These options take one or two particle names recognised by the spectrum, along with two numbers used to define the range to be cut.  They are best explained by way of an example entry in the \yaml{Rules} section of a \GB \YAML file:
\begin{lstyaml}
  - capability: unimproved_MSSM_spectrum
    function: get_CMSSM_spectrum
    options:
      mass_cut: [["h0_1",120,130],["~e-_1",100,1000]]
      mass_ratio_cut: [["~e-_1","|~chi0_1|",1,1.01]]
\end{lstyaml}
The above code invalidates any points that do not have:
\begin{itemize}
 \item the mass of the lightest Higgs between 120 and 130 GeV,
 \item the mass of the lightest slepton between 100 and 1000 GeV, and
 \item the mass of the lightest slepton within 1\% of the mass of the lightest neutralino.
 \end{itemize}
Note in particular the use of absolute value signs for the neutralino mass, which may be negative.  This notation can be used wherever a particle name is given, in order to define the cut rule in terms of the absolute value of the particle mass.  Also note that two such cut rules are given for \yaml{mass_cut}, illustrating the fact that these options can be used to apply arbitrarily many cuts simultaneously.

\renewcommand\metavar\metavars
\newcommand\descwidth{10cm}
\newcommand\bewidth{1.8cm}
\newcommand\depwidth{3.5cm}
\begin{table*}[tp]
\centering
\scriptsize{
\begin{tabular}{l|p{\descwidth}|l}
 \textbf{Capability}
      &  \multirow{2}{*}{\parbox{\descwidth}{\textbf{Function} (\textbf{Return Type}):
             \\  \textbf{Brief Description}}} &   \textbf{Dependencies}
             \\ & &
\\ \hline
 \cpp{SMINPUTS}
      & \multirow{2}{*}{\parbox{\descwidth}{\cpp{get\_SMINPUTS}
              (\cpp{SMInputs}):
              \\  Provide Standard Model parameters in SLHA2 input conventions.}}
  &
  \\ & &
   \\ \hline
 \cpp{qedqcd\_subspectrum}
      & \multirow{2}{*}{\parbox{\descwidth}{\cpp{get\_QedQcd\_spectrum}
              (\cpp{const SubSpectrum*}):
              \\  Create \cpp{QedQcdWrapper} version of \cpp{SubSpectrum} from \cpp{SMInputs} structures.}}
  & \cpp{SMINPUTS}
  \\ & &
     \\ \hline
 \cpp{SM\_spectrum}
      & \multirow{2}{*}{\parbox{\descwidth}{\cpp{get\_SM\_spectrum}
              (\cpp{Spectrum}):
              \\  Create \cpp{Spectrum} object from \cpp{QedQcdWrapper} and SM Higgs parameters.}}
  &  \cpp{SMINPUTS}
  \\ & &
  \\ \hline
    \cpp{Higgs_Couplings}
      & \multirow{2}{*}{\parbox{\descwidth}{\cpp{SM_higgs_couplings} (\cpp{HiggsCouplingsTable}):
              \\  Construct a table of SM Higgs couplings.}}
  & \cpp{Higgs_decay_rates}
  \\ & &
  \\ \hline
\end{tabular}
}
\caption{
SM \capabilities{} provided by \specbit. The \capability{} \lstinline{SM_spectrum} is understood to provide running parameters in the \MSbar scheme, along with pole masses.
\label{tab:specbitsmcap}
}
\end{table*}

\renewcommand\metavar\metavarf


\renewcommand\metavar\metavars

\renewcommand\descwidth{6.7cm}
\begin{table*}[tp]
\centering
\scriptsize{
\begin{tabular}{l|p{\descwidth}|l|l}
  \textbf{Capability}
      &  \multirow{2}{*}{\parbox{\descwidth}{\textbf{Function} (\textbf{Return Type}):
             \\  \textbf{Brief Description}}}
        & \textbf{Dependencies} & \multirow{2}{*}{\parbox{\bewidth}{\textbf{Backend} \\ \textbf{requirements}}}
\\ & & &
\\ \hline
  \cpp{unimproved\_MSSM}
   & \multirow{3}{*}{\parbox{\descwidth}{\cpp{get\_CMSSM\_spectrum}(\cpp{Spectrum}):
      \\ Make an MSSM spectrum object with \FlexibleSUSY using CMSSM boundary conditions.}}
  & \cpp{SMINPUTS} & \FlexibleSUSY
    \\ \ \cpp{_spectrum} & & &
    \\ & & &
    \\ & & &
    \\ \cmidrule{2-4}
     & \multirow{3}{*}{\parbox{\descwidth}{\cpp{get\_MSSMatMGUT\_spectrum}(\cpp{Spectrum}):
      \\  Make an MSSM spectrum object with \FlexibleSUSY using soft breaking masses input at the GUT scale.}}
  & \cpp{SMINPUTS} & \FlexibleSUSY
    \\ & & &
    \\ & & &
  \\ \cmidrule{2-4}
    & \multirow{3}{*}{\parbox{\descwidth}{\cpp{get\_MSSMatQ\_spectrum} (\cpp{Spectrum}):
      \\ Make an MSSM spectrum object with \FlexibleSUSY from soft breaking masses input at (user-defined) scale $Q$.}}
  & \cpp{SMINPUTS} & \FlexibleSUSY
    \\ & & &
    \\ & & &
  \\ \cmidrule{2-4}
     & \multirow{3}{*}{\parbox{\descwidth}{\cpp{get\_MSSM\_spectrum\_SPheno}(\cpp{Spectrum}):
      \\  Make an MSSM spectrum object with \SPheno for any supported MSSM model.}}
  & \cpp{SMINPUTS} & \SPheno
    \\ & & &
    \\ & & &
  \\ \cmidrule{2-4}
    & \multirow{6}{*}{\parbox{\descwidth}{\cpp{get\_MSSM\_spectrum\_from\_SLHAfile} (\cpp{Spectrum}):
      \\  Construct an MSSM spectrum from an SLHA file.
    Wraps it up in \cpp{MSSMSimpleSpec}; i.e. no RGE running possible.
    Designed for testing against benchmark points, but may be a useful last
    resort for interacting with ``difficult'' spectrum generators.}}
  &  &
    \\ & & &
    \\ & & &
    \\ & & &
    \\ & & &
    \\ & & &
  \\  \cmidrule{2-4}
     & \multirow{3}{*}{\parbox{\descwidth}{\cpp{get\_unimproved_MSSM\_spectrum\_as\_SLHAea}
              (\cpp{SLHAstruct}):
      \\  Convert an unimproved MSSM spectrum into an \cpp{SLHAstruct} object.}}
   & \cpp{unimproved\_MSSM\_spectrum} &
    \\ & & &
    \\ & & &
    \\ & & &
      \\ \cmidrule{2-4}
     & \multirow{4}{*}{\parbox{\descwidth}{\cpp{get\_unimproved\_MSSM\_spectrum\_as\_map}
              (\cpp{map\_str\_dbl}):
              \\  Convert an unimproved MSSM spectrum into a \cpp{std::map}, for saving via the \GB printer system.}}
   & \cpp{unimproved\_MSSM\_spectrum} &
    \\ & & &
    \\ & & &
    \\ & & &
  \\ \hline
   \cpp{MSSM\_spectrum}
      & \multirow{3}{*}{\parbox{\descwidth}{\cpp{get\_MSSM\_spectrum\_as\_map}
              (\cpp{map\_str\_dbl}):
              \\  Convert an MSSM spectrum into a \cpp{std::map}, for saving via the \GB printer system.}}
  & \cpp{MSSM\_spectrum} &
  \\ & & &
  \\ & & &
  \\  \cmidrule{2-4}
     & \multirow{2}{*}{\parbox{\descwidth}{\cpp{get\_MSSM\_spectrum\_as\_SLHAea}
              (\cpp{SLHAstruct}):
      \\  Turn an MSSM spectrum into an \cpp{SLHAstruct} object.}}
  & \cpp{MSSM\_spectrum} &
    \\ & & &

  \\ \hline
    \cpp{SM_subspectrum}
      & \multirow{4}{*}{\parbox{\descwidth}{\cpp{get\_SM\_SubSpectrum\_from\_MSSM\_Spectrum}\\
              (\cpp{const SubSpectrum*}):
              \\  Extract the ``low-energy'' \cpp{SubSpectrum} from an MSSM \cpp{Spectrum} object.}}
  &  \cpp{MSSM\_spectrum} &
  \\ & & &
  \\ & & &
  \\ & & &

  \\ \hline
    \cpp{SMlike_Higgs}
      & \multirow{3}{*}{\parbox{\descwidth}{\cpp{most_SMlike_Higgs_MSSM}
              (\cpp{int}):
              \\  Return the PDG code of the most SM-like of the neutral Higgs bosons in the MSSM.}}
  &  \cpp{MSSM\_spectrum} &
  \\ \ \cpp{_PDG_code} & & &
  \\ & & &

  \\ \hline
    \cpp{Higgs_Couplings}
      & \multirow{4}{*}{\parbox{\descwidth}{\cpp{MSSM_higgs_couplings_pwid}
              (\cpp{HiggsCouplingsTable}):
              \\  Construct a table of Higgs couplings from Higgs partial decay widths.}}
  &  \cpp{MSSM_spectrum} &
  \\ & & \cpp{SMlike_Higgs_PDG_code} &
  \\ & & \cpp{Reference_SM_Higgs_decay_rates} &
  \\ & & \cpp{Reference_SM_other_Higgs_decay} &
  \\ & & \ \cpp{_rates} &
  \\ & & \cpp{Reference_SM_A0_decay_rates} &
  \\ & & \cpp{Higgs_decay_rates} &
  \\ & & \cpp{h0_2_decay_rates} &
  \\ & & \cpp{A0_decay_rates} &
  \\ & & \cpp{H_plus_decay_rates} &
  \\ & & \cpp{t_decay_rates} &

  \\ \cmidrule{2-4}
      & \multirow{3}{*}{\parbox{\descwidth}{\cpp{MSSM_higgs_couplings_FH}
              (\cpp{HiggsCouplingsTable}):
              \\  Construct a table of Higgs couplings from couplings calculated by \FH.}}
  &  \cpp{MSSM_spectrum} &
  \\ & & \cpp{SMlike_Higgs_PDG_code} &
  \\ & & \cpp{Reference_SM_Higgs_decay_rates} &
  \\ & & \cpp{Reference_SM_other_Higgs_decay} &
  \\ & & \ \cpp{_rates} &
  \\ & & \cpp{Reference_SM_A0_decay_rates} &
  \\ & & \cpp{Higgs_decay_rates} &
  \\ & & \cpp{h0_2_decay_rates} &
  \\ & & \cpp{A0_decay_rates} &
  \\ & & \cpp{H_plus_decay_rates} &
  \\ & & \cpp{t_decay_rates} &
  \\ & & \cpp{FH_Couplings_output} &

    \\ \hline
    \cpp{FH_Couplings}
      & \multirow{2}{*}{\parbox{\descwidth}{\cpp{FH\_Couplings}
              (\cpp{fh_Couplings}):
        \\  Higgs couplings, as computed by \FH}}
    & \cpp{prec\_HiggsMasses} & \FeynHiggs
  \\ \ \cpp{_output} & & &

   \\ \hline
    \cpp{prec_HiggsMasses}
      & \multirow{3}{*}{\parbox{\descwidth}{\cpp{FH\_HiggsMasses}
              (\cpp{fh_HiggsMassObs}):
        \\  Higgs masses and mixings with theoretical uncertainties, as computed by \FH.}}
    &  \cpp{FH\_MSSMMasses} &\FeynHiggs
  \\ & & &
  \\ & & &

   \\ \hline
    \cpp{FH_MSSMMasses}
      & \multirow{2}{*}{\parbox{\descwidth}{\cpp{FH\_MSSMMasses}
              (\cpp{fh_MSSMMassObs}):
              \\  SUSY masses and mixings, as computed by \FH.}}
  &  & \FeynHiggs
  \\ & & &

   \\ \hline
\end{tabular}
}
\caption{
MSSM \capabilities{} provided by \specbit. The \capabilities{} \lstinline{MSSM_spectrum} and \lstinline{unimproved_MSSM_spectrum} are understood to provide MSSM couplings and soft masses in the \DR scheme, along with pole masses, with low-energy Standard Model inputs defined in the \MSbar scheme. Note also that module functions exist in \precisionbit with \capability{} \lstinline{MSSM_spectrum}, which depend on the \lstinline{unimproved_MSSM_spectrum}; see Table \ref{tab:precisionbit:BE} and Sec.~\ref{precisionspectrum} for more details.
\label{tab:specbitmssmcap}}
\end{table*}

\renewcommand\metavar\metavarf


\renewcommand\metavar\metavars
\renewcommand\descwidth{7.5cm}
\renewcommand\bewidth{1.6cm}

\begin{table*}[tp]
\centering
\scriptsize
\begin{tabular}{l|p{\descwidth}|l|l}
  \textbf{Capability}
      & \multirow{2}{*}{\parbox{\descwidth}{\textbf{Function} (\textbf{Return Type}):
             \\  \textbf{Brief Description}}}
          & \textbf{Dependencies}
          & \textbf{Options}
          \\ & & & (\textbf{Type})
\\ \hline
\cpp{SingletDM_spectrum}
      & \multirow{3}{*}{\parbox{\descwidth}{\cpp{get_SingletDM_spectrum_simple} (\cpp{Spectrum}):
              \\  Create simple container \cpp{Spectrum} object from \cpp{SMInputs} structures, SM Higgs parameters, and the SingletDM parameters.}}
  & \cpp{SMINPUTS} &
  \\ & & &
  \\ & & &
  \\  \cmidrule{2-4} &
      \multirow{3}{*}{\parbox{\descwidth}{\cpp{get_SingletDM_spectrum_FS}
              (\cpp{Spectrum}):
              \\ Create \cpp{Spectrum} object with RGE facilities for the scalar singlet model using \FlexibleSUSY. }}
  & \cpp{SMINPUTS} & \FlexibleSUSY
  \\ & & &
  \\ & & &
  \\  \cmidrule{2-4} &
   \multirow{2}{*}{\parbox{\descwidth}{\cpp{get_SingletDM_spectrum_as_map}
              (\cpp{map_str_dbl}):
              \\ Convert a SingletDM spectrum into a \cpp{std::map} for printing.}}
  & \cpp{SingletDM_spectrum} &
  \\ & & &

  \\ \hline
    \cpp{Higgs_Couplings}
      & \multirow{4}{*}{\parbox{\descwidth}{\cpp{SingletDM_higgs_couplings_pwid} (\cpp{HiggsCouplingsTable}):
              \\  Construct a table of Higgs couplings from Higgs partial decay widths.}}
  &  \cpp{SingletDM_spectrum} &
  \\ & & \cpp{Reference_SM_Higgs} &
  \\ & & \ \cpp{_decay_rates} &
  \\ & & \cpp{Higgs_decay_rates} &
  \\ \hline
  \cpp{vacuum_stability}
      & \multirow{4}{*}{\parbox{\descwidth}{\cpp{find_min_lambda}
              (\cpp{triplet<double>}):
              \\ Run the quartic Higgs coupling to high scales, returning a \cpp{triplet} containing the expected lifetime of the electroweak vacuum, the instability scale, and the perturbativity of the couplings up to the requested scale.}}
  &  \cpp{SingletDM_spectrum} & \cpp{set_high_scale}
  \\ & & & \cpp{(double)}
  \\ & & & \cpp{check_perturb_pts}
  \\ & & & \cpp{(int)}
  \\ & & &
  \\ \hline
  \cpp{check_perturb_min}
  &  \multirow{4}{*}{\parbox{\descwidth}{\cpp{get_check_perturb_min_lambda}
              (\cpp{double}):
              \\  Check that dimensionless couplings are sufficiently small up to the smaller of the Planck scale and the scale at which the quartic Higgs coupling is minimised \textit{and} non-positive.}}
  & \cpp{vacuum_stability} &
  \\ \ \cpp{_lambda} & & &
  \\ & & &
  \\ & & &
  \\ \hline
  \cpp{VS_likelihood}
      & \multirow{3}{*}{\parbox{\descwidth}{\cpp{get_likelihood}
              (\cpp{double}):
              \\  Get the \textit{a priori} likelihood of no electroweak vacuum decays having occurred in the past light cone.}}
  & \cpp{vacuum_stability} &
  \\ & & &
  \\ & & &
  \\ \hline
  \cpp{expected_lifetime}
     & \multirow{3}{*}{\parbox{\descwidth}{\cpp{get_expected_lifetime}
              (\cpp{double}):
              \\  Return the expected lifetime of the electroweak vacuum in units of years.}}
  & \cpp{vacuum_stability} &
  \\ & & &
  \\ & & &
  \\ \hline
\end{tabular}
\caption{
\doublecross{Capabilities}{capability} provided by \specbit for the SingletDM model. The \capability{} \lstinline{SingletDM_spectrum} is understood to provide running parameters in the \MSbar scheme, along with pole masses, with low-energy Standard Model inputs also defined in the \MSbar scheme.
\label{tab:specbitsingletdmcap}
}
\end{table*}

\renewcommand\metavar\metavarf


\subsection{Interface details for \GB module writers (\textsf{C{\smaller ++}} API for \cpp{Spectrum} and related classes)}
\label{sec:APIs}

In Sec.\ \ref{user_interface} we provided details for a base-level user of \GB to simply operate the code as written. However, one of the goals of \GB is to provide a framework into which researchers can add their own calculations while maintaining easy use of all pre-existing \GB functions. This section is intended as a guide for users of this kind.  This information will also be needed by users running \specbit externally to \GB. Here we describe details of the interfaces to objects provided by the \specbit module, and how they should be used to facilitate physics calculations of other modules.

In Sec.\ \ref{sec:basic_spec_usage} we demonstrate the most common methods for accessing spectrum information. In Secs.\ \ref{sec:spectrum_structure} and \ref{sec:subspectrum_structure}, we outline the class structure used to store spectrum information, and explain the details of its interface. The helper class \cpp{SMInputs} is discussed last, in Sec. \ref{sec:sminputs}.

\subsubsection{Basic spectrum access}
\label{sec:basic_spec_usage}
The spectrum information is stored in a \cpp{Spectrum} object, the full structure of which will be described in
Sec.\ \ref{sec:spectrum_structure}.  However here we briefly
describe the basic user interface.

To access a \cpp{Spectrum} object in a new module function, the module function should be declared to have a \dependency{} on the appropriate spectrum.  As described in section 3.2.1 of the \GB paper \cite{gambit}, the spectrum object can then be obtained by dereferencing the safe pointer returned by the \specbit module function, e.g.
\begin{lstcpp}
using namespace Pipes::@\metavar{my\_module\_function}@;
const Spectrum& spec = *Dep::SingletDM_spectrum;
\end{lstcpp}
One can now access the spectrum information using \cpp{spec}.  The spectrum information is primarily accessed using a string accessor plus a tag, as in
\begin{lstcpp}
double mh0 = spec.get(Par::Pole_Mass,"h0_1");
\end{lstcpp}
Note that \cpp{Par} is simply a namespace, so the tags can be brought into scope
with the \cpp{using} keyword, and then used more succinctly, i.e.\
\begin{lstcpp}
using namespace Par;
double mh0 = spec.get(Pole_Mass,"h0_1");
\end{lstcpp}
The tag in the first argument of the getter specifies the type
of information to look for.  The second argument is a string that, in
combination with the tag, tells the getter exactly what information is
requested.  To simplify access to quantities that are naturally
grouped together in vectors (e.g.~$CP$-even Higgs or neutralino masses in
the MSSM), the \cpp{get} function may also be called with an index:
\begin{lstcpp}
double mh0 = spec.get(Pole_Mass,"h0",1);
\end{lstcpp}

As well as a \cpp{get} function, \cpp{Spectrum} objects contain a \cpp{has}
function with an almost identical function signature, which can be used to check
if a quantity exists in the spectrum before attempting to retrieve it:
\begin{lstcpp}
bool has_mh0_1 = spec.has(Pole_Mass,"h0",1);
\end{lstcpp}
where \cpp{has_mh0_1} is \cpp{true} if the quantity exists in the spectrum, and
\cpp{false} if it does not.

It is also possible to \cpp{set} the values of spectrum parameters, but not via the
base-level \cpp{Spectrum} object interface. For this more advanced usage, see
Sec.\ \ref{sec:subspectrum_structure}.  Most users will not need to do this.

This form of string
getter allows the spectrum information to be accessed in a simple, uniform
way across all models and all \GB modules.  The user simply needs
to know the specific tags and string names used for each piece of
information.  Particle names, for example, can be found in the \GB particle database (see Sec.\ 10.1 of Ref.\ \cite{gambit}).

For convenience, in Appendix \ref{app:subspec_contents} we give tables listing the tags, strings, and indices needed to access data via the getters, checkers, and setters for the SM, the
MSSM, and the scalar singlet model.

Note that only information tagged with \cpp{Pole_Mass} can be retrieved via the
base-level \cpp{Spectrum} interface. For other information (and to run RGEs, when
available) one must use the equivalent \cpp{SubSpectrum} interface, which is described in
Sec.\ \ref{sec:subspectrum_structure}.

\subsubsection{\cpp{Spectrum} class structure}
\label{sec:spectrum_structure}
\begin{figure}[t]
\centering
\includegraphics[width=0.7\columnwidth]{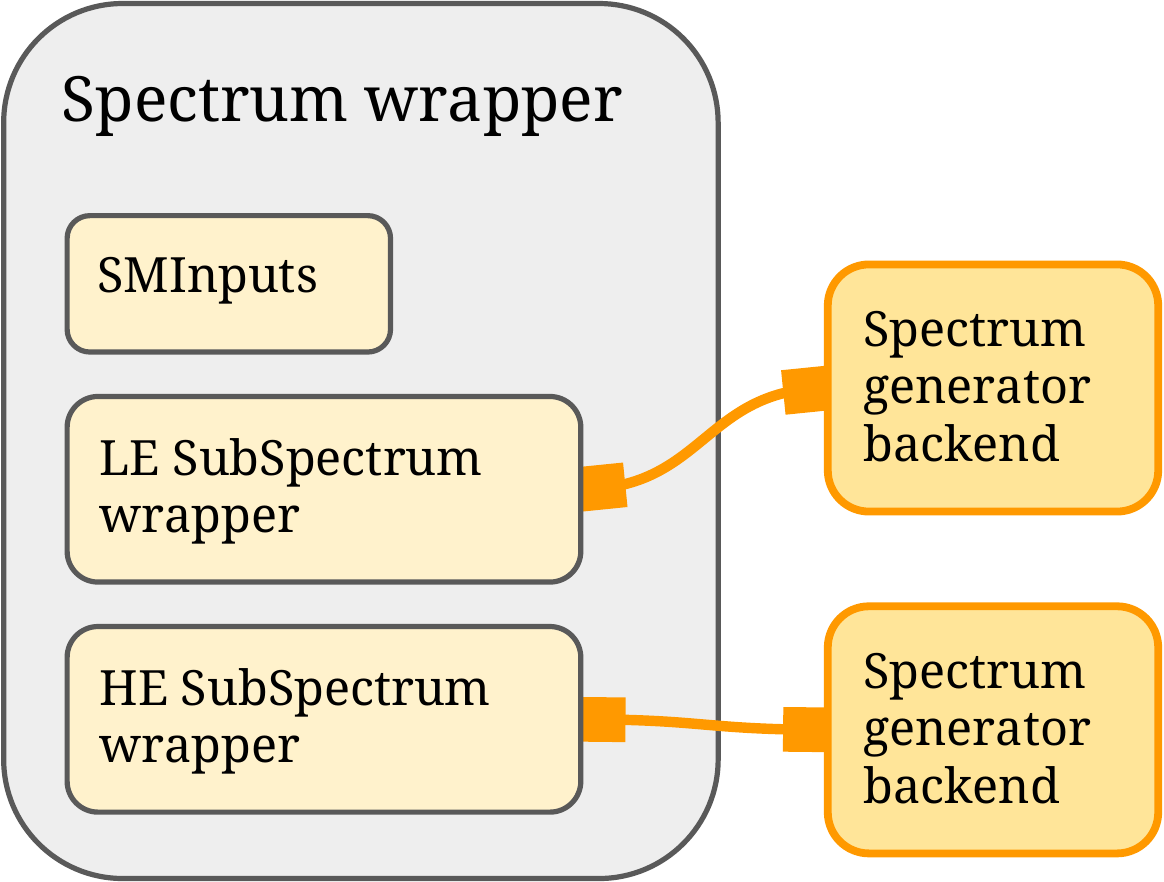}
\caption{Schematic representation of a \cpp{Spectrum} object. Basic Standard Model ``input'' information is contained in a structure call \cpp{SMInputs} {\color[HTML]{B5A650}(upper-left light-yellow box)}, the content of which mirrors the correspondingly named SLHA2 block. Low scale (generally $<m_Z$) spectrum information is wrapped by a ``low energy'' \cpp{SubSpectrum} object {\color[HTML]{B5A650}(centre-left light-yellow box)}. High scale (generally $>m_Z$) spectrum information is wrapped in a ``high energy'' \cpp{SubSpectrum} object {\color[HTML]{B5A650}(lower-left light-yellow box)}. This split mirrors a common requirement of spectrum generators that Standard Model input parameters be first supplied at a common scale, then run by the ``low energy'' \cpp{SubSpectrum} object to a matching scale, after which the main spectrum generation for the BSM model is performed and wrapped into the ``high energy'' \cpp{SubSpectrum} object. The outer \cpp{Spectrum} object provides a uniform interface to these underlying structures and makes it easy to retrieve common parameters such as pole masses without the need to interact directly with the subspectra.
\label{fig:spectrum_structure}
}
\end{figure}

At the centre of the \specbit module is a virtual interface class named \cpp{Spectrum}, for accessing typical spectrum generator output in a generalised and standardised way. Wherever possible, SLHA2 conventions \cite{Skands:2003cj,Allanach:2008qq} are used as the standard. The objects accessed by this interface can thus be considered as an in-memory representation of the data that one would typically find in the SLHA2 ASCII output of a spectrum generator.

Each \cpp{Spectrum} object contains three main data members, illustrated in Fig. \ref{fig:spectrum_structure}: an \cpp{SMInputs} object, which contains SM input parameters (closely mirroring the \cpp{SMINPUTS} block defined by SLHA2); and two \cpp{SubSpectrum} objects, one labelled \cpp{LE} for `low-energy' and which typically contains low-scale SM information, and the other labelled \cpp{HE} for `high-energy', which typically contains higher-scale model information such as Higgs-sector or BSM data. The reason for this separation is that, typically, one requires different calculations for accurately evaluating and evolving low-energy SM parameters (like quark masses) and high-energy ones like the MSSM. The code to do each calculation will be wrapped in a separate \cpp{SubSpectrum} structure.  This wrapper also provides access to running parameters and RGE-running facilities, if these are available in a backend code wrapped by a particular \cpp{SubSpectrum}. We describe the \cpp{SubSpectrum} interface in Sec.\ \ref{sec:subspectrum_structure}. Details needed for writing such a class are found in Sec.\ \ref{sec:specbit_adding_new_models}. The purpose of the host \cpp{Spectrum} object is to provide a consistent interface to the underlying subspectra and SM input parameters.

\begin{figure*}[t]
\centering
\begin{lstcpp}
class Spectrum
{
   public:
      SubSpectrum& get_LE();       // Access "low-energy" SubSpectrum
      SubSpectrum& get_HE();       // Access "high-energy" SubSpectrum
      SMInputs&    get_SMInputs(); // Access SMInputs associated with this Spectrum

      std::unique_ptr<SubSpectrum> clone_LE() const; // Copy "low-energy" SubSpectrum
      std::unique_ptr<SubSpectrum> clone_HE() const; // Copy "high-energy" SubSpectrum
      // (whole Spectrum object can be copied using standing copy constructor/assignment methods)

      void RunBothToScale(double scale); // Run both LE and HE SubSpectrum to same scale
                                         // (Default units: GeV)
      // Get/check/set functions for parameter retrieval/input with zero or one indices.
      bool   has(const Par::Tags partype, const std::string& mass, const int index=NONE) const;
      double get(const Par::Tags partype, const std::string& mass, const int index=NONE) const;

      SLHAstruct getSLHAea(int slha_version) const; // Create SLHAea object from LE+HE SubSpectra in SLHA standard slha_version. HE takes precedence.
      void writeSLHAfile(int slha_version, str fname) const; // Output spectrum contents as an SLHA file, conforming to SLHA standard slha_version.
};
\end{lstcpp}
\caption{Simplified class declaration of primary \cpp{Spectrum} class. Not shown are various overloads, constructors and assignment operators, helper functions, and all private members. Additionally, the index arguments for the has/get functions are implemented via operator overloads rather than optional arguments as is shown.  For the full class declaration please see {\termplainstyle Elements/include/gambit/Elements/spectrum.hpp} in the \GB source code.}
\label{fig:spectrum_object}
\end{figure*}

In Fig.\ \ref{fig:spectrum_object} we provide a simplified overview of the contents of the \cpp{Spectrum} class, for reference. We will now discuss the various functions and data members shown in this figure.
\paragraph*{Data member getters}
The central members of a \cpp{Spectrum} object are accessed via three functions:
\begin{description}
\item[{\CPPidentifierstyle get\_LE()}/{\CPPidentifierstyle get\_HE()}] Accesses the hosted \cpp{SubSpectrum} object identified as `low-energy' or `high-energy'.
\item[\CPPidentifierstyle get\_SMInputs()] Accesses the hosted \cpp{SMInputs} object.
\item[{\CPPidentifierstyle clone\_LE()}/{\CPPidentifierstyle clone\_HE()}] Creates a copy of the hosted \cpp{SubSpectrum} object identified as `low-energy' or `high-energy'.
\end{description}
The `clone' functions exist because by design, it is not possible to perform actions that modify contents of \cpp{Spectrum} objects provided to \GB module functions as dependencies. This protection extends to the \cpp{SubSpectrum} objects, so in order to perform actions that modify the spectrum (like RGE running) one must first copy the whole \cpp{Spectrum} or `clone' the relevant \cpp{SubSpectrum}.
\paragraph*{Copying \cpp{Spectrum} objects}
Because \GB prevents the results of module functions from being modified by other module functions, if a user wishes to modify the contents of a \specbit-provided \cpp{Spectrum} object, the user must first copy it. The copy constructor (and associated constructors) of the \cpp{Spectrum} object are designed to perform a `deep' copy.  To make a copy, one can simply use the copy constructor or assignment operators, i.e.
\begin{lstcpp}
Spectrum spec_copy(spec);  //This is a deep copy
Spectrum spec_copy = spec; //Also a deep copy
\end{lstcpp}
\paragraph*{RGE running}
There is little in the underlying object that can be modified from the \cpp{Spectrum} interface, except via the \cpp{RunBothToScale} function. This function runs the renormalisation-scale-dependent parameters of the underlying \cpp{SubSpectrum} objects to the same scale. Note that the underlying RGEs may not be valid beyond certain scales on both the high and low ends, so some caution and knowledge of the underlying objects is required in order to safely use this feature (See `RGE running' in the following subsection).

\subsubsection{\cpp{SubSpectrum} objects}
\label{sec:subspectrum_structure}
\begin{figure}[t]
\centering
\includegraphics[width=0.85\columnwidth]{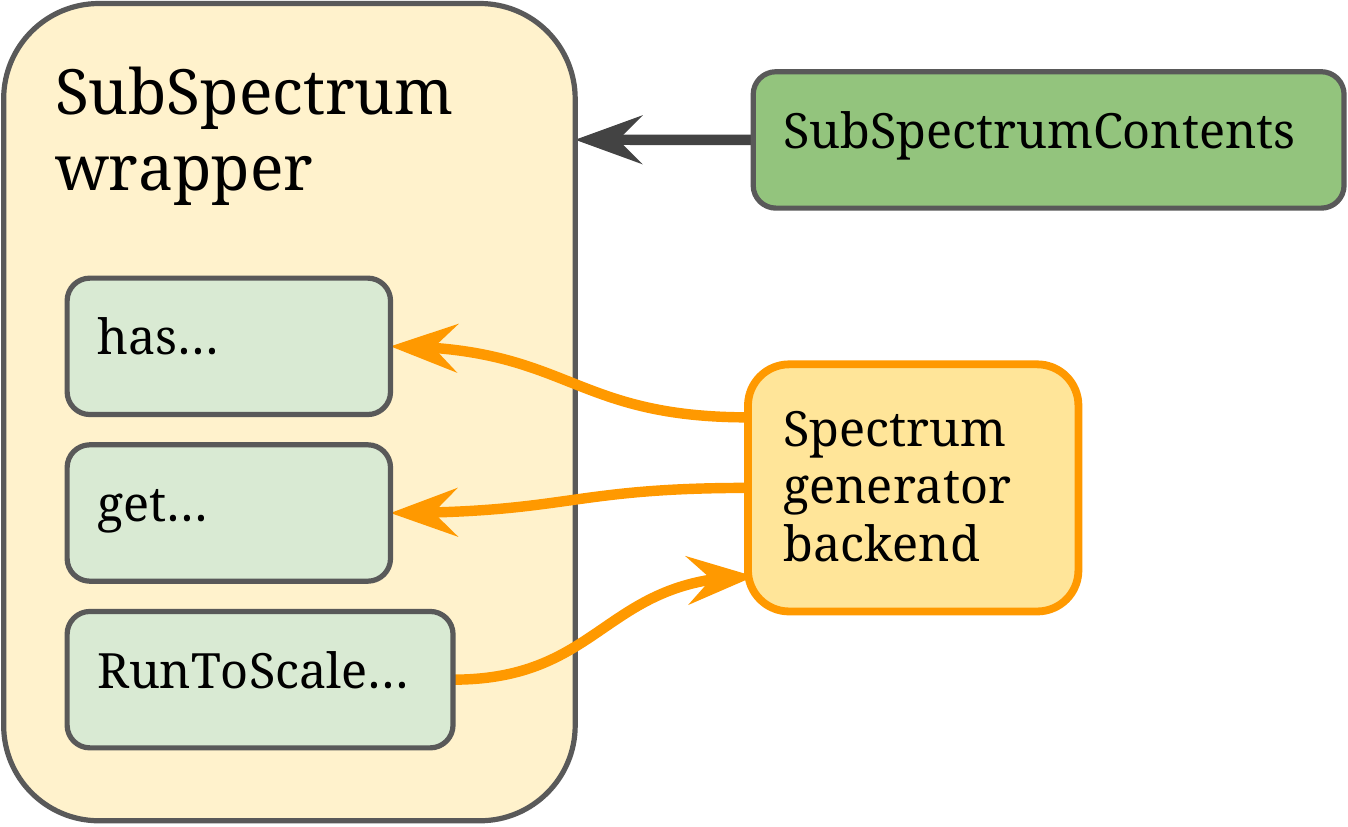}
\caption{Schematic representation of a \cpp{SubSpectrum} object. These objects interface directly to spectrum information or external spectrum generators {\color[HTML]{FA9800}(yellow box)}. Accessor functions (e.g.\ \cpp{has}, \cpp{get}) are used to access spectrum parameters, and, in the case where a spectrum generator is connected, the \cpp{RunToScale} function can be called to run the whole spectrum to a different scale {\color[HTML]{78A86E}(pale-green boxes)}. To standardise parameters that should be retrievable from generated spectra for each model, we define a separate \cpp{SubSpectrumContents} object for each model. Each wrapper is then associated with some \cpp{SubSpectrumContents} {\color[HTML]{18A802}(green box)}, and is required to provide all the contents defined therein.
\label{fig:subspectrum_structure}
}
\end{figure}

While the \cpp{Spectrum} interface provides fast and easy access to pole masses, it does not allow access to more detailed spectrum information, such as the values of couplings, mixings, and running mass parameters. For this one must use the \cpp{SubSpectrum} interface. As we covered in Sec.\ \ref{sec:spectrum_structure}, the simplest way to obtain a \cpp{SubSpectrum} object is to extract it from a host \cpp{Spectrum} object using the \cpp{get_LE()}/\cpp{get_HE()} functions or \cpp{clone_LE()}/\cpp{clone_HE()} functions:
\begin{lstcpp}
const SubSpectrum& mssm = spec.get_HE();
unique_ptr<SubSpectrum> copy = spec.clone_HE();
\end{lstcpp}
\begin{figure*}[t]
\centering
\begin{lstcpp}
class SubSpectrum
{
  public:
    std::unique_ptr<SubSpectrum> clone() const;   // Clone (copy) this SubSpectrum object

    virtual void writeSLHAfile(int slha_version, const str& fname) const; // Write out spectrum to SLHA file (if possible) in SLHA standard slha_version
    virtual SLHAstruct getSLHAea(int slha_version) const;         // Get spectrum in SLHAea format (if possible)
    virtual void add_to_SLHAea(int slha_version,SLHAstruct&) const;// Add spectrum data to an SLHAea object (if possible) in SLHA standard slha_version

    // Limits to RGE running; warning/error raised if running beyond these is attempted.
    // If these aren't overridden in the derived class then effectively no limit on running will exist.
    // These are public so that module writers can use them to check what the limits are.
    virtual double hard_upper() const {return DBL_MAX;}
    virtual double soft_upper() const {return DBL_MAX;}
    virtual double soft_lower() const {return 0.;}
    virtual double hard_lower() const {return 0.;}
    // Returns the renormalisation scale of parameters. By default should be in units of GeV.
    virtual double GetScale() const;
    void RunToScale(double scale, int behaviour = 0); // Run spectrum to scale (Default units: GeV).

    // Get/check/set functions for parameter retrieval/input with zero, one, and two indices
    virtual bool   has(const Par::Tags, const str&, int index1=NONE, int index2=NONE) const;
    virtual double get(const Par::Tags, const str&, int index1=NONE, int index2=NONE) const;
    virtual void   set(const Par::Tags, const double, const str&, int index1=NONE, int index2=NONE);
    void  set_override(const Par::Tags, const double, const str&, int index1=NONE, int index2=NONE);
};
\end{lstcpp}
\caption{Simplified class definition of \cpp{SubSpectrum} class, which is the virtual base class for interacting with spectrum generator output, including an interface for running RGEs. Not shown are various overloads, constructors, helper functions, and all private members. The index arguments for the \cpp{has}/\cpp{get}/\cpp{set} functions are implemented via operator overloads rather than the optional arguments shown. For the full class definition please see {\termplainstyle Elements/include/gambit/Elements/subspectrum.hpp} in the \GB source code.}
\label{fig:subspectrum_object}
\end{figure*}

Following the same pattern as the \cpp{Spectrum} interface, spectrum information can be accessed via the \cpp{SubSpectrum} interface using a \cpp{get} function, for which the arguments are a tag, a string identifier, and zero to two indices:
\begin{lstcpp}
using namespace Par;
double mass_su1 = mssm.get(Pole_Mass,"~u",1);
\end{lstcpp}

Let us now examine the contents of the \cpp{SubSpectrum} interface class. This class is a virtual interface for derived classes individually designed to wrap the data from external spectrum generator codes. The virtualisation allows module writers to treat spectrum data in their own module functions in a generic way, with the selection of the actual spectrum generator to use being a decision left to runtime. The wrapping of external codes is then decoupled from the use of the data required by module functions, meaning that replacement wrappers can be written for new external spectrum generator codes, and the new wrapper activated, without necessitating any modification of downstream module functions.

For reference, a simplified outline of the \cpp{SubSpectrum} class is shown in Figure \ref{fig:subspectrum_object}. For the precise class declaration please see the file
\begin{itemize}[topsep=3pt]
\item[] \term{Elements/include/gambit/Elements/subspectrum.hpp}
\end{itemize}
in the \GB source tree.
\paragraph*{Getters, checkers and setters} 
Like the \cpp{Spectrum} interface, the \cpp{SubSpectrum} interface also provides a \cpp{has} function, which accepts the same input arguments as the \cpp{get} function, and which returns a \cpp{bool} that indicates whether or not a given \cpp{SubSpectrum} object contains the requested parameter:
\begin{lstcpp}
bool has_mass_su1 = mssm.has(Pole_Mass,"~u",1);
\end{lstcpp}
Unlike the \cpp{Spectrum} interface, the \cpp{SubSpectrum} interface also provides \cpp{set} functions.  These can be used to modify or overwrite parameter data contained in the \cpp{SubSpectrum} object.  For example,
\begin{lstcpp}
double newmass  = 500; // Units generally in
double newmass2 = 600; // powers of GeV
mssm.set(Pole_Mass,newmass,"~u",1);
mssm.set_override(Pole_Mass,newmass2,"~u",1);
\end{lstcpp}
The \cpp{set} and the \cpp{set_override} functions differ in an important way. The \cpp{set} function, if permitted by the underlying wrapper class, will directly change a value in the underlying spectrum generator code, and thus may, for example, impact renormalisation group running. On the other hand, the \cpp{set_override} function does not see the underlying spectrum generator code at all, and will simply store a replacement value that will be preferentially retrieved when using the \cpp{get} function. This can be useful for storing, for example, precision calculations of pole masses in such a way that they can be seamlessly used by other \GB module functions, via the standard interface. We emphasise that the \cpp{set_override} value will take ultimate priority, and in fact the underlying value known to the spectrum generator code will become masked, and irretrievable. So in the example above, a call to \cpp{mssm.get(Pole_Mass,"~u",1)} after the call to \cpp{mssm.set_override} would retrieve the value 600.

There is also an optional final \cpp{bool} argument to \cpp{set_override} (not shown in Figure \ref{fig:subspectrum_object}), which indicates whether the function is permitted to add to the contents of a \cpp{SubSpectrum}. The default is \cpp{false}, leading to an error if a user tries to override a part of the \cpp{SubSpectrum} contents that does not already exist. By setting this \cpp{true}, one can insert arbitrary new information into the \cpp{SubSpectrum}, which will become retrievable with the \cpp{get} method.  This functionality should be used sparingly, as the writers of other module functions generally will not be aware of the existence of the additional data. If you find yourself wanting to add large amounts of new data to a spectrum, it is generally better to define a new \cpp{SpectrumContents}, and create a new spectrum wrapper class that conforms to this expanded contents set (see Sec. \ref{sec:subspectrum_contents}).
\paragraph*{RGE running}
The \cpp{SubSpectrum} interface provides the main connection to RGE running facilities in any underlying spectrum-generator code; the \cpp{RunBothToScale} function of the \cpp{Spectrum} interface is just a thin wrapper over this. These facilities do not exist in all \cpp{SubSpectrum} objects, so one must make sure, via the \GB \cross{dependency resolution} process, that an appropriate spectrum is supplied to module functions that require running facilities. If a module function attempts to call any of the following functions when no underlying RGEs are connected, then \specbit will throw an error and terminate.


The two most fundamental RGE-related functions are the \cpp{RunToScale} and \cpp{GetScale} functions. The latter function takes no arguments and simply returns the energy scale at which any running parameters accessible by the \cpp{get} function are defined, for example couplings and running masses. The \cpp{RunToScale} function accepts a new energy scale as input, and when called will run the underlying RGE code and recalculate the running parameters.

The precise behaviour of the \cpp{RunToScale} function can be altered via the optional integer `\cpp{behaviour}' argument. This controls how attempts to run parameters beyond the known accurate range of the underlying RGEs are handled. There exist two sets of limits in the \cpp{SubSpectrum} interface; upper and lower ``soft'' limits, and upper and lower ``hard'' limits. These specify the energy ranges that affect the action of the \cpp{behaviour} switch, and can be checked with the following functions:
\begin{lstcpp}
virtual double hard_upper() const
virtual double soft_upper() const
virtual double soft_lower() const
virtual double hard_lower() const
\end{lstcpp}
Note that when writing a new \cpp{SubSpectrum} wrapper these functions need to be overridden to specify valid ranges of RGE evolution, otherwise the default limits (which effectively allow unlimited running) will be used.

The ``behaviour'' flag acts in conjunction with the hard and soft limits according to the following rules:
\begin{description}
\item[{\CPPidentifierstyle behaviour = 0} (default)] -- If running beyond soft limit requested, halt at soft limit (parameter evolution will simply stop silently at this limit).
\item[{\CPPidentifierstyle behaviour = 1}] -- If running beyond soft limit requested, throw warning; beyond hard limit, throw error.
\item[{\CPPidentifierstyle behaviour = <anything else>}] -- Ignore limits and attempt running to requested scale (errors may still be thrown by the backend spectrum generator code if it sets its own hard-coded limits).
\end{description}
\paragraph*{SLHAea output}
Strictly speaking, the \cpp{SubSpectrum} objects do not rely on SLHA, and can interface to the spectrum codes for any model.  However, SLHA is an existing standard for the MSSM, so we provide the spectrum in this standard. The \cpp{SubSpectrum} interface thus provides three functions that convert \cpp{SubSpectrum} contents into SLHA2 format, making use of the \textsf{SLHAea} \Cpp library\footnote{http://fthomas.github.io/slhaea/}:
\begin{description}
\item[\CPPidentifierstyle void writeSLHAfile(int slha\_version, const str\& fname)] Writes out the contents of the \cpp{SubSpectrum} in SLHA format with the name \term{fname}. If \cpp{slha\_version} is set to $1$ then it will conform to SLHA1 standard, while if it is set to $2$ then it will conform to the SLHA2 standard. Ideally this should only be used for debugging purposes, as minimising disk access is preferable in high-performance computing applications. However, some backend codes do not have proper APIs and can only be run via the command line, or have otherwise hard-coded requirements for input SLHA files.
\item[\CPPidentifierstyle SLHAstruct getSLHAea(int slha\_version)] Writes the contents of the \cpp{SubSpectrum} into an \cpp{SLHAea::Coll} object (for which \cpp{SLHAstruct} is an alias). This is an in-memory string-based representation of an SLHA file, which is in the SLHA standard \cpp{SLHA\_version} while negating any need to write to disk. Block contents can be easily accessed via \cpp{[]} operators (see the \textsf{SLHAea} documentation for examples).
\item[\CPPidentifierstyle void add\_to\_SLHAea(SLHAstruct\&)] Similar to \cpp{getSLHAea}, however the spectrum content is added to a pre-existing \cpp{SLHAstruct}, passed in by reference via the input argument and modified in place.
\end{description}
SLHAea is very flexible and can be used to also create SLHA-like structures for other models.  Therefore this may also be exploited for non-minimal SUSY models and non-SUSY models too, though other than the MSSM and NMSSM there is no existing standard to fix the precise form and ensure that other codes will understand the results.  However, this feature is also rather useful for auto-generated programs, where one can construct the interface between the different generated codes to work with new SLHA blocks.

\subsubsection{The \cpp{SMInputs} class}
\label{sec:sminputs}

The three data members of a \cpp{Spectrum} object are two \cpp{SubSpectrum} objects, plus an \cpp{SMInputs} object. The latter object does not strictly contain information about the calculated spectrum, and the data it contains is not accessed via the \cpp{get} function of the host \cpp{Spectrum} object. Rather, it contains information that in most cases was used as \textit{input} to the external spectrum generator code. It is a simple \cpp{struct} that directly mirrors the \term{SMINPUTS}, \term{VCKMIN} and \term{UPMNSIN} blocks defined by the SLHA2 standard (see Figure \ref{fig:sminputs}).

Because \cpp{SMInputs} is a structure, all data members (the elements of the SLHA2 blocks) have public access, and so parameter values can be accessed directly (see Figure \ref{fig:sminputs} for the variable names). The structure contains only two functions, \cpp{getSLHAea} and \cpp{add_to_SLHAea}. The former simply returns the contents of the structure as an SLHAea object (which can be easily written to disk, for example), while the second adds the contents of the structure to an existing SLHAea object. The latter feature is mainly used internally by \cpp{Spectrum} objects to construct a full SLHA2 representation of their contents.

\begin{figure}
   \centering
   \begin{lstcpp}
   struct SMInputs
   {
      // SLHA1
      double alphainv;  // 1:
      double GF;        // 2:
      double alphaS;    // 3:
      double mZ;        // 4:
      double mBmB;      // 5:
      double mT;        // 6:
      double mTau;      // 7:
      // SLHA2
      double mNu3;      // 8:
      // double mW;     // 9:
      double mE;        // 11:
      double mNu1;      // 12:
      double mMu;       // 13:
      double mNu2;      // 14:

      double mD;        // 21:
      double mU;        // 22:
      double mS;        // 23:
      double mCmC;      // 24:

      // Block VCKMIN
      struct CKMdef {
        double lambda;
        double A;
        double rhobar;
        double etabar; };
      CKMdef CKM;

      // Block UPMNSIN
      struct PMNSdef {
        double theta12; //
        double theta23; //
        double theta13; //
        double delta13; //
        double alpha1;  //
        double alpha2;  //
      };
      PMNSdef PMNS;

      // Retrieve contents as an SLHAea object
      SLHAstruct getSLHAea(int getSLHAea) const;

      // Add contents to existing SLHAea object;
      void add_to_SLHAea(int slha_version,SLHAstruct& slha) const;
   };
   \end{lstcpp}
   \caption{ Definition of the \cpp{SMInputs} class. No simplification has been performed; this is the full class definition as found in {\termplainstyle Elements/include/gambit/Elements/sminputs.hpp} (with condensed comments). }
   \label{fig:sminputs}
\end{figure}

\subsubsection{Extra overloads for \cpp{get}/\cpp{set}/\cpp{has} functions}
\label{sec:sminputs}

In the discussions of the interface for \cpp{Spectrum} and \cpp{SubSpectrum} objects, we have mentioned only one set of function signatures for the \cpp{get}, \cpp{set} and \cpp{has} functions used to access spectrum data, however several overloads of these functions are available. Most of these are designed to streamline the interaction between \cpp{Spectrum} objects and the \GB particle database (see Sec 10.1 of Ref.\ \cite{gambit}), and they allow particle data to be retrieved according to PDG codes and, where appropriate, antiparticle string names (see \term{Models/src/particle_database.cpp} for the full set of definitions known to \GB).

For \cpp{Spectrum} objects, the extra function signatures are as follows:
\begin{lstcpp}
// PDG code + context integer
bool   has(@\metavar{tag}@, @\metavar{PDGcode}@, @\metavar{context}@)
double get(@\metavar{tag}@, @\metavar{PDGcode}@, @\metavar{context}@)

// PDG code + context integer (as std::pair)
bool   has(@\metavar{tag}@, std::pair<int,int>)
double get(@\metavar{tag}@, std::pair<int,int>)

// Short name plus index (as std::pair)
bool   has(@\metavar{tag}@, const std::pair<str,int>)
double get(@\metavar{tag}@, const std::pair<str,int>)
\end{lstcpp}

For \cpp{SubSpectrum} objects the above extra function signatures also exist, plus matching \cpp{set} and \cpp{set_override} functions:
\begin{lstcpp}
// PDG code + context integer
void   set(@\metavar{tag}@, @\metavar{value}@, @\metavar{PDGcode}@, @\metavar{context}@)
void   set_override(@\metavar{tag}@, @\metavar{value}@, @\metavar{PDGcode}@, @\metavar{context}@)

// PDG code + context integer (as std::pair)
void   set(@\metavar{tag}@, @\metavar{value}@, std::pair<int,int>)
void   set_override(@\metavar{tag}@, @\metavar{value}@, std::pair<int,int>)

// Short name plus index (as std::pair)
void   set(@\metavar{tag}@, @\metavar{value}@, const std::pair<str,int>)
void   set_override(@\metavar{tag}@, @\metavar{value}@, std::pair<str,int>);
\end{lstcpp}

When PDG codes + context integers are used with these functions to specify a particle, the \GB particle database will be used to translate the code into a string identifier, which will then be checked against the string keys defined in the underlying \cpp{SubSpectrum} wrapper to do the data lookup. \emph{When writing new wrappers it is therefore important to ensure that particle string names match the strings registered in the \GB particle database}.

Similarly, whenever any of these interface functions are called with a string name (potentially also with one index), the underlying wrapper will first be searched for a match to that string. If a match cannot be found, then the \GB particle database will be used to try to translate the string name into a short name plus an index (e.g. \cpp{"~chi_0"} will be translated to the pair \cpp{("~chi",0)}) or vice versa. If there is still no match found with any spectrum contents, the \GB particle database will again be used to try to translate the string name into the string for a matching antiparticle (e.g. \cpp{"~chi+_0"} will be translated to \cpp{"~chi-_0"}) and the search will be tried again (including translating from string name to short name plus index and vice versa). If no match can be found, an error will be thrown and the scan aborted (except in the case of the \cpp{has} functions, which will simply return \cpp{false}).

These mechanisms together provide considerable convenience when using the \cpp{Spectrum} interface in module functions that do not know precisely how the backend spectrum generator has been connected to a given wrapper. As an example, any of the following calls to a \cpp{Spectrum} object containing a \cpp{SubSpectrum} that conforms to the \cpp{MSSM} spectrum contents (see Appendix \ref{app:subspec_contents}) will retrieve the same information (the lightest selectron/anti-selectron pole mass):

\begin{lstcpp}
@\metavar{spectrum}@.get(Pole_Mass,"~e-",1);
@\metavar{spectrum}@.get(Pole_Mass,"~e-_1");
@\metavar{spectrum}@.get(Pole_Mass,std::make_pair("~e-",1));
@\metavar{spectrum}@.get(Pole_Mass,1000011,0);
@\metavar{spectrum}@.get(Pole_Mass,std::make_pair(1000011,0));

@\metavar{spectrum}@.get(Pole_Mass,"~e+",1);
@\metavar{spectrum}@.get(Pole_Mass,"~e+_1");
@\metavar{spectrum}@.get(Pole_Mass,std::make_pair("~e+",1));
@\metavar{spectrum}@.get(Pole_Mass,-1000011,0);
@\metavar{spectrum}@.get(Pole_Mass,std::make_pair(-1000011,0));
\end{lstcpp}

There is one final set of overloads to be mentioned: the \cpp{safeget} overloads. These act as drop-in replacements for any of the \cpp{get} functions, however they will throw a \GB error (causing the program to halt and report a problem) if the retrieved value is not-a-number (NaN).

\subsection{Adding support for new models and/or codes}
\label{sec:specbit_adding_new_models}
Adding spectrum generators for models not shipped with \GB is an inevitably technical task, as it involves writing an interface to new code.  We have however endeavoured to make the procedure as orderly as possible. To spare more casual users the full technical details, we have placed these in Appendix \ref{sec:adding_new_models}.  Here we give only a brief overview of the procedure, along with a check-list of tasks to be completed and links to the sections in the Appendix that explain the corresponding details. Some of the jargon used in the check-list has not been necessary to introduce up until now, but rather than explain it here we instead refer readers to the full details in the Appendix.

The bulk of the work is in writing a \cpp{SubSpectrum} wrapper class, which will wrap the new spectrum generator and be usable as we describe in Sec.\ \ref{sec:APIs}. There are also several peripheral tasks to be completed, such as writing a \modulefunction{} to properly construct the new \cpp{SubSpectrum} and package it into a \cpp{Spectrum} object, potentially writing new \cpp{SubSpectrumContents} definitions (which enforce consistency between wrappers that ostensibly share a physics model), and making additions to the \GB model hierarchy (discussed in Ref.\ \cite{gambit}). These required tasks are:

\begin{enumerate}
 \item {\bf Choose a \cpp{SubSpectrumContents} definition} to which the new wrapper will conform, or write a new one (Sec. \ref{sec:subspectrum_contents}). \label{item:contents}
 \item {\bf Write a \cpp{SpecTraits} specialization} to define helper types for the new \cpp{SubSpectrum} wrapper (Sec. \ref{sec:parameter_box_wrapper}, but see also the extensions with the \cpp{Model} and \cpp{Input} typedefs used in Secs.\ \ref{sec:model_member_functions} and \ref{sec:non_class_functions}). \label{item:traits}
 \item {\bf Write the \cpp{SubSpectrum} wrapper class} (Sec. \ref{sec:parameter_box_wrapper}).  This step can be further broken down into: \label{item:wrapper}
 \begin{itemize}
 \item {\bf Write the \cpp{get}/\cpp{set} interface functions} to the external code, which return values conforming to the requirements of the chosen \cpp{SubSpectrumContents}, and connect them to the \cpp{get}/\cpp{set} interface via the \cpp{fill_getter_maps} and \cpp{fill_setter_maps} functions of the wrapper class. Sections \ref{sec:parameter_box_wrapper}, \ref{sec:model_member_functions} and \ref{sec:non_class_functions} cover the various allowed types of interface functions and how to properly connect them to the wrapper interface.
 \item {\bf Write the special \cpp{get_Model} and \cpp{get_Input} functions} in the wrapper (if needed by the chosen type of interface function; see sections \ref{sec:model_member_functions} and \ref{sec:non_class_functions})
 \item {\bf Write the RGE interface functions} if RGE running capabilities are desired (Sec. \ref{sec:interfacing_to_rge}).
 \end{itemize}
 \item {\bf Write the module function(s)} that construct and return a \cpp{Spectrum} interface object connected to the new wrapper (Sec. \ref{sec:wrapper_module_function}). \label{item:modfunc}
\end{enumerate}
To illustrate the process concretely, we provide a full worked example of the writing of a complicated wrapper for a \FlexibleSUSY spectrum generator in Appendix \ref{app:wrapper_worked_examples}.

\subsection{Advanced spectrum usage: vacuum stability}
\label{Sec:VacStab}
In this Section, we present a sophisticated example of how the \cpp{Spectrum} object and \specbit module functions can be used, where we calculate stability of the electroweak vacuum for simple Higgs sectors.  These functions are included in \specbit \textsf{\gambitversion} and will be used in a follow-up to the study of the scalar singlet dark matter global fit \cite{SSDM}.  To explain what these module functions do and why, we first present a brief review of vacuum stability.  We then give the details of the corresponding observable and likelihood functions implemented in \GB.

The stability of the SM electroweak vacuum in the absence of new physics below the Planck scale ($M_{\text{Pl}}\approx 1.2\times 10^{19}$ GeV) has been studied extensively \cite{Sher1989,Elias-Miro2012,Alekhin2012,Bezrukov:2012sa,Degrassi2012a,Masina2013,Branchina2013,Buttazzo2013,DiLuzio:2014bua,Nielsen:2014spa,Andreassen:2014gha,Espinosa:2015qea,Bednyakov:2015sca} (see also references within, and Refs.\ \cite{Lindner1986,Schrempp1996,Altarelli1994}).  Although the first bounds on the Higgs mass from vacuum stability were estimated over three decades ago \cite{Cabibbo1979,Hung1979}, the experimentally measured $125$\,GeV Higgs \cite{Aad2012,Chatrchyan2012} has created increased interest in this topic, as this value results in the existence of a high-energy true vacuum state.  If the Higgs field tunnelled from the electroweak vacuum to the global minimum at any point in space, a bubble of low-energy vacuum would form, a process known as \textit{bubble nucleation}.  This bubble would propagate outwards at very nearly the speed of light \cite{Sher1989}, converting all space in its future light-cone into the true vacuum state, having catastrophic results.

The lifetime of the SM electroweak vacuum does exceed the lifetime of the Universe.  It thus exists in a \textit{meta-stable} state, and the SM is not inconsistent with reality.  Yet there is still a non-zero probability of decay, such that a transition to the true vacuum is at some point inevitable.

If new physics does exist below the Planck scale, this may dramatically decrease or increase the lifetime of the false vacuum.  For example, as shown in Refs.~\cite{Branchina2013,Branchina2015,DiLuzio:2015iua}, using effective couplings suppressed by $1/M_{\text{Pl}}$, high-energy physics could indeed reduce the lifetime of the electroweak false vacuum down to a fraction of a second.  However, even the most minimal extensions to the SM can have a notable impact on the stability of the electroweak vacuum.

Examples of the most recent calculations of the vacuum stability of the SM are those of Ref.~\cite{Buttazzo:2013uya} and Ref.~\cite{Bednyakov:2015sca}, using three-loop RGEs and two-loop threshold corrections to the quartic Higgs coupling $\lambda$ at the weak scale.  This is referred to as next-to-next-to leading order (NNLO).  Before these results the state of the art was next-to leading order (NLO) using two-loop RGEs and one-loop threshold corrections at the weak scale \cite{Casas1995,Casas1996,Isidori2001,Burgess2002,Isidori2008,ArkaniHamed2008,Bezrukov2009,Hall2010,Ellis2009}.   These vacuum stability studies were however performed in a flat background spacetime.  Studies have also been done with the SM coupled to gravity in a curved spacetime \cite{Loebbert2015,Czerwinska2015}, motivated by the possibility that the region of the true minimum (typically $> 10^{10}$\,GeV) may be influenced by the gravitational field.

There have been numerous studies of vacuum stability in BSM theories, such as scalar extensions of the Higgs sector \cite{Gonderinger2010,Drozd2011,Chen2012,Belanger2013a,Khan2014,Alanne2014,Han2015a,Kanemura2015}.  In these scenarios the addition of a scalar particle can have a positive contribution to the running of the quartic Higgs coupling, which for certain values of the new couplings can completely remove the high energy global minimum \cite{Khan2014}. See Ref.\ \cite{SSDM} for a discussion of vacuum stability in a scalar singlet-extended Higgs-sector model.

Within \GB we currently use \FlexibleSUSY \cite{Athron:2014yba} to solve the RGEs up to the Planck scale.
For a generic model we are limited to two-loop RGEs and one-loop threshold corrections when using SARAH output.  However, higher-order RGEs may be added where these are known, such as in the SM.  Therefore even for the scalar singlet model, where we use generated RGEs, this is comparable to the precision of the most recent studies \cite{Khan2014,Han2015a}.  Additionally, as with the \GB spectrum object itself, the vacuum stability module functions may be used in conjunction with any spectrum generator with RGE running functionality that can be interfaced to \GB.

The details of the likelihood function and the relevant physics used to classify a parameter point as either stable, meta-stable or unstable are presented in the following section.
\subsubsection{Likelihood details}\label{sec:vs_likelihood}
In this section we outline the details of the derivation leading to the likelihood function used in the \GB vacuum stability module functions.  The electroweak vacuum obtains a high energy global minimum when the running quartic Higgs coupling becomes negative at a high scale.  For a large renormalisation scale $\mu$ the Higgs potential can be approximated with an effective potential of the form
\begin{align}
V(h)=\frac{\lambda(\mu)}{4}h^4\label{eqn:potential}
\end{align}
for $h\gg v$, where $\mu=\mathcal{O}(h)$.  The scale at which the quartic coupling becomes negative, and thus the size of the potential barrier between the electroweak vacuum and this global minimum, will affect the likelihood of a quantum mechanical tunnelling to this lower-energy state.

If there exists a high-energy global minimum, then there is a non-zero probability of decay to this state, and thus the electroweak vacuum has a finite lifetime.

There are three possible cases for electroweak vacuum stability, outlined below.
\begin{itemize}
\item Stable: If $\lambda(\mu)>0$ for all $\mu<M_{\text{Pl}}$ then the electroweak vacuum is the only minimum of the Higgs potential (up to $M_{\text{Pl}}$) and is therefore absolutely stable.

\item Meta-stable: If there exists a $\mu_0<M_{\text{Pl}}$ such that $\lambda(\mu_0)<0$ \textit{and} the lifetime of the electroweak vacuum state is longer than the age of the universe.

\item Unstable: If there exists a $\mu_0<M_{\text{Pl}}$ such that $\lambda(\mu_0)<0$ \text{and} the lifetime of the electroweak vacuum state is less than the age of the universe.
\end{itemize}

Due to the shape of the resulting likelihood function, the distinction between meta-stability and instability is very well defined, with the expected lifetime of the electroweak vacuum being extremely sensitive to changes in the model parameters.

To determine the stability of a model we calculate the rate of quantum mechanical tunnelling through an arbitrary potential barrier, following the derivation in Ref.~\cite{Coleman1977} obtain an analogous field-theoretic solution.  In this context, a tunnelling event corresponds to a bubble nucleation in which the Higgs field decays to the lower energy state.  We invoke the thin-wall approximation \cite{Kolb1990}, in which the equation of motion of the bubble of radius $R$, nucleated at time $t=0$, is $\left(|\vec{x}|^2-t^2\right)=R^2$.

From quantum mechanics, one can derive the rate of tunnelling for a particle through a potential barrier, which can then be generalised to a rate of bubble nucleation per unit volume per unit time of
 \begin{align}
\Gamma\approx\Gamma_0 e^{-B/\hbar}\label{eqn:T},
\end{align}
where $\Gamma_0$ depends on the size of the past light-cone and $B$ is determined by the shape and size of the potential barrier.

Consider a scalar field $\phi$. For bubble nucleation in flat 4-dimensional spacetime we can take $\phi$ to be a function of $\rho=\left(\tau^2+|\vec{x}|^2\right)^{1/2}$ only (where $\tau=it$), as there exists an $O(4)$ symmetry.  $B$ is given by the action \cite{Coleman1977}
\begin{align}
B=S_E\equiv2\pi^2\int_0^{\infty}\rho^3{\rm d}\rho \left[\frac{1}{2}\left(\frac{{\rm d}\phi}{{\rm d}\rho}\right)^2+V\right]\label{eqn:EucAction}
\end{align}
satisfying the Euclidean equation of motion for $\phi$,
\begin{align}
\frac{{\rm d}^2\phi}{{\rm d}\rho^2}+\frac{3}{\rho}\frac{{\rm d}\phi}{{\rm d}\rho}=\frac{{\rm d} V}{{\rm d} \phi}.\label{eqn:Euclidean_spherical}
\end{align}
The boundary conditions of this equation of motion can be combined into the single requirement that $\lim_{\rho\rightarrow\infty}\phi(\rho)=q_0$. The solution to Eq.~\ref{eqn:Euclidean_spherical} is known as the \text{bounce} solution, as it describes, in Euclidean time, a solution that is in the unstable vacuum at $t\rightarrow -\infty$, the global vacuum at $t=0$, and bounces back to the starting point at $t\rightarrow \infty$.

The process outlined above for a simple scalar field can be applied to the Higgs field, with the use of the approximation $V(h) \approx \lambda(h) h^4/4$.  If we perform this calculation at tree level, we can obtain an approximation for the coefficient $B$. Taking $\lambda(h)=\lambda$, $\lambda<0$, we have a solution \cite{Isidori2001} to the analogue of Eq.~\ref{eqn:Euclidean_spherical} as
\begin{align}
h(\rho)=\sqrt{\frac{2}{\left|\lambda\right|}}\frac{2R}{\rho^2+R^2}.
\end{align}
where $R$ is a dimensional factor associated with the size of the bounce.  This can be a relevant quantity such as the height of the barrier, or the change in renormalisation scale between adjacent minima; we shall use the latter.  From Eq.~\ref{eqn:EucAction} we then obtain the action
\begin{align}
S_E=\frac{8\pi^2}{3\left|\lambda\right|},
\end{align}
The validity of the approximation $V\approx \lambda h^4/4$, with $\lambda<0$, is not immediately evident, as $h=0$ is an unstable maximum, and indeed any value of $h>0$ is unstable.  However, as stated in Ref.~\cite{Isidori2001}, the bounce solution is not a constant field configuration, so requires a non-zero kinetic energy.  Because of this, the required bounce solution is suppressed even in the absence of a potential barrier, and thus this is still a valid approximation \cite{Lee1986}.

Now that we have an expression for $B$ we need to determine the pre-factor.  The explicit form of this pre-factor, $\Gamma_0$, was first calculated in Ref.~\cite{Callan1977} by taking account of one-loop quantum corrections.  Yet because the behaviour of the exponential in $\Gamma$ dominates to such a large extent, the value of $\Gamma_0$ need only be an approximation, so we follow the analysis of Ref.~\cite{Sher1989} and determine $\Gamma_0$ by a dimensional reasoning.

If we take the Planck constant and the speed of light to be unity ($\hbar=c=1$), then the rate of decay per unit time per unit volume, $\Gamma/V$ has units of $[\textrm{length}]^{-4}$.  Thus $\Gamma_0$ must have units of $[\textrm{length}]^{-4}$ or $[\textrm{energy}]^4$.  The characteristic scale relevant in this problem is the width or height of the potential barrier, thus we take $\Gamma_0\approx \Lambda_B^4=1/R^4$, where we henceforth take $\Lambda_B$ to be the energy at which $\lambda(\mu)$ is at a minimum\footnote{If the minimum value of the running quartic coupling $\lambda_{\textrm{min}}<0$ is achieved after $M_{\textrm{Pl}}$, but $\lambda(M_{\textrm{Pl}})<0$, we take $\lambda_{\textrm{min}}=\lambda(M_{\textrm{Pl}})$.}.

$\Gamma/V$ is the rate of decay per unit time per unit volume.  As we are ultimately interested in the probability of the Universe having decayed in our past light cone, we multiply $\Gamma/V$ by the volume of the past light cone $\sim$$T_U^4$, where $T_U$ is the age of the universe.\footnote{This can be computed more rigorously for a standard FLRW cosmology, see Ref.\ \cite{Buttazzo2013}, however to the level of detail required here this result is equivalent.}  Thus we obtain a rate
\begin{align}
\Gamma \approx (T_u\Lambda_B)^4 e^{-S_E},\label{eqn:prob_decay}
\end{align}
and we obtain the total expected decay rate in the past light cone
\begin{align}
p\approx \left(e^{140}\frac{\Lambda_B}{M_{\textrm{Pl}}}\right)^4\exp\left({-\frac{8\pi^2}{3\left|\lambda(\Lambda_B)\right|}}\right)\label{eqn:prob1}.
\end{align}
Here $\lambda(\Lambda_B)$ is the minimum value of the quartic Higgs coupling, and we have expressed the age of the Universe as  $T_u\approx e^{140}/M_{\textrm{Pl}}$.\footnote{Eq.~\ref{eqn:prob1} may be expressed in different units.  For example, Ref.~\cite{Sher1989} expresses the age of the universe in ``units of the electroweak scale", as $T_u\approx e^{101}$, where the mass of the $Z$ boson is set as $m_{z}=1$.  Similarly Ref.~\cite{Degrassi2014} expresses the probability in what would equivalently be called ``units of the Planck scale", but leave the original factor of $M_{\textrm{Pl}}$ in the expression; this is the style we follow.}

The arguments used to arrive at Eq.~\ref{eqn:prob1} were based only on dimensional analysis.  Although this quantity is now dimensionless it cannot be immediately interpreted as a probability, instead it should be interpreted as the expected value for the random variable $k$, where $k$ is the number of decay events that occurred in the time given (in this case $\approx 10$ billion years).\footnote{By a decay event we mean the decay of the Universe at our position due to a decay at some point in our past light cone, and thus a decay of the observable Universe.  Of course, more than one decay event does not physically make sense, so all situations where $k\ge1$ are effectively equivalent -- the Universe has decayed to the true vacuum.}  To model the probability that the Universe has actually decayed in the given time interval, we use a Poisson distribution, $\Theta(k ;\lambda)=(\lambda^k/k!) e^{-\lambda}$.  Because we want the likelihood that no decay has occurred in our past light-cone, we calculate the probability that $k=0$, which is given by
\begin{align}
\mathcal{L}=\exp\left[-\left(e^{140}\frac{\Lambda_B}{M_{\textrm{Pl}}}\right)^4\exp\left({-\frac{8\pi^2}{3\left|\lambda(\Lambda_B)\right|}}\right)\right].	\label{eqn:prob2}
\end{align}
The likelihood given by Eq.~\ref{eqn:prob2} is typically either extremely small or exactly one, being extremely sensitive to the value of $\lambda(\Lambda_B)$.  This results in an almost step-function transition from a meta-stable to an unstable Universe when actual Lagrangian-level model parameters are varied.

The likelihood in Eq.~\ref{eqn:prob2} is difficult to study, due to its double exponential behaviour.  In \GB we return the log-likelihood, so this likelihood contains some useful information even within the stable/meta-stable region of the parameter space.  In most cases though, this function will either entirely rule out a point or contribute zero to the total log-likelihood. This is indeed useful for ruling out large regions of a parameter space quickly.

\subsubsection{Code Description}\label{sec:vs_code}

The vacuum stability calculations in \GB are separated into several module functions, shown in Table\ \ref{tab:specbitsingletdmcap}.  As renormalisation group running requires a model spectrum object, these include \cpp{Spectrum} \dependencies{}.  The spectrum must be capable of running to different energy scales, and as such cannot be a simple container spectrum.

The module function \cpp{find_min_lambda}, which has \capability{} \cpp{vacuum_stability}, runs the quartic Higgs coupling to determine the scale at which it reaches a turning point.  This function produces a \cpp{struct} of type \cpp{dbl_dbl_bool}, which contains the expected lifetime of the Universe (\cpp{double}), the scale of the instability (\cpp{double}), and whether or not the couplings remain perturbative up to the scale returned (\cpp{bool}).  If a turning point is not reached before some user-specified high scale (set via \YAML option \yaml{set_high_scale}, which defaults to $M_\mathrm{Pl}$), the high scale is returned instead of the scale of the turning point.  These calculations can be applied to any model with a Higgs field that does not mix with other scalars.

Derived from the results of \cpp{find_min_lambda} is the physical observable
\begin{itemize}
\item  \cpp{expected_lifetime}: the expected lifetime of the electroweak vacuum in years
\end{itemize}
and the likelihood functions
\begin{itemize}
\item \cpp{VS_likelihood}: the log-likelihood of vacuum decay not having yet occurred, given the current age of the Universe
\item \cpp{check_perturb_min_lambda}: the perturbativity of the running couplings at the instability scale.  To allow this function to be used as a log-likelihood, it returns 0 for models where the couplings remain perturbative up to the instability scale, and invalidates the model otherwise.
\end{itemize}

In \cpp{VS_likelihood}, the expected age is compared to the current age of the Universe, which is currently hard-coded to $13.6\times10^9$ years.  If $\lambda$ does not have a turning point before $M_{\textrm{Pl}}$, then the lifetime returned is $1\times 10^{300}$ years.

The function \cpp{check_perturb_min_lambda} is an efficient way to check that the couplings of a theory remain perturbative up to $M_\mathrm{Pl}$ or some alternative scale, chosen via the option \yaml{set_high_scale} of function \cpp{find_min_lambda}.  By performing this check directly in \GB, we avoid any need to rely on similar checks implemented in some (but not all) spectrum generators.  The actual calculation is performed simultaneously with the vacuum stability check, and is thus extracted by \cpp{check_perturb_min_lambda} from the output of the \cpp{vacuum_stability} capability.  The result is a flag set \cpp{true} only if all spectrum parameters tagged \cpp{dimensionless} are less than $(4\pi)^{1/2}$ at a set of points evenly distributed in log space between the electroweak scale and the smaller of $\Lambda_B$ and the high scale chosen for \cpp{find_min_lambda}.  The default is to test 10 points; this can be reset with the \YAML option \yaml{check_perturb_pts} of \cpp{find_min_lambda}.

A simple example of the \GB input \YAML commands for using the vacuum stability likelihood, and outputting the Higgs pole mass and the expected age of the universe as observables, is given below.
\begin{lstcpp}
    - capability: print_SingletDM_spectrum
      purpose: Observable

    - capability: VS_likelihood
      purpose: LogLike

    - capability: expected_lifetime
      purpose: Observable
\end{lstcpp}

\begin{figure*}[t]
\centering
\includegraphics[width=0.75\textwidth]{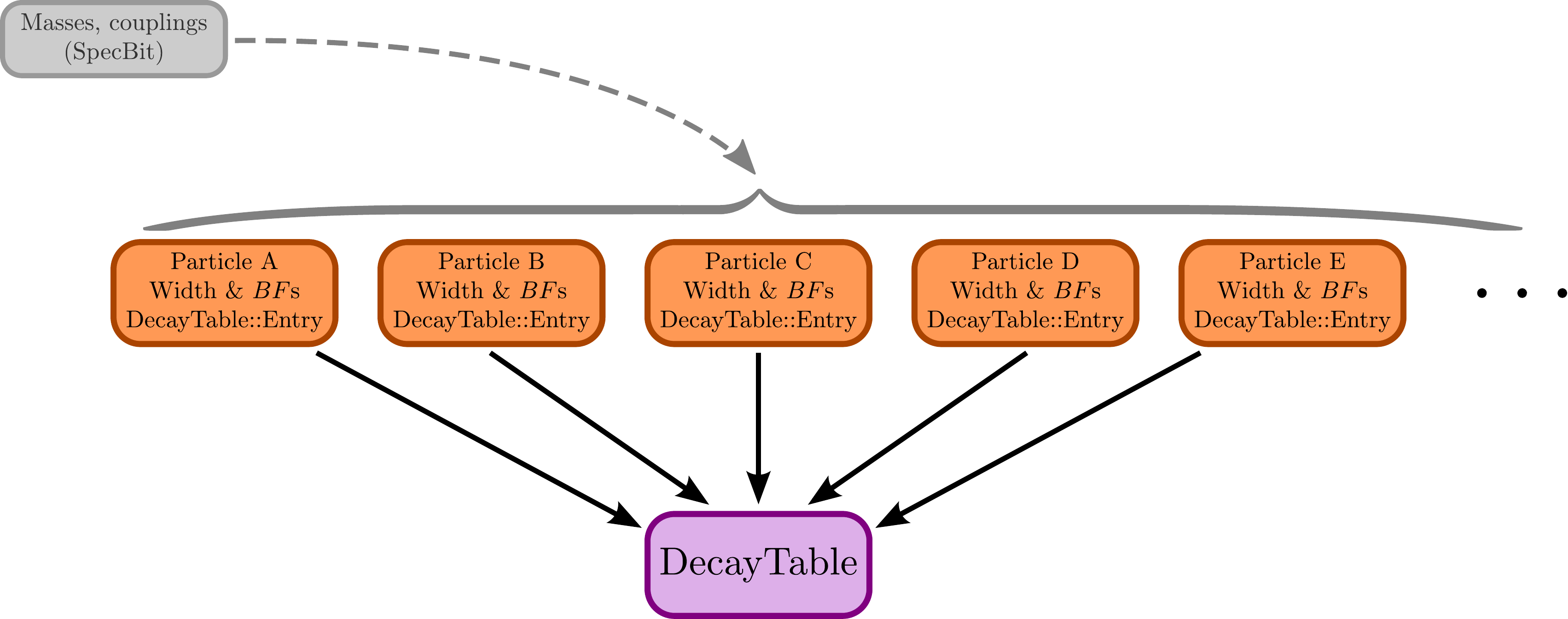}
\caption{Schematic representation of the structure of \decaybit.  Masses and couplings are provided from \specbit{} {\color[HTML]{7A7A7A}(grey box)}, and used to calculate decay widths and branching fractions ($BF$s) for each particle individually, sometimes with with the assistance of backends such as \SUSYHIT and \FH.  The widths and $BF$s are gathered into a single \protect\cpp{DecayTable::Entry} for each particle {\color[HTML]{FF7621}(orange boxes)}.  These are then collected into a single \GB\ \cpp{DecayTable} {\color[HTML]{78157C}(purple box)}, containing all decays of the entire spectrum, and eventually passed on to other \GB modules.}
\label{fig:decaybit}
\end{figure*}

\subsection{Higgs couplings}
\label{sec:higgs_couplings}

In addition to providing particle spectra to the rest of \GB, \specbit can also produce Higgs couplings, which it provides as a compact \cpp{HiggsCouplingsTable}.  The Higgs couplings are required by \higgsbounds/\higgssignals, which are discussed in the ColliderBit manual, Ref.\ \cite{ColliderBit}. These objects carry all decays of neutral and charged Higgs bosons present in the theory under investigation, their $CP$ eigenvalues, a list of all particles in the theory that Higgses can decay invisibly to, values of the SM-normalised effective couplings of each neutral Higgs to $WW$, $ZZ$, $t\bar{t}$, $b\bar{b}$, $c\bar{c}$, $s\bar{s}$, $\tau^+\tau^-$, $\mu^+\mu^-$, $Z\gamma$ and $h\gamma$, and additional comparison decays of SM Higgses with the same masses as the neutral BSM Higgses.  \specbit obtains all decays from \decaybit.  It makes the $CP$ and invisible state identifications within its own functions.  The effective couplings can be obtained in two different ways.  The first is to build them directly from couplings provided by dedicated routines in external calculators, as in \cpp{MSSM_higgs_couplings_FH} (see Table \ref{tab:specbitmssmcap}), which retrieves couplings from \FH (see Sect.\ \ref{decays:MSSM} for details of this interface).  Alternatively, they can be estimated on the basis of the partial width approximation, where the squared effective coupling is taken as the ratio of the partial widths to the relevant final state of the BSM and equivalent SM-like Higgs.  Examples of the partial width approach are \cpp{MSSM_higgs_couplings_pwid} (Table \ref{tab:specbitmssmcap}) and \cpp{SingletDM_higgs_couplings_pwid} (Table \ref{tab:specbitsingletdmcap}).

Note that all Higgs couplings available in \GB presently refer to loop-corrected values; tree-level couplings could be added if required, but should be differentiated from the present couplings by assigning them an alternative capability (e.g.\ \cpp{Higgs_Couplings_tree_level}).

\section{DecayBit}
\label{Sec:DecayBit}

All particle decay widths used in \GB are provided by the dedicated
module \decaybit.  These range from known decays of SM particles, to
modifications of SM decay widths and branching fractions due to new
physics (e.g. top and Higgs decays), to decay widths of new particles
present in BSM theories.

These data are needed by \colliderbit \cite{ColliderBit} for predicting Higgs decay rates and calculating corresponding LHC Higgs likelihoods. They also allow \colliderbit to connect predictions for LHC production of BSM particles with observable final states, by simulating particle collisions and subsequent decays of the particles produced in the initial interaction.  The same data are required by \darkbit \cite{DarkBit} to simulate cascade decays of final states produced in dark matter annihilation and decay, allowing it to make predictions for cosmic ray fluxes in dark matter indirect detection experiments.  Decay widths of SM and BSM particles are also important for predicting the relic density of dark matter, through their participation in thermal freeze-out, or if dark matter is produced non-thermally from decays of some heavy resonance.

Although \decaybit provides uncertainties on some decay widths and branching fractions (currently only those that we take from the PDG \cite{PDB}), these are not utilised as yet by other \GB modules (indeed not even by \higgssignals nor \higgsbounds).  Future likelihood functions are expected to make use of these data, and future versions of \decaybit are anticipated to also feature calculation of uncertainties on partial and total widths of BSM states.

The overall structure of \decaybit is outlined in Fig.\ \ref{fig:decaybit}, showing the modular calculation of each particle's decays, and their combination into a complete \cpp{DecayTable} for the model under investigation.  Here we begin by describing the decays currently implemented in \decaybit (Sec.\ \ref{decays}), including the details of the calculations, input data and external codes it employs.  We then go through the specific module functions it offers (Sec.\ \ref{decaybit:functions}), and the details of the code (Sec.\ \ref{code:decaybit}), covering the associated data structures and utility functions, how to add support for new channels and models, and an explicit worked example of the computation of Higgs boson decays.

\subsection{Supported decays}
\label{decays}

\subsubsection{Standard Model}
\label{decays:SM}

\decaybit contains total widths and associated uncertainties for decays of the $W$, $Z$, $t$, $b$, $\tau$ and $\mu$, as well as their antiparticles, and the most common mesons $\pi^0$, $\pi^\pm$, $\eta$, $\rho^0, \rho^\pm$ and $\omega$.  It also includes partial widths to all observed, distinct final states for $W$, $Z$, $t$, $b$, $\tau$, $\mu$, $\pi^0$ and $\pi^\pm$, including many-body final states in many cases (although we do not include such states where the branching fraction would otherwise be contained within a lower-multiplicity channel).  We take these data directly from the Particle Data Group compilation of experimental results \cite{PDB}.  These data are used in GAMBIT in two scenarios. The first is whenever a decay has no relevant BSM contribution, e.g.\ if BSM contributions are heavily loop-suppressed, or if the particle sector of the scan is simply the SM, as when scanning e.g.\ the DM halo model only.  The second is in cases where the only significant BSM modification to the decay of the SM particle is to open a new decay channel, meaning that the new decay can simply be added to the existing SM decays.  Where neither of these scenarios apply, i.e. where BSM corrections are relevant for SM $\to$ SM decays, \decaybit accounts for the corrections via an appropriately model-dependent calculation.

We group together hadronic channels for $W$ and $Z$ decays, assigning them a final state consisting of two generic hadrons.  Note that `hadron' is recognised as a generic particle by the \GB particle database; see Sec 10.1 of Ref.\ \cite{gambit}.

\decaybit also features various calculations of Higgs boson partial and total decay widths.  For a pure SM Higgs, the user can choose to either calculate decays at the predicted value of the Higgs mass with \FH, or extract them from precomputed tables contained in \decaybit.  The latter are the predicted SM Higgs partial and total widths given as a function of Higgs mass in Ref.\ \cite{YellowBook13}.  Using these tables wherever possible is preferable to simply calling \HDECAY, as they include additional higher-order corrections for 4-body final states obtained using \prophecy \cite{Bredenstein:2006rh,Bredenstein:2006ha}.

\subsubsection{Scalar singlet}
\label{decays:SS}

When $\ms < m_h/2$, the Higgs can decay to two singlet scalars.  The partial width for this process is
\beq
  \Gamma_{h\rightarrow{\sss SS}} = \frac{\lhs^2 v_0^2}{32\pi m_h}\left(1 -4 \ms^2/m_h^2\right)^{1/2}.
  \label{Gamma_SS}
\eeq
Whenever this channel is kinematically open, \decaybit calculates the partial width and adds it to the existing set of SM Higgs decays, rescaling the decay branching fractions (BFs) and total width obtained for a pure SM Higgs.

The decay $h\rightarrow SS$ would be entirely invisible at the LHC, so would
contribute to the invisible width of the Higgs boson.  Assuming SM-like
couplings, as is the case for the scalar singlet model, 95\% confidence level
(CL) upper limits on the Higgs invisible width from LHC and Tevatron data are
presently at the level of 19\% \cite{Belanger:2013xza}.  We implement this as a
complete likelihood function in \decaybit, by interpolating the function shown
in Fig.~8 of~\cite{Belanger:2013xza}.

\subsubsection{MSSM}
\label{decays:MSSM}

\decaybit will calculate decays of all sparticles and additional Higgs bosons in the MSSM, including branching fractions to SM and SUSY final states.  It can also calculate SUSY corrections to decays of the top quark and the SM-like Higgs.

The decays of each Higgs may come (independently) from backend functions in either \HDECAY via \SUSYHIT, or \FeynHiggs.  For the calculation of effective Higgs couplings by \specbit (see Sec.\ \ref{sec:higgs_couplings}), \decaybit will also compute decays of an SM Higgs with the same mass as any of the MSSM Higgs bosons, using either \FH or by interpolating its own internal tables.  For masses between 80\,GeV and 1\,TeV, these tables come directly from Ref.\ \cite{YellowBook13}.  Between 1\,GeV and 80\,GeV, and between 1\,TeV and 16\,TeV, we supplement them with additional results that we obtained by running \HDECAY \textsf{6.51} \cite{Djouadi:1997yw}.  For Higgs masses above 1\,TeV, in order to obtain finite results we disabled electroweak corrections and decays to $s\bar{s}$, $b\bar{b}$, $gg$ and $\mu^+\mu^-$ (which are negligible at large masses, where $W^+W^-$, $ZZ$ and $t\bar{t}$ final states dominate).  Because of these settings, and the absence of any corrections obtained with \prophecy \cite{Bredenstein:2006rh,Bredenstein:2006ha} in our supplemental calculations, small discontinuities can be expected at the transitions between different tables.

Top quark decays are available only by calling \FeynHiggs.  Sparticle
decays are available only from \SDECAY via \SUSYHIT.  \decaybit can
calculate Higgs and top widths for arbitrary MSSM-63 models when
employing \FeynHiggs, but is limited to MSSM-20 models for sparticle
and Higgs decays when using \SUSYHIT, due to the fact that \SUSYHIT is
not SLHA2 compliant as models with greater freedom rely on the more
general SLHA2 format allowing full family mixing amongst the
sfermions.

We backend \SUSYHIT 1.5 in a similar fashion to other codes, but we patch it via the \GB build system when it is automatically downloaded, in order to make a number of crucial modifications.  Here we go through the most important of these.

\renewcommand\metavar\metavars
\renewcommand\descwidth{6cm}

\begin{table*}[tp]
\centering
\scriptsize{
\begin{tabular}{l|p{\descwidth}|l|l}
  \textbf{Capability}
      & \multirow{2}{*}{\parbox{\descwidth}{\textbf{Function} (\textbf{Return Type}):
             \\  \textbf{Brief Description}}}
          & \textbf{Dependencies}
          & \textbf{Options} (\textbf{Type})
          \\ & & &
  \\ \hline
  \cpp{t}\metavar{X}\cpp{_decay_rates}
      & \multirow{2}{*}{\parbox{\descwidth}{ \cpp{t}\metavar{X}\cpp{_decays(DecayTable::Entry)}:
              \\ Computes all SM decays of the $t$/$\bar{t}$ quark.}}
          & \cpp{} & \cpp{}
          \\ & & &
  \\ \hline
  \cpp{mu_}\metavar{Y}\cpp{_decay_rates}
      & \multirow{2}{*}{\parbox{\descwidth}{ \cpp{mu_}\metavar{Y}\cpp{_decays(DecayTable::Entry)}:
              \\ Computes all SM decays of the $\mu^+$/$\mu^-$ lepton.}}
          & \cpp{} & \cpp{}
          \\ & & &
  \\ \hline
  \cpp{tau_}\metavar{Y}\cpp{_decay_rates}
      & \multirow{2}{*}{\parbox{\descwidth}{ \cpp{tau_}\metavar{Y}\cpp{_decays(DecayTable::Entry)}:
              \\ Computes all SM decays of the $\tau^+$/$\tau^-$ lepton.}}
          & \cpp{} & \cpp{}
          \\ & & &
  \\ \hline
  \cpp{pi_}\metavar{Z}\cpp{_decay_rates}
      & \multirow{2}{*}{\parbox{\descwidth}{ \cpp{pi_}\metavar{Z}\cpp{_decays(DecayTable::Entry)}:
              \\ Computes all SM decays of the $\pi^0$, $\pi^+$ and $\pi^-$ mesons.}}
          & \cpp{} & \cpp{}
          \\ & & &
          \\ & & &
  \\ \hline
  \cpp{rho_}\metavar{Z}\cpp{_decay_rates}
      & \multirow{2}{*}{\parbox{\descwidth}{ \cpp{rho_}\metavar{Z}\cpp{_decays(DecayTable::Entry)}:
              \\ Computes all SM decays of the $\rho^0$, $\rho^+$ and $\rho^-$ mesons.}}
          & \cpp{} &
          \\ & & &
          \\ & & &
  \\ \hline
  \cpp{eta_decay_rates}
      & \multirow{2}{*}{\parbox{\descwidth}{ \cpp{eta_decays(DecayTable::Entry)}:
              \\ Computes all SM decays of the $\eta$ meson.}}
          & \cpp{} & \cpp{}
          \\ & & &
  \\ \hline
  \cpp{omega_decay_rates}
      & \multirow{2}{*}{\parbox{\descwidth}{ \cpp{ompega_decays(DecayTable::Entry)}:
              \\ Computes all SM decays of the $\omega$ meson.}}
          & \cpp{} & \cpp{}
          \\ & & &
  \\ \hline
  \cpp{Z_decay_rates}
      & \multirow{2}{*}{\parbox{\descwidth}{ \cpp{Z_decays(DecayTable::Entry)}:
              \\ Computes all SM decays of the $Z$ boson.}}
          & \cpp{} & \cpp{}
          \\ & & &
  \\ \hline
  \cpp{W_}\metavar{Y}\cpp{_decay_rates}
      & \multirow{2}{*}{\parbox{\descwidth}{ \cpp{W_}\metavar{Y}\cpp{_decays(DecayTable::Entry)}:
              \\ Computes all SM decays of the $W^+$/$W^-$ gauge boson.}}
          & \cpp{} & \cpp{}
          \\ & & &
          \\ & & &
  \\ \hline
  \cpp{Reference_SM}
      & \multirow{4}{*}{\parbox{\descwidth}{ \cpp{Ref_SM_Higgs_decays_table}\\\cpp{(DecayTable::Entry)}:
              \\ Computes all decays of the Higgs boson in the SM, by interpolating in the tables of Ref.\ \cite{YellowBook13}.}}
          & \cpp{mh} & \cpp{higgs_minmass(double)}
          \\ \ \cpp{_Higgs_decay_rates} & & & \cpp{higgs_maxmass(double)}
          \\ & & &
          \\ & & &
  \\ \cmidrule{2-4}
      & \multirow{3}{*}{\parbox{\descwidth}{ \cpp{Ref_SM_Higgs_decays_FH}\\\cpp{(DecayTable::Entry)}:
              \\ Computes all decays of the Higgs boson in the SM, using Higgs couplings obtained from \FH.}}
          & \cpp{SMlike_Higgs_PDG_code} &
          \\ & & \cpp{FH_Couplings_output} &
          \\ & & \cpp{SLHA_pseudonyms} &
          \\ & & &
          \\ & & &
  \\ \hline
  \cpp{Higgs_decay_rates}
      & \multirow{2}{*}{\parbox{\descwidth}{ \cpp{SM_Higgs_decays(DecayTable::Entry)}:
              \\ Returns all decays of the SM Higgs boson.}}
          & \cpp{Reference_SM} &
          \\ & & \ \cpp{_Higgs_decay_rates} &
  \\ \hline
\end{tabular}
}
\caption{SM decay functions provided by \decaybit.  None of these functions has any backend requirements.  Here \metavar{X} is either empty or equal to \cpp{bar}, \metavar{Y} $\in\{$\cpp{plus}\metavar{,} \cpp{minus}$\}$ and \metavar{Z} $\in\{$\cpp{0}\metavar{,} \cpp{plus}\metavar{,} \cpp{minus}$\}$. All functions with \metavar{X}=\cpp{bar}, \metavar{Y}=\cpp{minus} or \metavar{Z}=\cpp{minus} have a dependency on their $CP$-conjugate equivalent, with \metavar{X} empty, \metavar{Y}=\cpp{plus} or \metavar{Z}=\cpp{plus}.
\label{tab:decaybit:SM}}
\end{table*}

\renewcommand\descwidth{8.5cm}

\begin{table*}[tp]
\centering
\scriptsize{
\begin{tabular}{l|p{\descwidth}|l}
  \textbf{Capability}
      & \multirow{2}{*}{\parbox{\descwidth}{\textbf{Function} (\textbf{Return Type}):
             \\  \textbf{Brief Description}}}
          & \textbf{Dependencies}
          \\ & &
  \\ \hline
  \cpp{Higgs_decay_rates}
      & \multirow{2}{*}{\parbox{\descwidth}{ \cpp{SingletDM_Higgs_decays(DecayTable::Entry)}:
              \\ Computes all decays of the Higgs boson, including to scalar singlets.}}
          & \cpp{SingletDM_spectrum}
          \\ & & \cpp{Reference_SM_Higgs_decay_rates}
  \\ \hline
  \cpp{lnL_Higgs_invWidth}
      & \multirow{2}{*}{\parbox{\descwidth}{ \cpp{lnL_Higgs_invWidth_SMlike(double)}:
              \\ Computes the log-likelihood of the Higgs invisible width.}}
          & \cpp{Higgs_decay_rates}
          \\ & &
  \\ \hline
\end{tabular}
}
\caption{Scalar singlet functions provided by \decaybit. Neither of these functions has any backend requirements nor options.
\label{tab:decaybit:SS}}
\end{table*}

\renewcommand\descwidth{7.5cm}

\begin{table*}[tp]
\centering
\scriptsize{
\begin{tabular}{l|p{\descwidth}|l|l}
  \textbf{Capability}
      & \multirow{2}{*}{\parbox{\descwidth}{\textbf{Function} (\textbf{Return Type}):
             \\  \textbf{Brief Description}}}
          & \textbf{Dependencies}
          & \multirow{2}{*}{\parbox{\bewidth}{\textbf{Backend} \\ \textbf{requirements}}}
          \\ & & &
  \\ \hline \vspace{-3.5mm}\\ \hline
  \cpp{t_decay_rates}
      & \multirow{2}{*}{\parbox{\descwidth}{ \cpp{FH_t_decays(DecayTable::Entry)}:
              \\ Computes MSSM decays of the $t$ quark.}}
          & \cpp{FH_Couplings_output} &
          \\ & & &
  \\ \hline \vspace{-3.5mm}
  \\ \hline
  \cpp{Higgs_decay_rates}
      & \multirow{2}{*}{\parbox{\descwidth}{ \cpp{MSSM_h0_1_decays(DecayTable::Entry)}:
              \\ MSSM decays of the lightest $CP$-even Higgs with \susyhit.}}
          & \cpp{SLHA_pseudonyms} & \susyhit
          \\ & & &
  \\ \cmidrule{2-4}
      & \multirow{2}{*}{\parbox{\descwidth}{ \cpp{FH_MSSM_h0_1_decays(DecayTable::Entry)}:
              \\ MSSM decays of the lightest $CP$-even Higgs with \FH.}}
          & \cpp{SLHA_pseudonyms} &
          \\ & & \cpp{FH_Couplings_output} &
  \\ \hline
  \cpp{h0_2_decay_rates}
      & \multirow{2}{*}{\parbox{\descwidth}{ \cpp{h0_2_decays(DecayTable::Entry)}:
              \\ Decays of the heaviest $CP$-even MSSM Higgs with \susyhit.}}
          & \cpp{SLHA_pseudonyms} & \susyhit
          \\ & & &
  \\ \cmidrule{2-4}
      & \multirow{2}{*}{\parbox{\descwidth}{ \cpp{FH_h0_2_decays(DecayTable::Entry)}:
              \\ Decays of the heaviest $CP$-even MSSM Higgs with \FH.}}
          & \cpp{SLHA_pseudonyms} &
          \\ & & \cpp{FH_Couplings_output} &
  \\ \hline
  \cpp{A0_decay_rates}
      & \multirow{2}{*}{\parbox{\descwidth}{ \cpp{A0_decays(DecayTable::Entry)}:
              \\ Decays of the $CP$-odd MSSM Higgs boson with \susyhit.}}
          & \cpp{SLHA_pseudonyms} & \susyhit
          \\ & & &
  \\ \cmidrule{2-4}
      & \multirow{2}{*}{\parbox{\descwidth}{ \cpp{FH_A0_decays(DecayTable::Entry)}:
              \\ Decays of the $CP$-odd MSSM Higgs boson with \FH.}}
          & \cpp{SLHA_pseudonyms} &
          \\ & & \cpp{FH_Couplings_output} &
  \\ \hline
  \cpp{H}\metavar{Y}\cpp{_decay_rates}
      & \multirow{2}{*}{\parbox{\descwidth}{ \cpp{H}\metavar{Y}\cpp{_decays}:
              \\ Decays of the MSSM charged Higgs with \susyhit.}}
          & \cpp{SLHA_pseudonyms} & \susyhit
          \\ & & &
  \\ \cmidrule{2-4}
      & \multirow{2}{*}{\parbox{\descwidth}{ \cpp{FH_H}\metavar{Y}\cpp{_decays}:
              \\ Decays of the MSSM charged Higgs with \FH.}}
          & \cpp{SLHA_pseudonyms} &
          \\ & & \cpp{FH_Couplings_output} &
  \\ \hline
  \cpp{Reference_SM_other}
      & \multirow{5}{*}{\parbox{\descwidth}{ \cpp{Ref_SM_other_Higgs_decays_table}\\\cpp{(DecayTable::Entry)}:
              \\ Computes all decays of an SM Higgs with mass equal to the mass of the \textit{least} SM-like $CP$-even Higgs boson in the MSSM, by interpolating in the tables of Ref.\ \cite{YellowBook13} and extensions computed with \HDECAY \textsf{6.51}.}}
          & \cpp{SMlike_Higgs_PDG_code} &
          \\  \ \cpp{_Higgs_decay_rates} & & \cpp{MSSM_spectrum} &
          \\ & & &
          \\ & & &
          \\ & & &
          \\ & & &
  \\ \cmidrule{2-4}
      & \multirow{3}{*}{\parbox{\descwidth}{ \cpp{Ref_SM_other_Higgs_decays_FH(DecayTable::Entry)}:
              \\ Computes all decays of an SM Higgs with mass equal to the mass of the \textit{least} SM-like $CP$-even Higgs boson in the MSSM, using Higgs couplings obtained from \FH.}}
          & \cpp{SMlike_Higgs_PDG_code} &
          \\ & & \cpp{FH_Couplings_output} &
          \\ & & \cpp{SLHA_pseudonyms} &
          \\ & & &
  \\ \hline
  \cpp{Reference_SM_A0}
      & \multirow{5}{*}{\parbox{\descwidth}{ \cpp{Ref_SM_A0_decays_table(DecayTable::Entry)}:
              \\ Computes all decays of an SM Higgs with mass equal to the mass of the $CP$-odd Higgs boson in the MSSM, by interpolating in the tables of Ref.\ \cite{YellowBook13} and extensions computed with \HDECAY \textsf{6.51}.}}
          & \cpp{SMlike_Higgs_PDG_code} &
          \\  \ \cpp{_decay_rates} & & \cpp{MSSM_spectrum} &
          \\ & & &
          \\ & & &
          \\ & & &
  \\ \cmidrule{2-4}
      & \multirow{3}{*}{\parbox{\descwidth}{ \cpp{Ref_SM_A0_decays_FH(DecayTable::Entry)}:
              \\ Computes all decays of an SM Higgs with mass equal to the mass of the $CP$-odd Higgs boson in the MSSM, using Higgs couplings obtained from \FH.}}
          & \cpp{SMlike_Higgs_PDG_code} &
          \\ & & \cpp{FH_Couplings_output} &
          \\ & & \cpp{SLHA_pseudonyms} &
          \\ & & &
  \\ \hline \vspace{-3.5mm}\\ \hline
  \cpp{gluino_decay_rates}
      & \multirow{2}{*}{\parbox{\descwidth}{ \cpp{gluino_decays(DecayTable::Entry)}:
              \\ Computes decays of the gluino.}}
          & \cpp{SLHA_pseudonyms} & \susyhit
          \\ & & &
  \\ \hline
  \cpp{chargino}\metavar{Y}\cpp{_}\metavar{A}\cpp{_decay_rates}
      & \multirow{2}{*}{\parbox{\descwidth}{ \cpp{chargino}\metavar{Y}\cpp{_}\metavar{A}\cpp{_decays(DecayTable::Entry)}:
              \\ Computes decays of the charginos.}}
          & \cpp{SLHA_pseudonyms} & \susyhit
          \\ & & &
  \\ \hline
  \cpp{neutralino_}\metavar{B}\cpp{_decay_rates}
      & \multirow{2}{*}{\parbox{\descwidth}{ \cpp{neutralino_}\metavar{B}\cpp{_decays(DecayTable::Entry)}:
              \\ Computes decays of the neutralinos.}}
          & \cpp{SLHA_pseudonyms} & \susyhit
          \\ & & &
  \\ \hline \vspace{-3.5mm}\\ \hline
  \cpp{sup}\metavar{X}\cpp{_}\metavar{C}\cpp{_decay_rates}
      & \multirow{2}{*}{\parbox{\descwidth}{ \cpp{sup}\metavar{X}\cpp{_}\metavar{C}\cpp{_decays(DecayTable::Entry)}:
              \\ Computes decays of the $\tilde{u}$ squarks.}}
          &  & \susyhit
          \\ & & &
  \\ \hline
  \cpp{sdown}\metavar{X}\cpp{_}\metavar{C}\cpp{_decay_rates}
      & \multirow{2}{*}{\parbox{\descwidth}{ \cpp{sdown}\metavar{X}\cpp{_}\metavar{C}\cpp{_decays(DecayTable::Entry)}:
              \\ Computes decays of the $\tilde{d}$ squarks.}}
          &  & \susyhit
          \\ & & &
  \\ \hline
  \cpp{scharm}\metavar{X}\cpp{_}\metavar{C}\cpp{_decay_rates}
      & \multirow{2}{*}{\parbox{\descwidth}{ \cpp{scharm}\metavar{X}\cpp{_}\metavar{C}\cpp{_decays(DecayTable::Entry)}:
              \\ Computes decays of the $\tilde{c}$ squarks.}}
          &  & \susyhit
          \\ & & &
  \\ \hline
  \cpp{sstrange}\metavar{X}\cpp{_}\metavar{C}\cpp{_decay_rates}
      & \multirow{2}{*}{\parbox{\descwidth}{ \cpp{sstrange}\metavar{X}\cpp{_}\metavar{C}\cpp{_decays(DecayTable::Entry)}:
              \\ Computes decays of the $\tilde{s}$ squarks.}}
          &  & \susyhit
          \\ & & &
  \\ \hline
  \cpp{stop}\metavar{X}\cpp{_}\metavar{A}\cpp{_decay_rates}
      & \multirow{2}{*}{\parbox{\descwidth}{ \cpp{stop}\metavar{X}\cpp{_}\metavar{A}\cpp{_decays(DecayTable::Entry)}:
              \\ Computes decays of the $\tilde{t}$ squarks.}}
          & \cpp{SLHA_pseudonyms} & \susyhit
          \\ & & &
  \\ \hline
  \cpp{sbottom}\metavar{X}\cpp{_}\metavar{A}\cpp{_decay_rates}
      & \multirow{2}{*}{\parbox{\descwidth}{ \cpp{sbottom}\metavar{X}\cpp{_}\metavar{A}\cpp{_decays(DecayTable::Entry)}:
              \\ Computes decays of the $\tilde{b}$ squarks.}}
          & \cpp{SLHA_pseudonyms} & \susyhit
          \\ & & &
  \\ \hline
  \cpp{selectron}\metavar{X}\cpp{_}\metavar{C}\cpp{_decay_rates}
      & \multirow{2}{*}{\parbox{\descwidth}{ \cpp{selectron}\metavar{X}\cpp{_}\metavar{C}\cpp{_decays(DecayTable::Entry)}:
              \\ Computes decays of the $\tilde{e}$ sleptons.}}
          &  & \susyhit
          \\ & & &
  \\ \hline
  \cpp{smuon}\metavar{X}\cpp{_}\metavar{C}\cpp{_decay_rates}
      & \multirow{2}{*}{\parbox{\descwidth}{ \cpp{smuon}\metavar{X}\cpp{_}\metavar{C}\cpp{_decays(DecayTable::Entry)}:
              \\ Computes decays of the $\tilde{\mu}$ sleptons.}}
          &  & \susyhit
          \\ & & &
  \\ \hline
  \cpp{stau}\metavar{X}\cpp{_}\metavar{A}\cpp{_decay_rates}
      & \multirow{2}{*}{\parbox{\descwidth}{ \cpp{stau}\metavar{X}\cpp{_}\metavar{A}\cpp{_decays(DecayTable::Entry)}:
              \\ Computes decays of the $\tilde{\tau}$ sleptons.}}
          & \cpp{SLHA_pseudonyms} & \susyhit
          \\ & & &
  \\ \hline
  \cpp{snu}\metavar{X}\cpp{_electronl_decay_rates}
      & \multirow{2}{*}{\parbox{\descwidth}{ \cpp{snu}\metavar{X}\cpp{_electronl_decays(DecayTable::Entry)}:
              \\ Computes decays of the $\tilde{\nu}_{e_\mathrm{L}}$ sneutrinos.}}
          &  & \susyhit
          \\ & & &
  \\ \hline
  \cpp{snu}\metavar{X}\cpp{_muonl_decay_rates}
      & \multirow{2}{*}{\parbox{\descwidth}{ \cpp{snu}\metavar{X}\cpp{_muonl_decays(DecayTable::Entry)}:
              \\ Computes decays of the $\tilde{\nu}_{\mu_\mathrm{L}}$ sneutrinos.}}
          &  & \susyhit
          \\ & & &
  \\ \hline
  \cpp{snu}\metavar{X}\cpp{_taul_decay_rates}
      & \multirow{2}{*}{\parbox{\descwidth}{ \cpp{snu}\metavar{X}\cpp{_taul_decays(DecayTable::Entry)}:
              \\ Computes decays of the $\tilde{\nu}_{\tau_\mathrm{L}}$ sneutrinos.}}
          & \cpp{SLHA_pseudonyms} & \susyhit
          \\ & & &
  \\ \hline
\end{tabular}
}
\caption{MSSM decay functions provided by \decaybit.  None have any options of note.  Here \metavar{A} and \metavar{B} are mass indices that run from 1 to 2 and 1 to 4 respectively, \metavar{C} is a helicity label (\cpp{l} or \cpp{r}), and \metavar{X} and \metavar{Y} are as in Table\ \protect\ref{tab:decaybit:SM} (i.e. empty/\cpp{bar} and \cpp{plus}/\cpp{minus}).
\label{tab:decaybit:MSSM}}
\end{table*}

\renewcommand\descwidth{5cm}

\begin{table*}[tp]
\centering
\scriptsize{
\begin{tabular}{l|p{\descwidth}|l|l}
  \textbf{Capability}
      & \multirow{2}{*}{\parbox{\descwidth}{\textbf{Function} (\textbf{Return Type}):
             \\  \textbf{Brief Description}}}
          & \textbf{Dependencies}
          & \textbf{Options} (\textbf{Type})
          \\ & & &
  \\ \hline
  \cpp{decay_rates}
      & \multirow{3}{*}{\parbox{\descwidth}{ \cpp{all_decays(DecayTable)}:
              \\ Computes all decays of all particles in a given model.}}
          & \metavar{X}\cpp{_decay_rates} & \cpp{drop_SLHA_file(bool)}
          \\ & & & \cpp{SLHA_output_filename(str)}
          \\ & & &
  \\ \cmidrule{2-4}
  \cpp{}
      & \multirow{2}{*}{\parbox{\descwidth}{ \cpp{all_decays_from_SLHA(DecayTable)}:
              \\ Reads decays from SLHA files.}}
          & & \cpp{SLHA_decay_filenames}
          \\ & & & \ \cpp{(std::vector<str>)}
  \\ \hline
  \cpp{SLHA1_violation}
      & \multirow{3}{*}{\parbox{\descwidth}{ \cpp{check_first_sec_gen_mixing(int)}:
              \\ Checks if first and second generation sfermions mix at more than a given \%.}}
          & \cpp{MSSM_spectrum} & \cpp{gauge_mixing_tolerance(double)}
          \\ & & &
          \\ & & &
  \\ \hline
  \cpp{SLHA_pseudonyms}
      & \multirow{5}{*}{\parbox{\descwidth}{ \cpp{get_mass_es_pseudonyms}\\\cpp{(DecayBit::mass_es_pseudonyms)}:
              \\ Computes nearest-match SLHA2 mass eigenstates for each SLHA1 sfermion weak/family eigenstate.}}
          & \cpp{MSSM_spectrum} & \cpp{gauge_mixing_tolerance(double)}
          \\ & & & \cpp{gauge_mixing_tolerance}
          \\ & & & \ \cpp{_invalidates_point_only(bool)}
          \\ & & &
          \\ & & &
  \\ \hline
\end{tabular}
}
\caption{Collector and helper module functions provided by \decaybit.  Here \metavar{X} refers to all particles that exist in a given model.  Note that \cpp{str} is a \GB-specific \cpp{typedef} of \cpp{std::string}.
\label{tab:decaybit:general}}
\end{table*}

\renewcommand\metavar\metavarf

\SUSYHIT is written as a standalone program, whereas \GB requires callable functions in a shared library.  We therefore convert the main program of \SUSYHIT to a function, and modify the makefile to produce a shared library.

Interfacing to \HDECAY and \SDECAY via \SUSYHIT means that we retain the standard \HDECAY and \SDECAY options chosen there: multi-body decays, loop-induced decays and 1-loop QCD corrections to 2-body decays are all enabled, and any running masses and couplings are calculated at the EWSB scale in the \DR scheme.  However, we directly inject the bottom quark pole mass calculated by \specbit into \SUSYHIT, rather than relying on its internal pole mass calculation.

For some rather specific models, the 1-loop QCD corrections $\Delta_{1}$ to the widths of the lighter sparticles are negative and larger than the tree-level results, causing \SUSYHIT to return negative decay widths.  We have so far found models where this is the case for gluinos, sbottoms, stops, sups, sdowns, neutralinos and charginos.  To rectify this, we have patched \SUSYHIT to implement a correction factor designed to approximately mimic resummation.  When the corrections are negative and larger than the tree-level result, instead of obtaining the 1-loop corrected widths as usual in \susyhit
\beq
  \Gamma_{\rm tot} = \Gamma_{\rm tree} (1 + \Delta_{1}),
  \label{oldgamma}
\eeq
we apply the correction as
\beq
  \Gamma_{\rm tot} = \frac{\Gamma_{\rm tree}}{1 - \Delta_{1}},
  \label{newgamma}
\eeq
motivated by a similar improvement implemented
in \textsf{NMSDECAY} \cite{NMSDECAY}.  Here $\Delta_{1} = \Delta_{X,1}
+ \Delta_{Y,1} + \ldots$ refers to the sum of 1-loop corrections to
the partial widths for all 2-body final states $X$, $Y$, etc.  Note
that of course this expression does not {\it actually} resum higher
order corrections correctly.  It is simply employed as an expression
that reproduces the leading order term of Eq.\ \ref{oldgamma} and
will be positive, so that the pathology of the negative width is
avoided.

With the original definition (Eq.\ \ref{oldgamma}) it is thus straightforward to define a 1-loop 2-body branching fraction $\mathcal{B}$ for final state $X$ as
\beq
  \mathcal{B}_X = \Gamma_{X,\rm tree} (1 + \Delta_{X,1}) / \Gamma_{\rm tot}.
  \label{oldbf}
\eeq
However, this no longer follows when $\Gamma_{\rm tot}$ is corrected by Eq.\ \ref{newgamma}.  In this case, we simply revert to determining branching fractions from the tree-level partial and total widths, but retain the more accurate total width obtained by Eq.\ \ref{newgamma} as the total decay width to be used elsewhere in \GB.

Correcting negative widths with Eq.\ \ref{newgamma} has a downside: the total width of a sparticle becomes discontinuous in the model parameters at $\Delta_{1} = -1$.  Whilst this is still an obvious improvement over having non-physical results for $\Delta_{1} < -1$, it is not ideal.  An alternative treatment would be to use Eq.\ \ref{newgamma} whenever $\Delta_{1} < 0$, which would ensure continuity, and to apply the correction to each partial width individually, allowing one-loop branching fractions to still be calculated when $\Delta_{1} < 0$.  However, this would modify the standard behaviour of \SUSYHIT in the regime $-1 < \Delta_{1} < 0$.  In the interests of keeping our modifications to external codes simple and minimal, we therefore consider it preferable to pay the price of a discontinuous width, restricting our modifications to changes that only have an impact when \SUSYHIT would otherwise return unphysical results.

A related issue is that at high SUSY-breaking scales, the \MSbar correction to the tree-level $t$ and $b$ pole masses required for calculating Higgs couplings in \susyhit also becomes large, leading to numerical instabilities.  To remedy this, we import the improved running \MSbar expressions for $m_t$ and $m_b$ from \HDECAY \textsf{6.51} to \susyhit.

Finally, we also implemented a diskless generalisation of the SLHA interface to \SUSYHIT.  This avoids the need to read and write files to disk, allowing \GB spectrum contents to be directly injected into \SUSYHIT common blocks in SLHA format, and decay data to be extracted directly from other common blocks and easily converted to the native \GB\ \cpp{DecayTable} format (described in Sec.\ \ref{code:decaybit}).

We connect to and use \FeynHiggs as a regular \GB backend, without
significant\footnote{We correct some minor bugs preventing compiler
safety checks.} modification.  \GB supports versions \textsf{2.11.2}
and later.  We adopt the settings for the real MSSM recommended in
the \FH documentation, but restrict neutral Higgs mixing to the
$CP$-even states, for compatibility with \higgsbounds/\higgssignals.
For version \textsf{2.12.0} we set \mbox{\fortran{loglevel = 3}} in
order to include NNLL corrections.\footnote{We make no attempt
within \GB to directly remedy any discrepancies seen between \FH and
other codes \cite{Athron:2016fuq,Staub:2017jnp}; we refer the reader
to those papers, and the first \GB MSSM results
papers \cite{CMSSM,MSSM} for further discussion.}  We pass MSSM models
to \FeynHiggs using the function \fortran{FHSetPara}, directly setting
the SM parameters in SLHA2 parametrisation, the $CP$-odd Higgs pole
mass, the $\mu$ parameter at the SUSY scale and the \DR soft terms
(sfermion mass parameter matrices, gaugino mass parameters and
trilinear couplings, also at the SUSY scale), from the outputs of
the \cpp{Spectrum} objects provided by \specbit.  We set
the \fortran{FHSetPara} arguments \fortran{Qt} and \fortran{Qb} to the
renormalisation scale of the \DR soft masses, so that \FH
interprets stop- and sbottom-sector parameters in the \DR scheme, and
converts them internally to the on-shell scheme.

\subsection{Available functions and options}
\label{decaybit:functions}

Here we give a compact census of the module functions in \decaybit.  These are also laid out in Tables \ref{tab:decaybit:SM}--\ref{tab:decaybit:general}.

Antiparticle equivalents of all functions described below are automatically created from the decays of their particle partners, assuming $CP$ symmetry. This need not be assumed when adding additional states or channels to \decaybit.

\subsubsection{Standard Model}
\label{decaybit:functions:SM}

The SM particle decay functions are listed in Table\ \ref{tab:decaybit:SM}.  These are assigned names
\begin{itemize}[topsep=3pt]
\item[] \metavar{particle\_designation}\cpp{\_decays}
\end{itemize}
and \capabilities{}
\begin{itemize}[topsep=3pt]
\item[] \metavar{particle\_designation}\cpp{\_decay\_rates},
\end{itemize}
 where \metavar{particle\_designation} is simply the particle name for neutral isosinglets, \metavar{particle\_designation} = \metavar{particle\_name}\_\{\cpp{plus}\metavar{,} \cpp{minus}\} for charged leptons and gauge bosons, \metavar{particle\_designation} = \{\metavar{particle\_name}\metavar{,} \metavar{particle\_name}\cpp{bar}\} for quarks, and \metavar{particle\_designation} = \metavar{particle\_name}\_\{\cpp{0}\metavar{,} \cpp{plus}\metavar{,} \cpp{minus}\} for isotriplet mesons.

The function
\begin{itemize}[topsep=3pt]
\item[] \cpp{Ref_SM_Higgs_decays_table}
\end{itemize}
constructs \cpp{Reference_SM_Higgs_decay_rates} from the tables of Ref.\ \cite{YellowBook13} (which are determined using \HDECAY \cite{Djouadi:1997yw,Spira:1997dg,Butterworth:2010ym} and \prophecy \cite{Bredenstein:2006rh,Bredenstein:2006ha}), using the pole mass provided by \specbit or \precisionbit (see Sec.\ \ref{decays:SM}).  The alternative function
\begin{itemize}[topsep=3pt]
\item[] \cpp{Ref_SM_Higgs_decays_FH}
\end{itemize}
constructs \cpp{Reference_SM_Higgs_decay_rates} from Higgs couplings obtained from \FH (by e.g.\ \cpp{SpecBit::FH_Couplings}; see Table \ref{tab:specbitmssmcap}).  The result of one or the other of these two functions is presented to the rest of \GB as the definitive \cpp{Higgs_decay_rates}, by function
\begin{itemize}[topsep=3pt]
\item[] \cpp{SM_Higgs_decays}.
\end{itemize}
The two functions are also used by other module functions to compare BSM Higgs decays to equivalent SM decays.

\subsubsection{Scalar singlet}

We highlight the parts of \decaybit specific to the scalar singlet model in Table\ \ref{tab:decaybit:SS}.  Only the \cpp{Higgs_decay_rates} are modified in this model.  The function
\begin{itemize}[topsep=3pt]
\item[] \cpp{SingletDM_Higgs_decays}
\end{itemize}
simply piggybacks off the SM Higgs decay calculation, and then rescales the total width and branching fractions to accommodate the additional $h\rightarrow SS$ contribution.  The SM-like Higgs invisible width likelihood is provided by the function
\begin{itemize}[topsep=3pt]
\item[] \cpp{lnL_Higgs_invWidth_SMlike},
\end{itemize}
which has \capability{} \cpp{lnL_Higgs_invWidth} and depends on the \cpp{Higgs_decay_rates}.

\subsubsection{MSSM}

Table\ \ref{tab:decaybit:MSSM} lists the full set of MSSM decay functions contained in \decaybit.  The naming scheme follows a similar structure of the SM decays, with function names
\begin{itemize}[topsep=3pt]
\item[] \metavar{particle\_designation}\cpp{\_decays}
\end{itemize}
and \capabilities{}
\begin{itemize}[topsep=3pt]
\item[] \metavar{particle\_designation}\cpp{\_decay\_rates}.
\end{itemize}
Here \metavar{particle\_designation} includes the particle name and any relevant indices: \cpp{plus} or \cpp{minus} for charged bosons, \cpp{bar} for sfermionic antiparticles, mass ordering indices for bosons, neutralinos and charginos and third-generation sfermions, and weak eigenstate indicators \cpp{l} or \cpp{r} for first and second-generation sfermions.  Despite the fact that \GB works exclusively in SLHA2 mass-ordered basis internally, we designate first and second-generation sfermion decays in the weak eigenstate basis because \susyhit is not SLHA2-compliant, and left-right mixing is typically small in the first two generations in the MSSM anyway. This will probably be generalised somewhat further in future revisions of \decaybit.  We discuss the conversion between mass, weak and family-mass eigenstates in more detail in Sec.\ \ref{code:decaybit:utils}.

\decaybit can compute MSSM Higgs sector decays using either \susyhit or \FH; it includes separate module functions for each of these options.\footnote{Note that the only module functions in Table\ \ref{tab:decaybit:MSSM} with an explicit backend requirement listed are those that make use of \susyhit.  The corresponding \FH functions instead depend on \cpp{FH_Couplings_output}, which can be obtained from \specbit's function \cpp{FH_Couplings} --- which exhibits the equivalent backend requirement on \FH (see Table \ref{tab:specbitmssmcap}).}  It also includes additional top decays via intermediate off-shell MSSM states, calculable via \feynhiggs.  Similar to the reference functions for the SM-like Higgs described in Sec.\ \ref{decaybit:functions:SM}, \decaybit also provides functions to compute reference decays of an SM Higgs with the same mass as the $A^0$ (capability \cpp{Reference_SM_A0_decay_rates}) and the `other', non SM-like, $CP$-even neutral MSSM Higgs (usually -- but not always -- the heavy Higgs; capability \cpp{Reference_SM_other_Higgs_decay_rates}).  These are necessary for constructing effective Higgs couplings from partial widths with \specbit (see Sec.\ \ref{sec:higgs_couplings}).

\subsubsection{Collectors and helpers}

\decaybit also ships with a number of other useful module functions, given in Table\ \ref{tab:decaybit:general}.  The most important are the collector functions
\begin{itemize}[topsep=3pt]
\item[] \cpp{all_decays} and
\item[] \cpp{all_decays_from_SLHA}.
\end{itemize}
These both have \capability{} \cpp{decay_rates} and return a complete \GB\ \cpp{DecayTable} object (see Sec. \ref{code:decaybit:decaytable} below).  This is a single object with all decays of all known particles in the models under investigation, to be used by the rest of \GB in subsequent physics calculations.  The first of these functions constructs the \cpp{DecayTable} from the results of the relevant individual decay module functions in \decaybit.  The second is a convenience/debug tool that simply bypasses the rest of \decaybit, reading some pre-computed decays from a user-supplied SLHA file and presenting them to the rest of the code as if they had been computed by \decaybit.

The other two module functions have to do with the conversion between sfermion weak/family and mass eigenstates.  The function
\begin{itemize}[topsep=3pt]
\item[] \cpp{check_first_sec_gen_mixing}
\end{itemize}
checks whether first or second generation weak eignestates are mixed beyond the level specified by the option \cpp{gauge_mixing_tolerance}, returning zero if all states are sufficiently pure and an integer between 1 and 6 otherwise, depending on which eigenstate shows the mixing.  This function is mostly intended for saving information about mixing for later inspection.  The actual sparticle decay functions all have \dependencies{} on the \cpp{SLHA\_pseudonyms}, which list the sparticle mass eigenstates containing the largest admixture of each weak eigenstate; the function that returns this list is
\begin{itemize}[topsep=3pt]
\item[] \cpp{get_mass_es_pseudonyms}.
\end{itemize}
The underlying weak/family--mass eigenstate conversion routines are discussed in more detail in Sec.\ \ref{code:decaybit:utils}.

\FloatBarrier

\subsection{Code description and interface details}
\label{code:decaybit}

\subsubsection{The \cpp{DecayTable}}
\label{code:decaybit:decaytable}

\GB includes a specific \cpp{DecayTable} container class for storing and passing decay information to the rest of the code.  A \cpp{DecayTable} contains one or more instances of the subclass \cpp{DecayTable::Entry}, each of which holds the full decay information for a single particle.  The primary job of \decaybit is to construct a \cpp{DecayTable::Entry} for each particle in a given model, and then to put them together into a complete \cpp{DecayTable} that other parts of the code can access.

Each \cpp{DecayTable::Entry} includes the width of the particle, the positive and negative uncertainties on that width, the branching fractions to different decay channels, the uncertainties on the branching fractions, and information about which code(s) (\GB, a backend or some combination) was responsible for computing the decay information.  The width and branching fraction uncertainties are not currently used in any calculations in \GB, and are only set for SM particles (where they are known experimentally from the Particle Data Group \cite{PDB}), but they may be set more completely and used in more advanced likelihood functions in the future, for example in Higgs physics.  Widths can be set and accessed directly with the \cpp{width_in_GeV} field.  Decay channels and branching fractions are added and retrieved with \cpp{set_BF()} and \cpp{BF()}.  The following simple example for the decays of the $W^+$ boson shows these in action:
\begin{lstcpp}
// Declare a set of decays for the
// @\cpppragma{$W^+$}@ DecayTable::Entry Wplus;

// Set the total width
Wplus.width_in_GeV = 2.085;

// Set branching fraction and uncertainty
// for @\cpppragma{$W \rightarrow e\,\nu_e$}@
Wplus.set_BF(0.1071, 0.0016, "e+", "nu_e");

// Find the partial width for @\cpppragma{$W \rightarrow e\,\nu_e$}@
double pw =  Wplus.width_in_GeV
           * Wplus.BF("e+","nu_e");
\end{lstcpp}

The functions \cpp{set_BF()} and \cpp{BF()} each have five different overloads, allowing the decay channel to be specified with any of the particle identifiers recognised by the \GB particle database (see Sec.\ 10.1 in Ref.\ \cite{gambit}).  These are:
\begin{enumerate}
\item a simple argument list of strings corresponding to the final state particles' long names (as in the example above)
\item an argument list of alternating PDG codes and context integers,
\item an argument list of alternating short name strings and integer particle indices,
\item a \cpp{std::vector} of long name strings, and
\item a \cpp{std::vector<std::pair<int,int>>} of PDG code -- context integer pairs.
\end{enumerate}
All of these forms allow an arbitrary number of final state particles to be specified for a given decay channel.  The \cpp{DecayTable::Entry} subclass also has a simple convenience method \cpp{has_channel()} that takes channels specified in any of the above forms, and returns a \cpp{bool} indicating whether the \cpp{Entry} already contains a field corresponding to that decay channel or not.

The individual entries in a full \cpp{DecayTable} can be accessed using semantics reminiscent of a regular \Cpp\ \cpp{std::map}, where the key is the decaying particle, referred to by any of the three possible designators recognised by the \GB particle database.  For example,
\begin{lstcpp}
// Declare a DecayTable
DecayTable BosonDecays

// Add entries (here are some I prepared earlier)
BosonDecays("W+")   = Wplus;
BosonDecays("W-")   = Wminus;
BosonDecays("Z0")   = Z;
BosonDecays("h0_1") = SM_higgs;

// Get the branching fraction for @\cpppragma{$W \rightarrow e\,\nu_e$}@
double enueBF = BosonDecays("W+").BF("e+","nu_e");

// Get underlying Entry for Higgs
// six different ways
DecayTable::Entry h;
h = BosonDecays("h0_1");
h = BosonDecays(25, 0);
h = BosonDecays("h0", 1);
h = BosonDecays.at("h0_1");
h = BosonDecays.at(25, 0);
h = BosonDecays.at("h0", 1);
\end{lstcpp}
In this example, a \cpp{DecayTable} is created for holding the decays of SM bosons.  The decays of the individual channels, each an instance of the \cpp{DecayTable::Entry} subclass, are then added to the table with the \cpp{()} method.  A branching fraction for $W \rightarrow e\,\nu_e$ is then extracted directly from the \cpp{DecayTable}.  The entire entry for Higgs decays is then extracted in six different ways, showing the standard bracket methods, the constant \cpp{at} access method (reminiscent of regular \Cpp\ STL container access), and the three different ways of referring to the particle whose decays are desired from the table.

The \cpp{DecayTable} machinery integrates seamlessly with SLHA and its generalisations, via \cpp{SLHAstruct} \textsf{SLHAea} objects.  Individual \cpp{DecayTable::Entry} instances can be constructed directly from SLHA DECAY blocks provided in \mbox{\cpp{SLHAea::Block}} format, and made to emit them with the method \cpp{getSLHAea_block}.  Likewise, a complete \cpp{DecayTable} can be constructed directly from an SLHA file or an \cpp{SLHAstruct} object containing DECAY blocks, and can be output to either format with \cpp{writeSLHA} and \cpp{getSLHAea}.

Full documentation of the programming interface provided by the \cpp{DecayTable} and \cpp{DecayTable::Entry} classes can be found in
\begin{itemize}[topsep=3pt]
\item[] \term{Elements/include/gambit/Elements/decaytable.hpp}.
\end{itemize}

\subsubsection{Utilities}
\label{code:decaybit:utils}

The module functions \cpp{check_first_sec_gen_mixing} and \cpp{get_mass_es_pseudonyms} (Table\ \ref{tab:decaybit:general}) rely on a series of helper functions that probe the sfermionic MSSM mixing matrices provided by \specbit, in order to determine the gauge composition of a given mass or family-specific mass eigenstate, or the converse for gauge eigenstates.  These functions are most relevant for \decaybit and backends such as \susyhit, but are needed also by \colliderbit and \darkbit.  They therefore ship as part of the main \GB distribution, in the \term{Elements} directory. Functions can be found here for returning: \begin{itemize}
\item the mass eigenstate with largest contribution from a given gauge eigenstate, plus full mixing and composition information if required (\mbox{\cpp{slhahelp::mass_es_from_gauge_es()}})
\item the gauge eigenstate with largest contribution from a given mass eigenstate, plus full mixing and composition information if required (\mbox{\cpp{slhahelp::gague_es_from_mass_es()}})
\item the (6-flavour) mass eigenstate that best matches a given family (2-flavour) mass state, plus full mixing and composition information if required (\mbox{\cpp{slhahelp::mass_es_closest_to_family()}})
\item the family (2-flavour) mass state that best matches a given (6-flavour) mass eigenstate, plus full mixing and composition information if required (\mbox{\cpp{slhahelp::family_state_closest_to_mass_es()}})
\item the two (6-flavour) mass eigenstates that best match a requested generation, as well as the implied (2-flavour) family mixing matrix between the family states. (\mbox{\cpp{slhahelp::family_state_mix_matrix()}})
\end{itemize}
Full documentation can be found in
\begin{itemize}[topsep=3pt]
\item[] \term{Elements/include/gambit/Elements/mssm_slhahelp.hpp}.
\end{itemize}

The \decaybit module proper contains just two utilities of note that are not actual module functions (in the \GB sense).

The first is the class \mbox{\cpp{DecayBit::mass_es_pseudonyms},} which is used as the return type for \cpp{get_mass_es_pse}-\cpp{udonyms} after obtaining the necessary content from the \cpp{slhahelp} functions.  This is essentially just a container of named strings, one for each SLHA1 MSSM weak/family eigenstate sfermion.

The second is the function \cpp{DecayBit::CP_conjugate()}, which takes as its lone argument a \cpp{DecayTable::Entry} and returns a corresponding \cpp{DecayTable::Entry} for the $CP$ conjugate of the original particle, assuming $CP$ symmetry.  The \metavar{X}=\cpp{bar} and \metavar{Y}=\cpp{minus} variants of the module functions in Tables\ \ref{tab:decaybit:SM} and \ref{tab:decaybit:MSSM} consist of nothing more than a call to this function, using their \dependencies{} on the corresponding \metavar{X}=empty and \metavar{Y}=\cpp{plus} variants.

\subsubsection{A worked example: SM-like Higgs decays}

To illustrate some of the workings of \decaybit more concretely, here we go through the example of decays of the Higgs in detail.

The rollcall header \term{DecayBit_rollcall.hpp} declares a series of functions that are each capable of generating a \cpp{DecayTable::Entry} containing the width and branching fractions for Higgs decays:
\begin{lstcpp}
#define CAPABILITY Reference_SM_Higgs_decay_rates
START_CAPABILITY

  #define FUNCTION Ref_SM_Higgs_decays_table
  START_FUNCTION(DecayTable::Entry)
  DEPENDENCY(mh, double)
  #undef FUNCTION

#undef CAPABILITY

#define CAPABILITY Higgs_decay_rates
START_CAPABILITY

  #define FUNCTION SM_Higgs_decays
  START_FUNCTION(DecayTable::Entry)
  DEPENDENCY(Reference_SM_Higgs_decay_rates,
   DecayTable::Entry)
  #undef FUNCTION

  #define FUNCTION SingletDM_Higgs_decays
  START_FUNCTION(DecayTable::Entry)
  DEPENDENCY(Reference_SM_Higgs_decay_rates,
   DecayTable::Entry)
  DEPENDENCY(SingletDM_spectrum, Spectrum)
  ALLOW_MODELS(SingletDM, SingletDMZ3)
  #undef FUNCTION

  #define FUNCTION MSSM_h0_1_decays
  START_FUNCTION(DecayTable::Entry)
  DEPENDENCY(SLHA_pseudonyms,
   DecayBit::mass_es_pseudonyms)
  BACKEND_REQ(cb_widthhl_hdec, (sh_reqd),
   widthhl_hdec_type)
  BACKEND_REQ(cb_wisusy_hdec, (sh_reqd),
   wisusy_hdec_type)
  BACKEND_REQ(cb_wisfer_hdec, (sh_reqd),
   wisfer_hdec_type)
  BACKEND_OPTION( (SUSY_HIT), (sh_reqd) )
  ALLOW_MODELS(MSSM63atQ, MSSM63atMGUT)
  #undef FUNCTION

#undef CAPABILITY
\end{lstcpp}
The first of these, \cpp{Ref_SM_Higgs_decays_table}, calculates the width and decay branching fractions of the pure SM Higgs, interpolating in the tables of Ref.\ \cite{YellowBook13} using the Higgs pole mass (provided by another module function as a \dependency{}).  This decay information is presented to the rest of the code with \capability{} \cpp{Reference_SM_Higgs_decay_rates}, to allow other functions to declare that they depend on knowing the pure SM decays as reference quantities, even if the \textit{actual} decays of the Higgs in the model being scanned turn out to be different to the SM case.  The function \cpp{SM_Higgs_decays} has a \dependency{} on this reference set of decay information, and simply presents it `as is' to the rest of the code as the definitive decays of the Higgs, as indicated by the \capability{} \cpp{Higgs_decay_rates}. The module function \cpp{SingletDM_Higgs_decays} gives the Higgs decays in the case of an SM-like Higgs that can also decay to two singlets.  This function also piggybacks off the SM reference decay information, simply adding in the $h\rightarrow SS$ decay and rescaling the SM branching fractions.  \cpp{MSSM_h0_1_decays} returns decays of the lightest MSSM Higgs.

The latter two functions have corresponding \cpp{ALLOW_MODELS} declarations, whereas \cpp{SM_Higgs_decays} is considered to be universally compatible with all models, and will be used as a fallback in any case where a model-specific Higgs decay function does not exist, i.e. assuming that the Higgs is unaltered from the SM.  The singlet variant also has a \mbox{\dependency{} on} the \cpp{SingletDM_spectrum}, which it uses to extract the $S$ mass and coupling in order to compute the invisible width.

The MSSM variant has no \dependency{} on a \cpp{Spectrum} object, because it instead relies on \mbox{\cpp{BACKEND_REQ}uirements}, which it demands to be filled from its only declared \cpp{BACKEND_OPTION}, \susyhit.  A spectrum object is not needed because the backend initialisation function for \susyhit has this \dependency{} itself, as that is where the pole masses, couplings and mixings are actually fed into \susyhit.  The function \cpp{MSSM_h0_1_decays} does have a \dependency{} on the \cpp{SLHA_pseudonyms} however, as these are needed to interpret the weak/family eigenstate decay widths returned by \susyhit, and to place them into the resulting \cpp{DecayTable::Entry} in terms of mass eigenstates.

The macro calls illustrated in these declarations are described in more detail in the full \GB paper \cite{gambit}.

The SM reference function computes the total width and branching fractions of the Higgs as follows:
\begin{lstcpp}
// Reference SM Higgs decays from Ref. @\cpppragma{\cite{YellowBook13}}@.
void Ref_SM_Higgs_decays_table(
       DecayTable::Entry& result)
{
  using namespace
    Pipes::Ref_SM_Higgs_decays_table;

  // Get the Higgs pole mass
  double mh = *Dep::mh;

  // Invalidate point if @\cpppragma{$m_h$}@ is outside the most
  // reliable part of the LHCHiggsXSWG @\cpppragma{\cite{YellowBook13}}@
  // tables.
  double minmass =
   runOptions->getValueOrDef<double>(90.0,
   "higgs_minmass");
  double maxmass =
   runOptions->getValueOrDef<double>(160.0,
   "higgs_maxmass");
  if (mh < minmass or mh > maxmass)
  {
    std::stringstream msg;
    msg << "Requested Higgs virtuality is " << mh
        << "; allowed range is "
        << minmass << "--" << maxmass << " GeV.";
    invalid_point().raise(msg.str());
  }

  // Set the contents of the Entry
  result.calculator = "GAMBIT::DecayBit";
  result.calculator_version = gambit_version;
  result.width_in_GeV = virtual_SMHiggs_widths
      ("Gamma",mh);
  result.set_BF(virtual_SMHiggs_widths("bb",mh),
                0.0, "b", "bbar");
  ...
}
\end{lstcpp}
First the Higgs mass is extracted from the \cpp{mh} \dependency{}, and checked against the allowed minimum and maximum Higgs mass.  These bounds are used to ensure that the SM-like Higgs is not outside the range over which the values covered by the tables of Ref.\ \cite{YellowBook13} are most reliable.  The calculator and version are then set, and the total width is extracted from the Ref.\ \cite{YellowBook13} tables with the function \cpp{virtual_SMHiggs_widths} (found in \term{Elements/src/virtual_higgs.cpp}).  Finally, branching fractions to different final states are set by calls to \cpp{set_BF()}, in a similar fashion to the first example of Sec.\ \ref{code:decaybit:decaytable}.

The body of \cpp{SM_Higgs_decays} simply rebrands this information from \cpp{Reference_SM_Higgs_decay_rates} to \cpp{Higgs_decay_rates}, indicating to the rest of \GB that it indeed represents the true Higgs decays in this case, and should be used as such in later calculations:
\begin{lstcpp}
// SM decays: Higgs
void SM_Higgs_decays (DecayTable::Entry& result)
{
  namespace Pipes::SM_Higgs_decays;
  result = *Dep::Reference_SM_Higgs_decay_rates;
}

\end{lstcpp}

The decays of the Higgs are harvested and assembled with the decays of all other particles into a single \cpp{DecayTable}, by the function \cpp{all_decays}:
\begin{lstcpp}
#define CAPABILITY decay_rates
START_CAPABILITY

  #define FUNCTION all_decays
  START_FUNCTION(DecayTable)
  DEPENDENCY(Higgs_decay_rates, DecayTable::Entry)
  ...
  MODEL_CONDITIONAL_DEPENDENCY(h0_2_decay_rates,
   DecayTable::Entry, MSSM63atQ, MSSM63atMGUT)
  ...
  #undef FUNCTION

#undef CAPABILITY
\end{lstcpp}
Here we see the generic \dependency{} upon \cpp{Higgs_decay_rates}; regardless of the model being scanned, \cpp{all_decays} requires a \cpp{DecayTable::Entry} that describes the decays of the Higgs boson.  Depending on which model is actually being scanned, this can end up being fulfilled by any one of the three functions with \capability{} \cpp{Higgs_decay_rates} described above.  The function \cpp{all_decays} also has a \cpp{MODEL_CONDITIONAL_DEPENDENCY} on the set of decays of each BSM particle, such as the second $CP$-even Higgs in the MSSM.  The function body of \cpp{all_decays} simply takes all the entries that it depends on, and adds them to a single \cpp{DecayTable} object, along the lines of the second example in Sec.\ \ref{code:decaybit:decaytable}.

\subsubsection{Adding support for new models and programs}

Supporting a new model in \decaybit is a matter of: \begin{enumerate}
\item writing additional module functions that can generate \cpp{DecayTable::Entry} objects for the novel field content,
\item adding new module functions that can compute new decays of existing particles in the new model, and
\item adding \dependencies{} on these new functions to \cpp{all_decays}.
\end{enumerate}

Adding a new backend capable of calculating decay rates and branching fractions is essentially the same as adding any new backend, and should follow the recipe presented in the main \GB paper \cite{gambit}.  After this, using the new backend from \decaybit requires writing new module decay functions designed to take advantage of it, or modifying existing ones to do so.  In some rare cases, the backended functions might have the same signature as those of an existing backend, in which case the backend simply needs to be selected at the \YAML file level, using the \yaml{Rules} section (see Sec.\ 6 of \cite{gambit}).

\begin{figure}[t]
\centering
\includegraphics[width=\columnwidth]{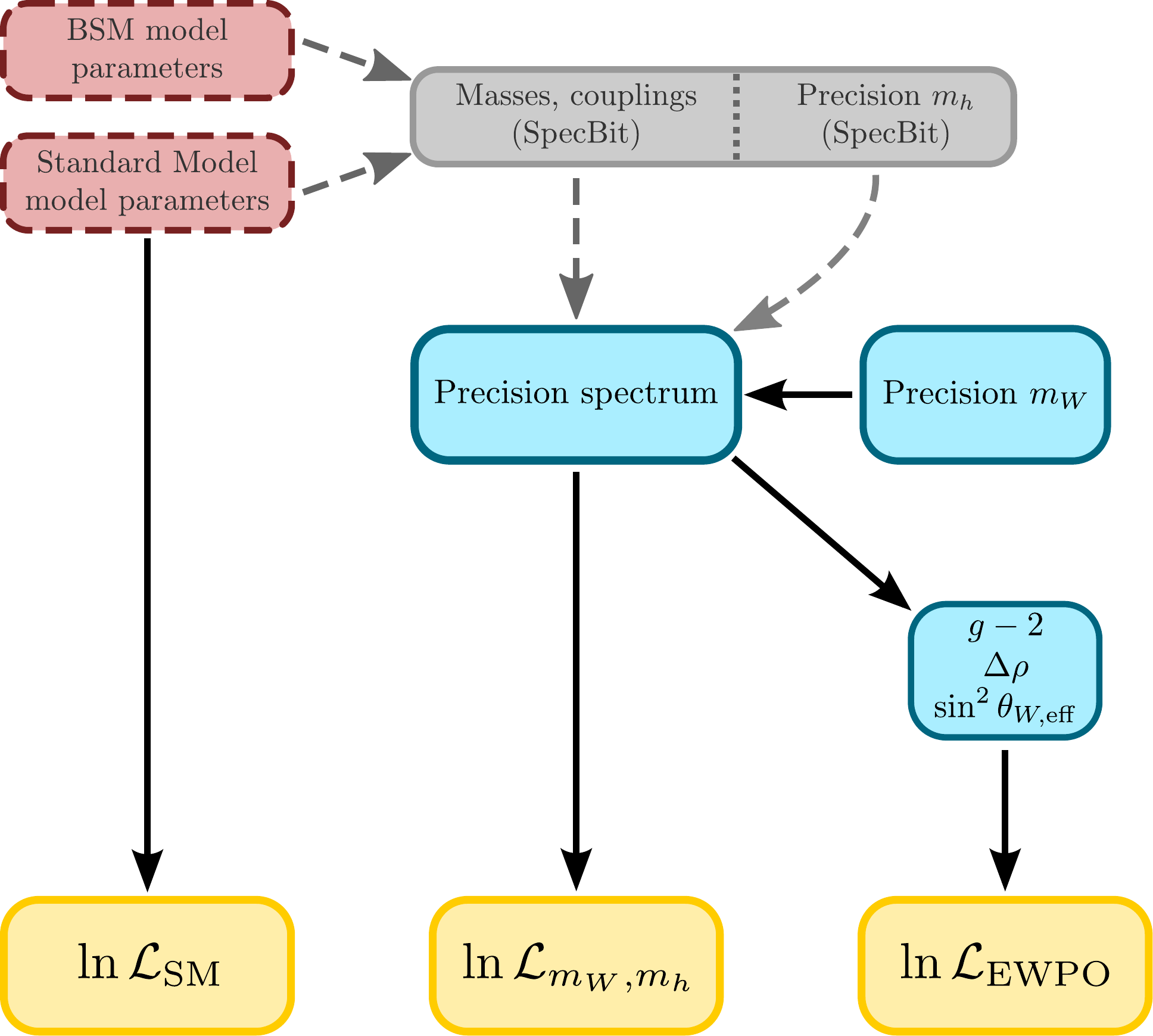}
\caption{Schematic representation of the structure of \precisionbit.  Model parameters {\color[HTML]{D64655}(red boxes)} are used by \specbit to provide masses and couplings {\color[HTML]{7A7A7A}(grey box)} to \precisionbit, and are also used by \precisionbit directly to compute most SM nuisance likelihoods {\color[HTML]{FA9800}(left yellow box)}.  \precisionbit uses the precision Higgs masses from \specbit{} {\color[HTML]{7A7A7A}(grey box right)}, along with its own precision calculation of the $W$ mass {\color[HTML]{4297FF}(upper-right blue box)}, to convert the spectrum provided by \specbit into a precision-updated version {\color[HTML]{4297FF}(upper-left blue box)}.  It then uses the improved spectrum for calculating likelihoods associated with the Higgs and $W$ masses {\color[HTML]{FA9800}(middle yellow box)}, as well as a suite of electroweak precision observables (EWPO) {\color[HTML]{4297FF}(lower blue box)}.  It then feeds the EWPO predictions to its own experiment-based EWPO likelihoods {\color[HTML]{FA9800}(right yellow box)}.}
\label{fig:precisionbit}
\end{figure}

%

\section{PrecisionBit}
\label{Sec:PrecBit}

The \precisionbit module serves a number of related purposes.  It provides so-called nuisance likelihoods, which describe the uncertainties on known quantities that are important inputs to other calculations, which one might like to vary within their experimentally-allowed ranges in the course of a scan.  This functionality is discussed in Sec.\ \ref{precisionbit:SM}.  Examples include the mass of the top quark and the strong coupling; similar quantities relevant only to dark matter observables are dealt with in \darkbit (e.g. the strange quark content of the proton and the local density of dark matter).

\precisionbit is also responsible for calculating BSM corrections to precision SM observables, such as the mass of the $W$ boson and the weak mixing angle $\theta_W$, and providing correspondingly corrected pole masses and couplings to the rest of \GB wherever appropriate.  It also provides likelihood functions for quantifying the agreement of the predicted corrections to the precision observables with experimental data.  We discuss \precisionbit's ability to compute MSSM precision corrections and likelihoods in Sec.\ \ref{precisionbit:MSSM}.  Except for precision mass corrections, \precisionbit does not (yet) support the calculation of precision corrections in non-supersymmetric models; these will be implemented as relevant when the corresponding models are implemented in \GB.

Fig.\ \ref{fig:precisionbit} shows how these different components of \precisionbit fit together, and how the two `halves' of the module are connected via \specbit and the input SM parameters.

\subsection{Standard Model nuisances}
\label{precisionbit:SM}

\precisionbit provides a number of SM nuisance likelihoods, related to
SM couplings, quark masses and the masses of the $Z$ and $W$
bosons. These likelihoods are generally approximated as $\chi^{2}$-like functions,
comparing experimentally determined central values (and errors) with
the SM input values provided by \GB.  Whenever they are available and relevant,
we add theoretical errors to the experimental ones in quadrature (i.e.\ assuming that they are uncorrelated).\footnote{Some members of the LHC Higgs Cross-section Working Group advocate a linear combination (see Sec.\ 12.5 of \cite{Dittmaier}), on the basis of an assumed flat prior on the theoretical systematic.  However, toy Monte Carlo simulation suggests that linear combination is not a statistically valid procedure, even for this choice of prior, except in the extreme limit of complete correlation between experimental and theoretical uncertainties.  Here we assume a Gaussian prior on the theoretical systematic anyway, so combination in quadrature with the (Gaussianly-distributed) experimental uncertainties is certainly well justified, as arguments associated with flat priors do not apply.}

\subsubsection{Couplings}

Likelihoods associated with the Fermi coupling constant ($G_{F}$) and
fine-structure constant ($\alpha_{\mathrm{em}}$) are constructed by
comparing model input values with those determined from electroweak
global fits, with $G_{F} = (1.1663787 \pm 0.0000006) \times 10^{-5}$ GeV$^{-2}$ and
$\alpha_{\mathrm{em}}(m_{Z})^{-1} = 127.940 \pm 0.014$
(\MSbar scheme)~\cite{PDB}. Similarly, the strong
coupling ($\alpha_{s}$) likelihood is calculated using a
$\chi^{2}$-averaged value of $\alpha_{s}(m_{Z}) = 0.1185 \pm 0.0005$ (\MSbar scheme) taken
from the sub-field of lattice determinations~\cite{PDB}. Uncertainties
are taken to correspond to $1\,\sigma$ confidence intervals with no
additional theoretical uncertainties. The \precisionbit functions
providing these likelihoods are summarised in Table~\ref{tab:precisionbit:SM}.

\renewcommand\metavar\metavars
\renewcommand\descwidth{7cm}

\begin{table*}[tp]
\centering
\scriptsize{
\begin{tabular}{l|p{\descwidth}|l|l}
  \textbf{Capability}
      & \multirow{2}{*}{\parbox{\descwidth}{\textbf{Function} (\textbf{Return Type}):
             \\  \textbf{Brief Description}}}
          & \textbf{Dependencies}
          & \textbf{Options (Type)}
          \\ & & &
  \\ \hline
\cpp{lnL_alpha_em}
      & \multirow{3}{*}{\parbox{\descwidth}{ \cpp{lnL_alpha_em_chi2(double)}:
              \\ Computes the log-likelihood of the fine-structure
              constant, $\alpha_\mathrm{em}(m_Z)$, in the \MSbar scheme.}}
          & \cpp{SMINPUTS} & \cpp{}
          \\ & & &
          \\ & & &
  \\ \hline
\cpp{lnL_alpha_s}
      & \multirow{3}{*}{\parbox{\descwidth}{ \cpp{lnL_alpha_s_chi2(double)}:
              \\ Computes the log-likelihood of the strong coupling
              constant, $\alpha_s(m_Z)$, in the \MSbar scheme.}}
          & \cpp{SMINPUTS} & \cpp{}
          \\ & & &
          \\ & & &
  \\ \hline
\cpp{lnL_GF}
      & \multirow{3}{*}{\parbox{\descwidth}{ \cpp{lnL_GF_chi2(double)}:
              \\ Computes the log-likelihood of the Fermi coupling
              constant, $G_{F}$.}}
          & \cpp{SMINPUTS} & \cpp{}
          \\ & & &
          \\ & & &
  \\ \hline
\cpp{lnL_light_quark_masses}
      & \multirow{2}{*}{\parbox{\descwidth}{ \cpp{lnL_light_quark_masses_chi2(double)}:
              \\ Computes the joint log-likelihood of the \MSbar masses of the $u$, $d$ and $s$ quarks at $\mu=2$\,GeV.}}
          & \cpp{SMINPUTS} & \cpp{mud_obs(double)}
          \\ & & & \cpp{mud_obserr(double)}
          \\ & & & \cpp{msud_obs(double)}
          \\ & & & \cpp{msud_obserr(double)}
          \\ & & & \cpp{ms_obs(double)}
          \\ & & & \cpp{ms_obserr(double)}
  \\ \hline
\cpp{lnL_mcmc}
      & \multirow{2}{*}{\parbox{\descwidth}{ \cpp{lnL_mcmc_chi2(double)}:
              \\ Computes the log-likelihood of the \MSbar mass $m_c(m_c)$.}}
          & \cpp{SMINPUTS} & \cpp{}
          \\ & & &
  \\ \hline
\cpp{lnL_mbmb}
      & \multirow{2}{*}{\parbox{\descwidth}{ \cpp{lnL_mbmb_chi2(double)}:
              \\ Computes the log-likelihood of the \MSbar mass $m_b(m_b)$.}}
          & \cpp{SMINPUTS} & \cpp{}
          \\ & & &
  \\ \hline
\cpp{lnL_t_mass}
      & \multirow{2}{*}{\parbox{\descwidth}{ \cpp{lnL_t_mass_chi2(double)}:
              \\ Computes the log-likelihood of the top quark pole mass.}}
          & \cpp{SMINPUTS} & \cpp{}
          \\ & & &
  \\ \hline
\cpp{lnL_Z_mass}
      & \multirow{2}{*}{\parbox{\descwidth}{ \cpp{lnL_Z_mass_chi2(double)}:
              \\ Computes the log-likelihood of the $Z$ boson pole mass}}
          & \cpp{SMINPUTS} & \cpp{}
          \\ & & &
  \\ \hline
\cpp{lnL_W_mass}
      & \multirow{2}{*}{\parbox{\descwidth}{ \cpp{lnL_W_mass_chi2(double)}:
              \\ Computes the log-likelihood of the $W$ boson pole mass.}}
          & \cpp{mw} & \cpp{}
          \\ & & &
  \\ \hline
\cpp{lnL_h_mass}
      & \multirow{2}{*}{\parbox{\descwidth}{ \cpp{lnL_h_mass_chi2(double)}:
              \\ Computes the log-likelihood of the $h$ boson pole mass.}}
          & \cpp{mh} & \cpp{}
          \\ & & &
  \\ \hline
\end{tabular}
}
\caption{\precisionbit module functions providing log-likelihood
  calculations related to SM couplings and particle masses. All
  functions depend on \cpp{SMINPUTS} with the exception of the
  likelihoods of the $W$ and $h$ masses, which depend directly on the calculated $W$
  and $h$ masses, respectively.  These are available from both SM and BSM spectra, as summarised
  in Table~\ref{tab:precisionbit:MW}.  Except in very simple scenarios, the more detailed Higgs
  likelihoods of \colliderbit \cite{ColliderBit} are generally to be preferred to \protect\cpp{lnL_h_mass_chi2}, as they
  simultaneously take account of correlated constraints on the mass and couplings.
\label{tab:precisionbit:SM}}
\end{table*}

\renewcommand\metavar\metavarf

\subsubsection{Masses}
\label{Sec:PrecBitMasses}
Likelihoods associated with the \MSbar light quark
($u,d,s$) masses at $\mu=2$\,GeV, $m_{c}(m_{c})$, and $m_{b}(m_{b})$, are calculated by
\precisionbit by comparing
input values from \GB with the corresponding PDG values and
uncertainties~\cite{PDB}, approximating the likelihoods as
$\chi^{2}$ functions. Similarly, the likelihood corresponding to the top quark
mass is evaluated using combined Tevatron and LHC
measurements~\cite{ATLAS:2014wva}. For the light quark masses a
single, joint likelihood is calculated with terms corresponding to the
ratio of $u$ and $d$ masses, the ratio of the $s$ mass to the sum of $u$
and $d$ masses, and $m_{s}$, respectively. Users can optionally
provide their own values for these three quantities.

The $W$ and $Z$ boson mass likelihoods are based on mass measurements and corresponding uncertainties from the Tevatron and LEP experiments~\cite{PDB}.  These strictly correspond to the mass parameter in a Breit-Wigner function rather than pole masses, but the difference is negligible for almost all purposes.  \precisionbit also provides a simple Higgs mass likelihood, using the ATLAS and CMS combination \cite{Aad:2015zhl} and is used directly as the PDG average \cite{PDG15}.  The module functions providing the $W$, $Z$ and $h$ mass likelihoods are
summarised in Table~\ref{tab:precisionbit:SM}. Unlike the other SM particle masses, the model-dependent $W$ and $h$ mass
\dependencies{} are satisfied by functions specific to different model
spectrum types, listed in  Table~\ref{tab:precisionbit:MW}. In this
case, if theoretical uncertainties are available, we add them in
quadrature to the experimental uncertainties when evaluating the
$\chi^{2}$ approximation to the mass likelihoods.  By summing in quadrature, we assume that the experimental and theoretical errors are uncorrelated, which is reasonable given that they come from entirely different sources.  The more detailed Higgs
likelihoods of \colliderbit \cite{ColliderBit} will usually be more useful
than the simple Higgs mass likelihood provided in \decaybit. The latter can offer a lightweight alternative when varying the mass of the SM Higgs as a nuisance parameter in simple scans, as in e.g.\ Ref.\ \cite{SSDM}.

\renewcommand\metavar\metavars
\renewcommand\descwidth{10cm}

\begin{table*}[tp]
\centering
\scriptsize{
\begin{tabular}{l|p{\descwidth}|l}
  \textbf{Capability}
      & \multirow{2}{*}{\parbox{\descwidth}{\textbf{Function} (\textbf{Return Type}):
             \\  \textbf{Brief Description}}}
          & \textbf{Dependencies}
          \\ & &
  \\ \hline
\cpp{mw}
      & \multirow{2}{*}{\parbox{\descwidth}{ \cpp{mw_from_SM_spectrum(triplet<double>)}:
              \\ Provides the $W$ mass and its uncertainties from an SM spectrum.}}
          & \cpp{SM_spectrum}
          \\ & &
    \\ \cmidrule{2-3}
      & \multirow{2}{*}{\parbox{\descwidth}{ \cpp{mw_from_SS_spectrum(triplet<double>)}:
              \\ Provides the $W$ mass and its uncertainties from a scalar singlet model spectrum.}}
          & \cpp{SingletDM_spectrum}
          \\ & &
    \\ \cmidrule{2-3}
      & \multirow{2}{*}{\parbox{\descwidth}{ \cpp{mw_from_MSSM_spectrum(triplet<double>)}:
              \\ Provides the $W$ mass and its uncertainties from an MSSM spectrum.}}
          & \cpp{MSSM_spectrum}
          \\ & &
  \\ \hline
\cpp{mh}
      & \multirow{2}{*}{\parbox{\descwidth}{ \cpp{mh_from_SM_spectrum(triplet<double>)}:
              \\ Provides the $h$ mass and its uncertainties from an SM spectrum.}}
          & \cpp{SM_spectrum}
          \\ & &
    \\ \cmidrule{2-3}
      & \multirow{2}{*}{\parbox{\descwidth}{ \cpp{mh_from_SS_spectrum(triplet<double>)}:
              \\ Provides the $h$ mass and its uncertainties from a scalar singlet model spectrum.}}
          & \cpp{SingletDM_spectrum}
          \\ & &
    \\ \cmidrule{2-3}
      & \multirow{3}{*}{\parbox{\descwidth}{ \cpp{mh_from_MSSM_spectrum(triplet<double>)}:
              \\ Provides the mass and associated uncertainties of the most SM-like Higgs boson in the MSSM.}}
          & \cpp{MSSM_spectrum}
          \\ & & \cpp{SMlike_Higgs_PDG_code}
          \\ & &
  \\ \hline
\end{tabular}
}
\caption{Summary of \precisionbit module functions providing the $W$ and $h$
  masses from different \GB\ \cpp{Spectrum} inputs.  These functions have no
  options nor backend requirements.
\label{tab:precisionbit:MW}}
\end{table*}

\renewcommand\metavar\metavarf

\subsection{MSSM precision observables}
\label{precisionbit:MSSM}

\precisionbit allows for users to calculate a collection of precision
observables using an array of external code packages. These
observables can be accessed directly through module functions, passed as
input to likelihood calculations, or used to update particle spectra
for use throughout \GB.

\subsubsection{External code interfaces}

Currently, \precisionbit makes use of the external code
packages \FeynHiggs \cite{Heinemeyer:1998yj}, \superiso \cite{Mahmoudi:2008tp} and \gmtwocalc \cite{gm2calc} for calculating precision
observables in the MSSM. The \GB interface to \FH was described already in
Sec.~\ref{decays:MSSM}, and the interface to \superiso is
covered in detail in the \flavbit paper \cite{FlavBit}.

The module function \cpp{GM2C_SUSY} calls library routines
from \gmtwocalc to calculate the anomalous magnetic moment of the muon
and translates the result and the errors from $a_\mu$ to
$(g-2)_\mu$.  \gmtwocalc can accept parameters and masses given in
SLHA conventions.  The module function exploits \GB helper functions,
discussed in Section \ref{code:decaybit:utils}, to convert between
SLHA and SLHA2, which \GB uses internally to store the spectrum.  The
inputs from the spectrum object, including the SMINPUTS parameters,
are then passed on to the \gmtwocalc object by the module
function \cpp{GM2C_SUSY}.  One important subtlety is that \gmtwocalc
also has two non-SLHA inputs, which provide input values for the fine
structure constant, $\alpha$, in the Thompson limit and another
$\alpha$ that includes SM fermion contributions to the
on-shell photon vacuum polarization.  These definitions differ from
the \MSbar $\alpha(m_Z)$, the inverse of which appears in entry 1 of
the SLHA input block \cpp{SMINPUTS}.  By default, \GB uses the default
values recommended by the \gmtwocalc authors.   It is possible to overwrite these values using the \YAML options \yaml{GM2Calc_extra_alpha_e_MZ} and \yaml{GM2Calc_extra_alpha_e_thompson_limit}.

\precisionbit also contains an experimental interface to the proprietary
code \textsf{SUSY-POPE} \cite{Heinemeyer:2006px,Heinemeyer:2007bw},
but this is not recommended for general use.  In addition, a new \FH
routine
\fortran{FHEWPO}, based on work used for Refs.\ \cite{Heinemeyer:2013dia,Stal:2015zca}, has very recently been released for beta testing in \FH \textsf{2.13.0-beta}. We understand that this beta is now to be considered
production-stable,\footnote{T.\ Hahn, personal communication} and so
expect to support it in a future version of \GB. Further calculations
of the EWPO may in future become available through the FlexibleTools
collaboration behind \FlexibleSUSY or through use of auto-generation
software, which is now being developed with \gambit.

The various interface functions to different external code packages within
\precisionbit are summarised in Table~\ref{tab:precisionbit:BE}, and
include  calculations of backend-specific data-types for input to
  other functions, accessor functions for specific observables, and
  functions for updating particle spectra to include precision
  corrections to the $W$ and Higgs-sector masses.

\renewcommand\metavar\metavars
\renewcommand\descwidth{8cm}

\begin{table*}[tp]
\centering
\scriptsize{
\begin{tabular}{l|p{\descwidth}|l|l}
  \textbf{Capability}
      & \multirow{2}{*}{\parbox{\descwidth}{\textbf{Function} (\textbf{Return Type}):
             \\  \textbf{Brief Description}}}
          & \textbf{Dependencies}
          & \multirow{2}{*}{\parbox{\bewidth}{\textbf{Backend}
             \\  \textbf{requirements}}}
          \\ & & &
  \\ \hline
\cpp{FH_Precision}
      & \multirow{2}{*}{\parbox{\descwidth}{ \cpp{FH_PrecisionObs(fh_PrecisionObs)}:
              \\ Calculate precision observables in the MSSM with \FeynHiggs.}}
          & \cpp{FH_Couplings_output} & \FeynHiggs
          \\ & & &
  \\ \hline
\cpp{deltarho}
      & \multirow{2}{*}{\parbox{\descwidth}{ \cpp{FH_precision_deltarho(double)}:
              \\ Retrieve parameter $\Delta\rho$ from \FH precision calculations.}}
          & \cpp{FH_Precision} &
          \\ & & &
  \\ \hline
\cpp{prec_mw}
      & \multirow{3}{*}{\parbox{\descwidth}{ \cpp{FH_precision_mw(double)}:
              \\ Retrieve $W$ mass in the MSSM from \FH precision calculations.}}
          & \cpp{FH_Precision} &
          \\ & & &
          \\ & & &
  \\ \hline
\cpp{prec_sinW2_eff}
      & \multirow{3}{*}{\parbox{\descwidth}{ \cpp{FH_precision_sinW2(double)}:
              \\ Retrieve effective leptonic weak mixing angle $\sin^{2}\theta_{W,\mathrm{eff}}$ from \FH precision calculations.}}
          & \cpp{FH_Precision} &
          \\ & & &
          \\ & & &
  \\ \hline
\cpp{edm_e}
      & \multirow{3}{*}{\parbox{\descwidth}{ \cpp{FH_precision_edm_e(double)}:
              \\ Retrieve electron electric dipole moment from \FH precision calculations.}}
          & \cpp{FH_Precision} &
          \\ & & &
          \\ & & &
  \\ \hline
\cpp{edm_n}
      & \multirow{3}{*}{\parbox{\descwidth}{ \cpp{FH_precision_edm_n(double)}:
              \\ Retrieve neutron electric dipole moment from \FH precision calculations.}}
          & \cpp{FH_Precision} &
          \\ & & &
          \\ & & &
  \\ \hline
\cpp{edm_hg}
      & \multirow{3}{*}{\parbox{\descwidth}{ \cpp{FH_precision_edm_hg(double)}:
              \\ Retrieve mercury electric dipole moment from \FH precision calculations.}}
          & \cpp{FH_Precision} &
          \\ & & &
          \\ & & &
  \\ \hline
\cpp{MSSM_spectrum}
      & \multirow{2}{*}{\parbox{\descwidth}{
          \cpp{make_MSSM_precision_spectrum_H_W(Spectrum)}:
              \\ Function to provide an updated MSSM spectrum with precision $W$
              and Higgs masses.}}
          & \cpp{unimproved_MSSM_spectrum} &
\\ & & \cpp{prec_mw} &
\\ & & \cpp{prec_HiggsMasses} &
  \\ \cmidrule{2-4}
      & \multirow{2}{*}{\parbox{\descwidth}{
          \cpp{make_MSSM_precision_spectrum_W(Spectrum)}:
              \\ Function to provide an updated MSSM spectrum with precision $W$
              mass.}}
          & \cpp{unimproved_MSSM_spectrum} &
\\ & & \cpp{prec_mw} &
\\ & & &
  \\ \cmidrule{2-4}
      & \multirow{2}{*}{\parbox{\descwidth}{
          \cpp{make_MSSM_precision_spectrum_none(Spectrum)}:
              \\ Function to present an unimproved MSSM spectrum as if it is a precision spectrum.}}
          & \cpp{unimproved_MSSM_spectrum} &
\\ & & &
\\ & & &
  \\ \hline
\cpp{muon_gm2}
      & \multirow{3}{*}{\parbox{\descwidth}{ \cpp{FH_precision_gm2(double)}:
              \\ Retrieve MSSM contribution to $g-2$ from \FH precision calculations.}}
          & \cpp{FH_Precision} &
          \\ & & &
          \\ & & &
  \\ \cmidrule{2-4}
\cpp{}
      & \multirow{2}{*}{\parbox{\descwidth}{ \cpp{GM2C_SUSY(double)}:
              \\ Calculate MSSM contribution to $g-2$ using \gmtwocalc.}}
          & \cpp{MSSM_spectrum} & \gmtwocalc
          \\ & & &
\\ & & &
  \\ \cmidrule{2-4}
\cpp{}
      & \multirow{2}{*}{\parbox{\descwidth}{ \cpp{SI_muon_gm2(double)}:
              \\ Calculate MSSM contribution to $g-2$ using \superiso.}}
          & \cpp{SuperIso_modelinfo} & \superiso
\\ & & &
  \\ \hline
\cpp{SP_PrecisionObs}
      & \multirow{2}{*}{\parbox{\descwidth}{ \cpp{SP_PrecisionObs(double)}:
              \\ Calculate precision observables in the MSSM with \textsf{SUSY-POPE} (experimental).}}
          & \cpp{} & \textsf{SUSY-POPE}
          \\ & & &
\\ & & &
  \\ \hline
\end{tabular}
}
\caption{\precisionbit module function interfaces to external code
  packages for calculation of precision observables. These functions
  include calculations of backend-specific data-types for input to
  other functions, retrieval functions for specific observables, and
  functions for updating particle spectra for use throughout \GB.
\label{tab:precisionbit:BE}}
\end{table*}

\renewcommand\descwidth{8.7cm}

\begin{table*}[tp]
\centering
\scriptsize{
\begin{tabular}{l|p{\descwidth}|l}
  \textbf{Capability}
      & \multirow{2}{*}{\parbox{\descwidth}{\textbf{Function} (\textbf{Return Type}):
             \\  \textbf{Brief Description}}}
          & \textbf{Dependencies}
          \\ & &
  \\ \hline
\cpp{lnL_sinW2_eff}
      & \multirow{2}{*}{\parbox{\descwidth}{ \cpp{lnL_sinW2_eff_chi2(double)}:
              \\ Computes the log-likelihood associated with $\sin^{2}\theta_{W.\mathrm{eff}}$.}}
          & \cpp{prec_sinW2_eff}
          \\ & &
  \\ \hline
\cpp{lnL_gm2}
      & \multirow{2}{*}{\parbox{\descwidth}{ \cpp{lnL_gm2_chi2(double)}:
              \\ Computes the log-likelihood associated with muon $g-2$.}}
          & \cpp{muon_gm2}
          \\ & & \cpp{muon_gm2_SM}
  \\ \hline
\cpp{lnL_deltarho}
      & \multirow{2}{*}{\parbox{\descwidth}{ \cpp{lnL_deltarho_chi2(double)}:
              \\ Computes the log-likelihood associated with $\Delta\rho$.}}
          & \cpp{deltarho}
          \\ & &
  \\ \hline
\cpp{muon_gm2_SM}
      & \multirow{2}{*}{\parbox{\descwidth}{ \cpp{gm2_SM_ee(triplet<double>)}:
              \\ Computes the SM contribution to $g-2$, based on $e^+e^-$ data.}}
          &
          \\ & &
  \\ \cmidrule{2-3}
      & \multirow{2}{*}{\parbox{\descwidth}{ \cpp{gm2_SM_tautau(triplet<double>)}:
              \\ Computes the SM contribution to $g-2$, based on $\tau^+\tau^-$ data.}}
          &
          \\ & &
  \\ \hline
\end{tabular}
}
\caption{\precisionbit module functions providing log-likelihood
  calculations for electroweak precision observables.
\label{tab:precisionbit:lnL}}
\end{table*}

\renewcommand\metavar\metavarf

\subsubsection{Electroweak precision observable likelihoods}
\label{ewpolikes}
Electroweak precision observables are well measured observables from
the electroweak sector of the SM. Due to the precise measurements they
can provide constraints on BSM physics.  For a pedagogical
introduction see Ref.\ \cite{Matchev:2004yw}, where $Z$-pole and non
$Z$-pole sets of electroweak precision observables are listed in Tables
1 and 2, respectively.
\precisionbit includes a collection of functions for using the
precision observables calculated by external code packages to
calculate likelihoods. Currently, likelihood functions are provided for
the effective leptonic weak mixing angle $\sin^{2}\theta_{W,\mathrm{eff}}$, the departure from 1 of the ratio of the Fermi constants implied by neutral and charged currents $\Delta\rho$, and the muon anomalous
magnetic moment $a_{\mu}$ (also referred to as $g-2 \equiv 2a_{\mu}$). For each of these observables, \precisionbit calculates likelihoods by combining experimental and theoretical uncertainties
in quadrature (i.e.\ assuming that theoretical and experimental errors are not correlated), and constructing likelihood functions based on the difference between the calculated and
experimentally-measured values. The module functions that provide these \capabilities{} are listed in
Table~\ref{tab:precisionbit:lnL}.

\precisionbit also offers two module functions able to provide the predicted SM contribution to $g-2$.  These are listed along with the likelihood functions in Table~\ref{tab:precisionbit:lnL}.  The first is calibrated to $e^+e^-$ data, giving $a_{\mu,\mathrm{SM}}  = (11659180.2 \pm 4.9) \times 10^{-10}$ \cite{1010.4180}, whereas the second comes from $\tau^+\tau^-$ data, and corresponds to $a_{\mu,\mathrm{SM}}  = (11659189.4 \pm 5.4) \times 10^{-10}$ \cite{1010.4180}.

For the $g-2$ likelihood, we combine errors from the predicted SM contribution in quadrature with the theoretical error arising from the BSM contribution, as these two theory errors are entirely independent of each other, and therefore uncorrelated.
In the likelihood function, we compare the sum of the SM and BSM contributions to the experimental measurement $a_\mu  = (11659208.9 \pm 6.3) \times 10^{-10}$ \cite{PDG10,gm2exp}, where the error is the sum in quadrature of statistical ($5.4\times10^{-10}$) and systematic ($3.3\times10^{-10}$) uncertainties.

Theoretical calculations of the MSSM contribution to $a_{\mu}$ can be taken from either \gmtwocalc, \superiso or \FeynHiggs.  While \superiso and \FeynHiggs include essentially the same one and two-loop corrections, and give more or less consistent results, \gmtwocalc includes additional two-loop contributions and an improved on-shell calculation of the one-loop effects, leading to significant improvements in precision for some parameter combinations.  \gmtwocalc calculates an uncertainty on its prediction, using the magnitudes of the two-loop Barr-Zee corrections \cite{barrzee} to estimate the magnitude of neglected higher-order contributions \cite{gm2calc}; \precisionbit adopts this estimate directly when employing \gmtwocalc. For \FH and \superiso, no such estimate is available. Following the discussion in Ref.\ \cite{stockinger06}, we assign a theoretical uncertainty of either 30\% or $6\times10^{-10}$ (whichever is greater) to the values of $g-2$ that \precisionbit obtains from \superiso and \FH.

\precisionbit's theoretical predictions of $\sin^{2}\theta_{W,\mathrm{eff}}$ and $\Delta\rho$ in the MSSM are currently provided exclusively by \FeynHiggs.  We assume a theoretical error, $\sigma^\mathrm{theo}_{\sin^{2}\theta_{W,\mathrm{eff}}}$, of $12\times10^{-5}$ on $\sin^{2}\theta_{W,\mathrm{eff}}$ \cite{hep-ph/0412214}.  The fractional MSSM corrections to both $\sin^{2}\theta_{W,\mathrm{eff}}$ and $m_W$ are approximately proportional to $\Delta\rho$ at tree level, with a constant of proportionality quite close to one in magnitude (within a few tens of percent; \cite{hep-ph/0412214}). Combined with the fact that $\Delta\rho \ll 1$ in general, to a good approximation the theoretical error on its value can be estimated from the fractional uncertainty on $\sin^{2}\theta_{W,\mathrm{eff}}$ or $m_W$; we therefore adopt a theory error on $\Delta\rho$ of
\begin{equation}
\sigma^\mathrm{theo}_{\Delta\rho} = \mathrm{max}(\sigma^\mathrm{theo}_{m_W}/m_W, \sigma^\mathrm{theo}_{\sin^{2}\theta_{W,\mathrm{eff}}}/\sin^{2}\theta_{W,\mathrm{eff}}),
\end{equation}
where $\sigma^\mathrm{theo}_{m_W}=10$\,MeV \cite{hep-ph/0412214}.

\subsubsection{Precision-updated MSSM spectrum}
\label{precisionspectrum}

As described in Table~\ref{tab:precisionbit:BE}, \precisionbit
includes a number of module function for updating existing MSSM spectra with
precision calculations of particle masses:
\begin{itemize}[topsep=3pt]
\item[] \cpp{make_MSSM_precision_spectrum_H_W()},
\item[] \cpp{make_MSSM_precision_spectrum_W()}, and
\item[] \cpp{make_MSSM_precision_spectrum_none()}.
\end{itemize}
These functions take an
\cpp{unimproved_MSSM_spectrum} as a dependency, update its mass spectrum with precision values, and return the improved version as an \cpp{MSSM_spectrum}.  The \cpp{H_W} version updates both the Higgs-sector and $W$ masses, the \cpp{W} version does only the $W$ mass, and \cpp{make_MSSM_precision_spectrum_none} simply transmutes an \cpp{unimproved_MSSM_spectrum} into an \cpp{MSSM_spectrum} without further modification.  The precision $W$ and Higgs-sector masses are typically provided via \dependencies{} fulfilled by functions that call \FeynHiggs (though they can of course in principle come from anywhere).

In addition to updating the masses, the \cpp{make_MSSM_precision_spectrum_H_W()} and \cpp{make_MSSM_prec}-\cpp{ision_spectrum_W()} functions also update the associated
theoretical uncertainties provided by the original spectrum object.  The uncertainty on the pole mass of the $W$ is adopted from its \dependency{} on the precision value of $m_W$.  When calculating this with \FH, we assume an uncertainty of 10\,MeV \cite{hep-ph/0412214} (as discussed in the context of the theory error on $\Delta\rho$ in Sec.\ \ref{ewpolikes}).\footnote{It is however worth noting that this is in fact no smaller than the value that we assign to the uncertainty on the $W$ mass calculation performed by \flexiblesusy, so the replacement has no effect if \flexiblesusy has been used to generate the original \cpp{unimproved_MSSM_spectrum}.}

For the MSSM Higgs bosons ($A^0$, $h^0$, $H^{0}$ and $H^{\pm}$), the methods by which \precisionbit arrives at the central value and uncertainties on the mass are configurable by the user, using the options
 \cpp{Higgs_predictions_source} and \cpp{Higgs_predictions_error_method}, respectively.  For the \cpp{Higgs_predictions_source}, there are three possible values:
\begin{itemize}
\item[\yamlvalue{1}:] Update each of the Higgs masses to
the value derived from the precision calculator (canonically \FH).  This is the default.
\item[\yamlvalue{2}:] Keep the masses unchanged from the original input spectrum.
\item[\yamlvalue{3}:] Assign the mean of the masses found in the original spectrum and the ones provided by the precision calculator.
\end{itemize}

Five options are available for assigning theoretical
uncertainties to the MSSM Higgs masses. These are based on different combinations of three
quantities with unique values for each Higgs state: the error
associated with the mass from the original input spectrum
($\Delta_{s}$), the error determined by the precision calculator
($\Delta_{p}$), and the difference of central values from the two
calculations ($\Delta_{g}$).  All but one of the options for using these quantities assigns them as additional errors beyond a so-called `range around chosen central' (RACC) calculation.  The RACC is defined such that the upper error $\sigma_+$ is the distance from the central value chosen with \cpp{Higgs_predictions_error_method} to the greater of the values predicted by the original spectrum and the precision calculator.  The lower error $\sigma_-$ is defined analogously as the distance from the central value chosen to the lesser of the two predictions.  With these definitions, the available options for the \cpp{Higgs_predictions_error_method} are:
\begin{itemize}
\item[\yamlvalue{1}:] Upper and lower uncertainties both set equal to the sum in quadrature of all three uncertainties: $\sigma_+^2 = \sigma_-^2 = \Delta_{s}^2 + \Delta_{p}^2 + \Delta_{g}^2$.
\item[\yamlvalue{2}:] RACC, with $\Delta_{s}$ added to the error associated with the distance to the spectrum generator value (the spectrum-generator `edge'), and $\Delta_{p}$ added to the precision-calculator edge. This is the default.
\item[\yamlvalue{3}:] RACC, with $\Delta_{g}/2$ added to both $\sigma_+$ and $\sigma_-$.
\item[\yamlvalue{4}:] RACC, with $\Delta_{g}/2$ added at the spectrum-generator edge, and $\Delta_{p}$ added at the precision-calculator edge.
\item[\yamlvalue{5}:] RACC, with $\Delta_{g}/2$ added at the precision-calculator edge, and $\Delta_{s}$ added at the spectrum-generator edge.
\end{itemize}

The resulting precision-improved spectrum is provided to the rest of \GB for physics calculations, and in the case of the Higgs bosons, computation of Higgs-mass likelihoods by \colliderbit.

\section{Examples}
\label{Sec:threebit}

\subsection{Example \YAML files}

In \term{yaml_files}, we give some example input \YAML files that run various functions from \specbit, \decaybit and \precisionbit within \GB.  The file \term{SpecBit_MSSM.yaml} runs a number of tests on the various MSSM spectra, using either \flexiblesusy or \SPheno from within \specbit.  The file \term{SpecBit_vacuum_stability.yaml} does a short (10 minutes runtime on a single core) scan of the Higgs and top mass, in order to map out the regions of SM vacuum (meta-)stability and perturbativity.  The \decaybit example \term{DecayBit_MSSM20.yaml} computes decay widths and branching fractions for a single example point in the \textsf{MSSM20atQ} model \cite{gambit}. \term{DecayBit_SingletDM} does the same for ten randomly-selected parameter combinations in the scalar singlet dark matter model.  \term{PrecisionBit_MSSM20.yaml} computes all precision observables and likelihoods contained in \precisionbit, for the same example \textsf{MSSM20atQ} point as \term{DecayBit_MSSM20.yaml}.

\subsection{\threebit}
As a convenience for users simply wishing to use some basic functionality of \specbit, \decaybit and \precisionbit from the command line, we also provide an additional driver program \threebit.  This program doubles as an example of how the three modules can be used in standalone mode, i.e. without the \GB \textsf{Core} or the other modules.

\threebit takes a single MSSM model as input via a minimal \YAML file \term{3bithit.in}, in either the weak-scale MSSM20atQ or GUT-scale NUHM2 parametrisation. The user can also specify SM parameters, following SLHA2 conventions.  The program then solves the RGEs and evaluates pole masses, calculates decay widths and branching fractions of all SM particles and their superpartners, computes electroweak precision observables, and finally outputs the results as a single SLHA file \term{3bithit.out.slha}.

The basic spectrum generation in \threebit is done using \FlexibleSUSY via \specbit, with the Higgs and $W$ masses replaced by precision calculations performed with \FH by \precisionbit.  The resulting SM-like Higgs mass and uncertainties emitted in the final SLHA file are based on the default settings in \cpp{PrecisionBit::make_MSSM_precision_spectrum_H_W()}:
\begin{lstcpp}
Higgs_predictions_source = 1
Higgs_predictions_error_method = 2
\end{lstcpp}
The Higgs-sector decays are computed by \decaybit using \FH, SM decays are based on native \decaybit functions, and sparticle decays come from \SDECAY, via \susyhit and \decaybit.  \threebit computes the SUSY contribution to the anomalous magnetic moment of the muon, along with its uncertainty, using calls to \gmtwocalc from \precisionbit.  All other electroweak precision observables (EWPO) come from \FH via \precisionbit, where errors are also assigned to them.

\threebit performs a similar function to \susyhit, but computes EWPO in addition to spectra and decays.  It also employs \FlexibleSUSY instead of \SUSPECT for spectrum generation, \FH for Higgs decays rather than \HDECAY, and our patched version of \SDECAY (see \ref{decays:MSSM} for details) for sparticle decays. Its source can be found in \term{DecayBit/examples/3bithit.cpp}, and it can be built with
\begin{lstterm}
make 3bithit
\end{lstterm}

\section{Summary}
\label{sec:summary}

In this paper we have introduced the \GB modules \specbit, \decaybit
and \precisionbit.  These are designed to flexibly integrate publicly
available programs for spectrum generation, the calculation of decay
widths and branching ratios, and additional precision calculations,
like the anomalous magnetic moment of the muon.  Together,
these modules provide a powerful way to synergise spectrum
generators, decay codes and additional precision calculations, allowing users to
extract the information produced by them in a common format.  We have illustrated
this use with the example standalone program \threebit (Sec.\ \ref{Sec:threebit}), which
provides a single interface to \FlexibleSUSY, \FeynHiggs, \SDECAY and \gmtwocalc.  At the
same time, these modules play a crucial role within the full framework
of \GB, where they provide essential information to other packages
and important components of the log-likelihood functions used to drive global fits.

\begin{acknowledgements}
We thank all our colleagues within \GB for helpful discussions, especially Nazilla Mahmoudi for her comments on the muon $g-2$
functionality of \superiso. We would also like to thank Margarete
M\"uhlleitner and Ramona Gr\"ober for responding to questions
regarding \SDECAY, and Sven Heynemeyer and Sebastian Pa{\ss}ehr for
doing so on \FeynHiggs. PA would like to
thank his \FlexibleSUSY collaborators Dominik St\"ockinger, Jae-hyeon
Park, Dylan Harries and especially Alexander Voigt, for many helpful
discussions. \gambitacknos
\end{acknowledgements}

\section*{Appendices} 
\appendix
\titleformat{\section}{\normalsize\bfseries}{\normalsize\bfseries\thesection}{1em}{}
\titleformat*{\subsection}{\normalsize}
\titleformat{\subsubsection}{\normalsize\itshape}{\thesubsubsection}{1em}{}

\section{Physics Models}
\label{Sec:PhysBack}
Here we give a compact physical definition of each of the models
currently supported by \specbit, \decaybit and \precisionbit,
complementing the slightly more code-focussed description in the
main \GB paper \cite{gambit}.  \GB is primarily designed for global
fits to extensions of the Standard Model (SM) of particle physics.
Many calculations in these extensions require SM quantities, and many
such extensions make minimal modifications to the SM; examples
are the scalar singlet dark matter model (Sec.\ \ref{ss}) and
effective weakly-interacting massive particle (WIMP) dark matter (see
Ref.\ \cite{DarkBit} for an example in \GB).  We therefore begin with
the SM itself.

\subsection{Standard Model}
The SM describes the interactions of
all observed fundamental particles.  These interactions are invariant
under the symmetries of the gauge group,
\begin{equation}
G_{SM} = SU(3)_\mathrm{C} \times SU(2)_\mathrm{L} \times U(1)_\mathrm{Y}.
\end{equation}
The Lagrangian density is given by,
\begin{equation}
\begin{split}
\mathcal{L}_{SM} = & \textstyle\sum_{i=1,3} i (\overline{Q}_{i}\slashed{D} Q_{i} + \overline{L}_{i}\slashed{D} L_{i} + \overline{u}_{Ri}\slashed{D} u_{Ri} \\
& + \overline{d}_{Ri}\slashed{D} d_{Ri} +  \overline{e}_{Ri}\slashed{D} e_{Ri}) \\
& + \textstyle\sum_{i,j=1,3} \left[\textstyle\sum_{a,b=1,2} (\mathbf{Y}_u)_{ij}\epsilon_{ab}Q^a_i (\phi^\dagger)^b u_{Rj}  \right. \\
 & \left.  + (\mathbf{Y}_d)_{ij}\overline{Q}_i\phi d_{Rj}  + (\mathbf{Y}_e)_{ij}\overline{L}_i\phi e_{Rj}+ h.c.\right] \\
& + |D_\mu \phi|^2 - \mu^2 |\phi|^2 - \lambda |\phi|^4 \\
&-\frac{1}{4} B_{\mu\nu}B^{\mu\nu} -\frac{1}{4}W_{\mu\nu}^A W^{\mu\nu A} - \frac{1}{4} G^B_{\mu\nu}G^{\mu\nu B}
\end{split}
\label{Eq:SM_Lagrangian}
\end{equation}
where $ u_{Ri}$,  $d_{Ri}$ and $e_{Ri}$ represent the right-handed up-type quark, down-type quark and lepton.  Similarly,
\begin{equation}
 \phi  = \begin{pmatrix} \phi^+ \\ \phi^0 \end{pmatrix},
\quad  Q_i = \begin{pmatrix} u_{Li} \\ d_{Li} \end{pmatrix},
\quad  L_i = \begin{pmatrix} \nu_{Li} \\ e_{Li} \end{pmatrix},
\label{Eq:Fermion_doublets}
\end{equation}
represent the Higgs, left-handed quark and left-handed lepton doublets
respectively. Here $i$ and $j$ specify the generation of the fields, $a$ and $b$ are $SU(2)_\mathrm{L}$ indices, and $\epsilon_{ab}$ is the anti-symmetric tensor.

The $B_{\mu\nu}$, $W^A_{\mu\nu}$ and $G^B_{\mu\nu}$ are field strength tensors for the $B_\mu$, $W^A_{\mu}$ and $G^B_{\mu}$ fields respectively, which are respectively associated with the $U(1)_\mathrm{Y}$, $SU(2)_\mathrm{L}$ and $SU(3)_\mathrm{C}$ gauge groups. The kinetic terms for the matter fields include $\slashed{D} = D^\mu \gamma_\mu$, where $D_\mu$ is the usual covariant derivative, which, for example, takes the following form for all fields transforming non-trivially under both $SU(2)_\mathrm{L}$ and $SU(3)_\mathrm{C}$:
\begin{equation}
D_\mu = \partial_\mu - i g_1 Q_Y B_\mu - ig_2 W_\mu^A T^A  -i g_3\lambda^B G_\mu^B.
\label{Eq:CovDer}
\end{equation}
Here $g_1$ and $Q_Y$ are the GUT-normalised gauge coupling and GUT-normalised charges\footnote{Hypercharge is often written without GUT normalisation, such that the charge for a field $\phi$ is $Y^\phi = \sqrt{5/3} Q_Y^\phi$ and will be written in the covariant derivative with the gauge coupling $g' = \sqrt{3/5} g_1$.} of $U(1)_Y$. Similarly $g_2$ and $T^A$ are the $SU(2)_\mathrm{L}$ gauge couplings and generators respectively, and $g_3$ and $\lambda^B$ are the $SU(3)_\mathrm{C}$ gauge couplings and generators respectively.

Electro-weak symmetry is broken when the Higgs field develops a VEV, $v$,
\begin{equation}
\langle\phi\rangle = \frac{1}{\sqrt{2}} \begin{pmatrix} 0 \\ v \end{pmatrix},
\label{Eq:vev}
\end{equation} so that field may be rewritten as
\begin{equation}
\phi^0 = \frac{1}{\sqrt{2}}( v + h^0 + i G^0).
\label{Eq:higgsfield}
\end{equation} where $h^0$ is the physical Higgs boson and $G^0$ is the unphysical Goldstone boson that corresponds to the longitudinal mode for the massive $Z^0$ boson.

The SM VEV gives masses to the SM particles.  The $Z$ and $W$ gauge bosons obtain tree-level masses
\begin{equation}
m_Z = \frac12\sqrt{(g^{\prime \, 2} + g_2^2)}\, v \quad \quad M_W = \frac12 g_2 v.
\label{Eq:GaugeBosonMasses}
\end{equation}
The weak mixing angle $\theta_W$ is given by
\begin{equation} \sin^2 \theta_W  \equiv \frac{g^{\prime \, 2}}{g^{\prime \, 2} + g_2^2} = 1 - \frac{M_W^2}{m_Z^2},
\label{Eq:sthW}
\end{equation} though it is important to note that the last equality depends on the tree-level relations given in Eq.~\ref{Eq:GaugeBosonMasses}, so this definition holds only within the renormalisation scheme in which the gauge couplings are defined.  An alternative definition can be made in terms of pole masses,
\begin{equation} \sin^2\theta_W^{\textrm{on-shell}} \equiv 1 - \frac{M_W^{\textrm{(pole)}\, 2}}{m_Z^{\textrm{(pole)}\,2}}.
\label{sthWpole}
\end{equation}
  Finally, one should note that neither of these correspond exactly to
the leptonic effective weak mixing angle, which is an important
precision observable.  This quantity is relevant to \precisionbit and
is discussed in Sec.\ \ref{ewpolikes}.

\subsection{Minimal Supersymmetric Standard Model}
\label{sec:MSSMLagrangian}
The MSSM superpotential is given by,
\begin{multline}
\hat{W} = \epsilon_{ab} \left\{ \textstyle\sum_{i,j=1,3} \left[(\mathbf{Y}_u)_{ij}\hat{Q}^a_i\hat{H}^b_u\hat{U}^c_j - (\mathbf{Y}_d)_{ij}\hat{Q}^a_i\hat{H}^b_d\hat{D}^c_j \right.\right. \\
          - \left.\left. (\mathbf{Y}_e)_{ij}\hat{L}^a_i\hat{H}^b_d\hat{E}^c_j\right] - \mu\hat{H}^a_u\hat{H}^b_d \right\},
\label{eq:superpot}
\end{multline}
where a hat indicates a superfield and $i$, $j$ specify the
generation, while $a$, $b$ are $SU(2)_\mathrm{L}$ indices. This includes the Higgsino mixing parameter $\mu$, which appears as a
dimension one coefficient of the up- and down-type Higgs superfields,
\begin{equation}\hat{H}_u =  \begin{pmatrix} \hat{H}_u^+ \\ \hat{H}_u^0 \end{pmatrix} \quad\quad \hat{H}_{d} = \begin{pmatrix} \hat{H}_d^0 \\ \hat{H}_d^- \end{pmatrix}. \end{equation} The chiral superfields have dimensionless up- and down-type Yukawa couplings, $\mathbf{Y}_u$ and $\mathbf{Y}_d $ with the Higgs superfields and the chiral superfields for the right-handed quarks.  Specifically left-handed quarks, $\hat{Q}$, interact with both the chiral superfields for the right-handed up quarks, $\hat{U}^c$, and the up-type Higgs superfield through $\mathbf{Y}_u$ and with down quarks, $\hat{D}^c$, and the down-type Higgs superfields through $\mathbf{Y}_d $.  Similarly the chiral superfield for left-handed leptons
$\hat{L}$ has Yukawa interactions, $\mathbf{Y}_e$ with the chiral
superfields for the right-handed leptons, $\hat{E}^c$ and
$\hat{H}_{d}$.

Terms which break supersymmetry without introducing quadratic divergences are given in the soft SUSY breaking Lagrangian,

\begin{subequations}
\label{lsoft}
\begin{align}
\mathcal{L}_\mathrm{soft} =& -\frac12\left[M_1\bar{\tilde{B}}^0\tilde{B}^0 + M_2\bar{\tilde{W}}_A\tilde{W}_A + M_3\bar{\tilde{g}}_b\tilde{g}_b\right] \label{gauginos} \\
&  -\frac{i}{2}\left[M_1'\bar{\tilde{B}}^0\gamma_5\tilde{B}^0 + M_2'\bar{\tilde{W}}_A\gamma_5\tilde{W}_A + M_3'\bar{\tilde{g}}_B\gamma_5\tilde{g}_B\right] \label{$CP$-violating gauginos} \\
&  - \Big[b{H}^a_u{H}_{da} + \mathrm{h.c.}\Big] - m^2_{H_u}|H_u|^2 - m^2_{H_d}|H_d|^2 \label{Higgses} \\
&  +\textstyle\sum_{i,j=1,3} \left\{ -\left[\tilde{Q}_i^\dagger(\mathbf{m}^\mathbf{2}_Q)_{ij}\tilde{Q}_j +
\tilde{d}_{\mathrm{R}i}^\dagger(\mathbf{m}^\mathbf{2}_d)_{ij}\tilde{d}_{\mathrm{R}j} \right. \right. \nonumber \\
&  + \tilde{u}_{\mathrm{R}i}^\dagger(\mathbf{m}^\mathbf{2}_u)_{ij}\tilde{u}_{\mathrm{R}j}
   + \tilde{L}_{i}^\dagger(\mathbf{m}^\mathbf{2}_L)_{ij}\tilde{L}_{j}
   + \tilde{e}_{\mathrm{R}i}^\dagger(\mathbf{m}^\mathbf{2}_e)_{ij}\tilde{e}_{\mathrm{R}j} \Big] \label{sfermion masses} \\
 & +- \epsilon_{ab}\left[(\mathbf{T}_u)_{ij}\tilde{Q}^a_iH^b_u\tilde{u}^\dagger_{\mathrm{R}j}
   - (\mathbf{T}_d)_{ij}\tilde{Q}^a_iH^b_d\tilde{d}^\dagger_{\mathrm{R}j} \right. \nonumber\\
&  \hspace{1cm} - \left. (\mathbf{T}_e)_{ij}\tilde{L}^a_iH^b_d\tilde{e}^\dagger_{\mathrm{R}j} + \mathrm{h.c.}\right] \label{trilinear couplings} \\
&  - \epsilon_{ab}\left[(\mathbf{C}_u)_{ij}\tilde{Q}^a_iH^{*b}_d\tilde{u}^\dagger_{\mathrm{R}j}
   - (\mathbf{C}_d)_{ij}\tilde{Q}^a_iH^{*b}_u\tilde{d}^\dagger_{\mathrm{R}j} \right. \nonumber \\
&  \hspace{1cm} - \left.\left. (\mathbf{C}_e)_{ij}\tilde{L}^a_iH^{*b}_u\tilde{e}^\dagger_{\mathrm{R}j} + \mathrm{h.c.}\right] \right\}. \label{trilinear_C_couplings}
\end{align}
\end{subequations}

As in the previous equation we have indices $i$, $j$ for generation
and $a$, $b$ for $SU(2)_\mathrm{L}$ and in addition $A \in \{1,2,3\}$
is the index for $SU(2)_\mathrm{L}$ gauge generators and
$B \in \{1,2,3...8\}$ is the index for the $SU(3)_\mathrm{C}$.  The
superpartners of the $U(1)_Y$ gauge field, $\tilde{B}^0$, the
$SU(2)_\mathrm{L}$ gauge fields, $\tilde{W}_A$, and the
$SU(3)_\mathrm{C}$ gauge fields, $\tilde{g}_B$ get soft masses, $M_1$,
$M_2$ and $M_3$ respectively, as well as potential contributions
from $M_1'$, $M_2'$ and $M_3'$.  The Higgs scalar fields, $H_u$ and
$H_d$ receive soft scalar mass contributions, $ m^2_{H_u}$,
$m^2_{H_d}$ respectively and are mixed by the soft bilinear mass
squared, $b$.  The left-handed squarks, $\tilde{Q}_i$ and sleptons,
$\tilde{L}_i$ and the right handed up-type squarks,
$\tilde{u}_{\mathrm{Ri}}$, down-type squarks, $\tilde{d}_{\mathrm{Ri}}$,
and charged sleptons $\tilde{e}_{\mathrm{Ri}}$ all have soft scalar
masses, $(\mathbf{m}^\mathbf{2}_Q)_{ij}$, $(\mathbf{m}^\mathbf{2}_L)_{ij}$,
$(\mathbf{m}^\mathbf{2}_u)_{ij}$, $(\mathbf{m}^\mathbf{2}_d)_{ij}$ and
$(\mathbf{m}^\mathbf{2}_e)_{ij}$ respectively. Finally we have soft trilinears
for the up-type squarks and Higgs, $(\mathbf{T}_u)_{ij}$, down-type squarks
and Higgs, $(\mathbf{T}_d)_{ij}$, and the sleptons and down-type Higgs,
$(\mathbf{T}_e)_{ij}$ and also the terms that are are soft when there is no
singlet in the model, $(\mathbf{C}_u)_{ij}$, $(\mathbf{C}_d)_{ij}$ and
$(\mathbf{C}_e)_{ij}$.  The latter are normally assumed to be zero, even
though the MSSM has no scalar singlets as it is difficult to generate
these.  The soft trilinears may also be rewritten in terms of the following,
\begin{equation}
(\mathbf{A}_f)_{ij} = \frac{(\mathbf{T}_f)_{ij}}{ (\mathbf{Y}_f)_{ij}} \quad f \in \{u,d,e\}.
\end{equation}

In the MSSM the neutral components of the up- and down-type scalar Higgs fields, $H_u^0$ and $H_d^0$ respectively, develop VEVs,
\begin{equation}
\langle H_d \rangle = \frac{1}{\sqrt{2}} \begin{pmatrix} v_d \\ 0 \end{pmatrix},
\quad \langle H_u \rangle = \frac{1}{\sqrt{2}} \begin{pmatrix} 0 \\ v_u \end{pmatrix},
\label{Eq:VEVS}
\end{equation}
and the bilinear soft term, $b$, mixes them, so that they must be
rotated to obtain the physical mass eigenstates, $h_i$
\begin{equation}
\begin{pmatrix} h_1 \\ h_2 \end{pmatrix} = \sqrt{2} \, Z_H \begin{pmatrix} {\mathcal Re}( H_d^0 )  - \frac{v_d}{\sqrt{2}}\\ {\mathcal Re}(H_u^0)  - \frac{v_u}{\sqrt{2}}  \end{pmatrix},
\label{Eq:CPEvenRot}
\end{equation} where the mixing matrix is commonly expressed in terms of a mixing angle $\alpha$,
\begin{equation}
Z_H = \begin{pmatrix} - \sin{\alpha} & \cos{\alpha} \\ \cos{\alpha} & \sin{\alpha} \end{pmatrix}.
\label{Eq:alphadef}
\end{equation} Similarly the pseudoscalar Higgs state, $A$, is formed from the imaginary parts of the Higgs scalar fields
\begin{equation}
\begin{pmatrix} G^0 \\ A^0 \end{pmatrix} = \sqrt{2} \, Z_A \begin{pmatrix} {\mathcal Im}( H_d^0 ) \\ {\mathcal Im}(H_u^0)  \end{pmatrix}.
\label{Eq:CPOddRot}
\end{equation} where $G^0$ is the neutral Goldstone boson.  At tree level it is straightforward to show that the mass eigenstates can be obtained with a rotation,
\beq
Z_A = \begin{pmatrix} -\sin\beta & \cos \beta \\ \cos\beta & \sin\beta \end{pmatrix}
\eeq
giving a massless Goldstone boson and an expression for the pseudoscalar mass,
\beq
m^2_A = \frac{b}{\sin\beta \cos\beta}.
\label{Eq:mA}
\eeq  The charged Higgs mass eigenstate, $H^+$ and charged Goldstone boson, $G^+$ are related to the charged scalar Higgs fields through,
\beq
\begin{pmatrix} G^+ \\ H^+ \end{pmatrix} =  Z_{H^+} \begin{pmatrix} H_u^+ \\ (H_d^-)^\dagger  \end{pmatrix}.
\eeq

The squarks are rotated into the mass eigenstates as,
\begin{equation}
\begin{pmatrix}\tilde{u}_1 \\\tilde{u}_2 \\\tilde{u}_3 \\\tilde{u}_4 \\\tilde{u}_5 \\\tilde{u}_6  \end{pmatrix} = Z_{\tilde{u}} \begin{pmatrix} \tilde{Q}^1_1 \\\tilde{Q}^1_2 \\\tilde{Q}^1_3 \\\tilde{u}^\dagger_{\mathrm{R}1} \\ \tilde{u}^\dagger_{\mathrm{R}2} \\ \tilde{u}^\dagger_{\mathrm{R}3}  \end{pmatrix},
\quad
\begin{pmatrix}\tilde{d}_1 \\\tilde{d}_2 \\\tilde{d}_3 \\\tilde{d}_4 \\\tilde{d}_5 \\\tilde{d}_6  \end{pmatrix} = Z_{\tilde{d}} \begin{pmatrix} \tilde{Q}^2_1 \\\tilde{Q}^2_2 \\\tilde{Q}^2_3 \\\tilde{d}^\dagger_{\mathrm{R}1} \\ \tilde{d}^\dagger_{\mathrm{R}2} \\ \tilde{d}^\dagger_{\mathrm{R}3}  \end{pmatrix}
\label{Eq:sqrot}
\end{equation}
and the sleptons and sneutrinos are rotated as,
\begin{equation}
\begin{pmatrix}\tilde{e}_1 \\\tilde{e}_2 \\\tilde{e}_3 \\\tilde{e}_4 \\\tilde{e}_5 \\\tilde{e}_6  \end{pmatrix} = Z_{\tilde{e}} \begin{pmatrix} \tilde{L}^2_1 \\\tilde{L}^2_2 \\\tilde{L}^2_3 \\\tilde{e}^\dagger_{\mathrm{R}1} \\ \tilde{e}^\dagger_{\mathrm{R}2} \\ \tilde{e}^\dagger_{\mathrm{R}3}  \end{pmatrix},
\quad
\begin{pmatrix}\tilde{\nu}_1 \\\tilde{\nu}_2 \\\tilde{\nu}_3   \end{pmatrix} = Z_{\tilde{\nu}} \begin{pmatrix} \tilde{L}^1_1 \\\tilde{L}^1_2 \\\tilde{L}^1_3  \end{pmatrix}.
\label{Eq:slrot}
\end{equation}

The neutralinos are mass eigenstates from combinations of the Bino  $\tilde{B}^0$, neutral Wino $\tilde{W_3}$ and the up- and down-type Higgsinos, $\tilde{H}^0_u$ and $\tilde{H}^0_d$,
\begin{equation}
\begin{pmatrix}\tilde{\chi}^0_1 \\\tilde{\chi}^0_2 \\\tilde{\chi}^0_3 \\\tilde{\chi}^0_4  \end{pmatrix} = Z_N \begin{pmatrix}\tilde{B}^0 \\\tilde{W}_3 \\\tilde{H}^0_d \\\tilde{H}^0_u  \end{pmatrix},
\label{Eq:ChiRot}
\end{equation}
\noindent where we follow the SLHA \cite{Skands:2003cj} convention and require the mixing matrix, $Z_N$, to be real, which means the masses of the neutralinos may be negative.  Note that this is the form of the masses and mixings that can be accessed from the spectrum object in \gambit and used for passing into interfaces to various backend codes that need the spectrum as an input.  We follow the SLHA convention of negative masses because this ensures the maximum number of codes can be supported without conversion during interfacing.  For any codes or internal calculations which expect neutralino masses and mixings in another convention, this must be converted in the interface.

The mass matrix of the charginos is diagonalised through a biunitary
transformation, such that the two component chargino mass eigenstates
are given by,
\begin{equation}
\begin{pmatrix}\chi^+_1 \\ \chi^+_2\end{pmatrix} = U_+ \begin{pmatrix}\tilde{W}_1 - i \tilde{W}_2  \\   \tilde{H}_u^+\end{pmatrix},
\label{Eq:ChaRot+}
\end{equation}
\begin{equation}
\begin{pmatrix}\chi^-_1 \\ \chi^-_2\end{pmatrix} = U_- \begin{pmatrix} \tilde{W}_1 + i \tilde{W}_2  \\   \tilde{H}_u^- \end{pmatrix}
\label{Eq:ChaRot-}
\end{equation}

\subsection{Scalar Singlet Dark Matter Model}
\label{ss}

The scalar singlet model is a minimalistic extension of the SM, with one new scalar charged only under a $\mathbb{Z}_2$ symmetry.  This symmetry guarantees that the scalar is a stable dark matter candidate, and restricts the most general permitted renormalisable Lagrangian density on symmetry arguments to be,
\begin{equation}
\mathcal{L}_\mathrm{SS} = \frac12 \mu_S^2 S^2 + \frac12\lhs S^2|H|^2 + \frac14\ls S^4 + \frac12\partial_\mu S \partial^\mu S, \label{eqn:SingletDM_Lagrangian}
\end{equation}
where $\mu_S^2$ is a mass squared term for the singlet, $\lhs$ the Higgs-portal coupling and $\ls$ the quartic self coupling.

The singlet mass has an additional contribution from EWSB through the Higgs-portal coupling, giving a mass,
\begin{equation}
\label{m_S_tree}
m_S = \sqrt{\mu_S^2 + \frac12{\lhs v^2}},
\end{equation}
where $v$ is the SM Higgs VEV in Eq.~\ref{Eq:vev}.  As the $\mathbb{Z}_2$ symmetry must remain unbroken for stable dark matter we demand that the scalar field, $S$, does not obtain a VEV and thus no mixing occurs between the Higgs and the new scalar field.  This requirement is satisfied for $\lambda_S>0$.

\section{Spectrum generators as backends: \SPheno}
\label{app:SPheno_backend}

One of the key features of \GB is its ability to interface with different external codes to perform relevant calculations for a specific scan. These codes, or backends, are linked with \GB in a non-trivial way in order to access their functions directly via memory, rather than through a shell. Details and descriptions of the backend system and how to create a new backend can be found in the main \GB paper \cite{gambit}.

Consequently, as with other calculations, the generation of a spectrum can be delegated to a backend library which provides the required capability. At the time of the first release, \GB comes along with \FlexibleSUSY \cite{Athron:2014yba} internally implemented as the de facto spectrum generator for MSSM models, but also allows the use of the backended spectrum generator \SPheno \cite{Porod:2003um,Porod:2011nf}. In this section we describe how \SPheno is added as a backend to \GB and how it provides the spectrum capability \cpp{unimproved_MSSM_spectrum}, to serve as a guide for the addition of other spectrum generators in the future, e.g. \SOFTSUSY \cite{Allanach:2001kg, Allanach:2009bv, Allanach:2011de, Allanach:2014nba} or \SUSPECT \cite{Djouadi:2002ze}. It is worth noting that \SPheno is written in \Fortran, so some of the details of the backending process will differ significantly from software written in other languages; \SOFTSUSY, for example, is written in \Cpp, so it would need to draw on the classloading tool \doublecrosssf{BOSS}{BOSS} (Backend-On-a-Stick Script, described in \cite{gambit}).

\subsection{Installation and import of variables/functions from the backend}

The general guideline to create a \GB backend from an external tool, as described in the main \GB paper \cite{gambit}, is to compile it into a shared library in order to provide access to the internal variables and functions at runtime. This thus requires patching the makefile provided by \SPheno to create this shared library. Effectively one needs to add the flag \cpp{-shared} to the rule \cpp{lib/libspheno.so}, as well as the flag \cpp{-fPIC} to the compilation rule for every source file.

Once compiled, this makes the variables and functions inside \SPheno available to use in \GB. However, the main Fortran program \cpp{SPheno} does not get a shared library symbol, and thus all the functions and subroutines inside, such as the one that calculates the spectrum \cpp{CalculateSpectrum}, are not made available to \GB. This is easily fixed by a further patch into the main file \cpp{SPheno3.f90}, to change the \cpp{Program} directive for a \cpp{Module} directive, and commenting all the executed lines in the program, but leaving the subroutines unscathed.

Therefore, all the variables and functions can be accessed by the backend system. These are imported by using the macros \cpp{BE_VARIABLE} and \cpp{BE_FUNCTION}, described in the Backends section of the \GB main paper \cite{gambit}, with the library symbols \cpp{``__module_MOD_var''} for a variable \cpp{var} in a module \cpp{module}. Note that Fortran is case insensitive, but all library symbols are written in small caps (except the MOD tag). The files \term{SpecBit/include/gambit/SpecBit/frontends/SPheno.hpp} and \term{SpecBit/src/frontends/SPheno.cpp} are then created to import these variables and functions, and to give any specifications that are necessary.

Along with the imported variables and functions, a set of convenience functions is created, by the macro \cpp{BE_CONV_FUNCTION}, see Ref.\ \cite{gambit}, specified in the frontend file \term{SPheno.cpp}. These are \cpp{run_SPheno}, \cpp{ReadingData}, \cpp{InitializeStandardModel}, \cpp{Spectrum_Out} and \cpp{ErrorHandling}. In particular the convenience function \cpp{run_SPheno} will effectively be the main function of the backend and will cover the executed lines commented out by the patch, avoiding those that perform input and output operations, which will be described in the section below.

Lastly, every backend gets a backend initialisation function, where any overall initialisations for the backend are performed, inside the namespace \cpp{BE_INI_FUNCTION} in the frontend source file. The only initialisation required by \SPheno at this stage is the information about the model which is done by the function \cpp{ModelInUse}.

\subsection{Input and output, warnings and errors}

\SPheno follows the SLHA conventions \cite{Skands:2003cj, Allanach:2008qq} for the format of the spectra it generates, and as such all its input and (successful) output happens in the form of files using the SLHA format. Reading and writing from files is inefficient and hence, for the most part, \GB avoids performing any of these tasks at the point level, for it would slow down the scans significantly. Therefore, all the input required and output provided by \SPheno is stored and obtained directly to and from the backend variables, as described above, via two convenience functions \cpp{ReadingData} and \cpp{Spectrum_Out}.

The convenience function \cpp{ReadingData} initialises all required \SPheno variables for the relevant model and scan. Firstly all the SM parameters are initialised in the convenience function \cpp{InitializeStandardModel} using the structure \cpp{SMInputs}, followed by some internal initialization of variables via the backended functions \cpp{InitializeLoopFunctions} and \cpp{Set_All_Parameters_0}. In addition all the run options, as described in detail in Sec.\ \ref{sec:SPheno_options}, are parsed at this step.

Secondly, all the information pertaining to the SLHA blocks \textsf{MINPAR}, \textsf{EXTPAR}, \textsf{MSL2}, \textsf{MSE2}, \textsf{MSQ2}, \textsf{MSU2}, \textsf{MSD2}, \textsf{TUIN}, \textsf{TDIN} and \textsf{TEIN} is pulled from the model parameters and stored in the corresponding \SPheno variables. As expected, not every variable is required to be filled for every model, e.g.~for the CMSSM, only the variables relating to the \textsf{MINPAR} block are filled, which are obtained from the model parameters \cpp{M0}, \cpp{M12}, \cpp{TanBeta}, \cpp{SignMu} and \cpp{A0}. Variables relating to other SLHA input blocks are either filled somewhere else, \textsf{SMINPUTS} in \cpp{InitializeStandardModel}, and \textsf{MODSEL} in the backend initialization function, as described above; or are ignored for they have no internal use as an input, such as \textsf{SPINFO}.

The output convenience function \cpp{Spectrum_Out} deals with the creation of a \cpp{Spectrum} object from the spectrum generated by \SPheno. As was mentioned in Sec.\ \ref{sec:MSSM_options}, the backended \SPheno version does not provide post-generation RGE running, like \FlexibleSUSY does. Rather, the spectrum is stored in a \cpp{SLHAstruct} variable and then transformed into a static \cpp{Spectrum} object via the function \cpp{spectrum_from_SLHAea}. By default the scale dependent output (SLHA blocks \textsf{MSOFT}, \textsf{GAUGE}, etc.) is given at the SUSY scale.

Warnings and error messages in \SPheno are also written into a file and, likewise, \GB diverts that output to avoid slowing down the scans. Internally this is done by initializing the variable \cpp{ErrCan} to 0, hereby forcing \SPheno to write any output into an unused buffer and not to a file. Error tracking is hence done by use of the integer variable \cpp{kont}, whose value depends on the specific error that occurred. For any non-zero value of \cpp{kont}, the convenience function \cpp{ErrorHandling} appends the corresponding message to the logger and raises an \cpp{invalid_point} exception. Specific details on which values of \cpp{kont} correspond to which error messages can be found in Appendix C of the \SPheno manual~\cite{Porod:2011nf}.

\subsection{Calculation of the spectrum}

The spectrum is calculated in \SPheno by using the backended function \cpp{CalculateSpectrum}, provided all required variables are initialised previously. After this function is run, all the output variables are filled (including the error tracking variable \cpp{kont}), and they are written into a \cpp{Spectrum} object, as described above.

Once the spectrum is calculated, it is returned to the function that had this backend requirement, by providing the capability \cpp{SPheno_MSSMspectrum}. The built-in function with this backend requirement is \cpp{get_MSSM_spectrum_SPheno}. This function provide the capability \cpp{unimproved_MSSM_spectrum}, effectively passing along the created \cpp{Spectrum} object to the next level in the dependency tree. The definition of the function \cpp{get_MSSM_spectrum_SPheno} is:

\begin{lstcpp}
  #define FUNCTION get_MSSM_spectrum_SPheno
 START_FUNCTION(Spectrum)
 ALLOW_MODELS(CMSSM, MSSM63atMGUT, MSSM63atQ)
 DEPENDENCY(SMINPUTS, SMInputs)
 BACKEND_REQ(SPheno_MSSMspectrum,
  (libSPheno), int, (Spectrum&, const Finputs&) )
 BACKEND_OPTION((SPheno, 3.3.8), (libSPheno))
  #undef FUNCTION
\end{lstcpp}
where \cpp{Finputs} is a data structure containing the input parameters and the run options.

All of the additional calculations available in the out-of-the-box version of \SPheno, such as low energy observables, branching ratios and cross sections, are disabled in this backended version, as their functionality is already covered by other \GB modules: \flavbit \cite{FlavBit}, \decaybit and \colliderbit \cite{ColliderBit}.

\section{List of \specbit~\cpp{SubSpectrumContents} definitions}
\label{app:subspec_contents}

Spectrum information that has been stored in \specbit may be accessed
via the \cpp{SubSpectrum} interface class. The stored information is given
in Tables \ref{tab:SM_contents} (for low energy SM
information), \ref{tab:MSSM_contents} (for the MSSM)
and \ref{tab:SingletDM_contents} (for the scalar singlet model). To extract this information \specbit has accessors that have one of the following forms,
\begin{lstcpp}
 @\metavar{subspectrum}@.get(@\metavar{tag}@,@\metavar{label}@)
 @\metavar{subspectrum}@.get(@\metavar{tag}@,@\metavar{label}@,@\metavar{index 1}@)
 @\metavar{subspectrum}@.get(@\metavar{tag}@,@\metavar{label}@,@\metavar{index 1}@,@\metavar{index 2}@)
 \end{lstcpp}
 The arguments \metavar{label} and \metavar{tag} are given in Tables \ref{tab:SM_contents}, \ref{tab:MSSM_contents} and \ref{tab:SingletDM_contents}  in columns 3 and 4 respectively.  The fifth column in these tables specifies the number of indices that are required as additional arguments, which will be either $0$, $1$ or $2$.  When necessary the last two columns in these tables provide the range each index runs over.  These indices specify the element in a vector or matrix of parameters that are structured this way.

If the \cpp{using namespace Par} directive is not in use in the scope, then \metavar{tag} must be prefixed with the namespace qualifer \cpp{Par}, e.g.\ \cpp{Par::}\metavar{tag}.

The \cpp{SubSpectrum} classes provide interfaces to specific backend spectrum generators, while \cpp{SubSpectrumContents} classes define what parameters a \cpp{SubSpectrum} wrapper must provide. In Table \ref{tab:list_of_subspectrum_contents} we list the \cpp{SubSpectrumContents} definitions which ship with the \specbit module in \gambitVer, and describe each of them in detail in the table referenced beside each list item.

\begin{table}[h]
\centering
\scriptsize{
\begin{tabularx}{\columnwidth}{l|X|l}
Name & Plain English Description  & Table \\[2mm]
\hline
\cpp{SM}              & SM pole masses and QED$\times$QCD running parameters & \ref{tab:SM_contents} \\[1.5mm]
\cpp{SM_slha}         & SM pole masses and SLHA SMINPUTS MSbar masses (light quarks) & \ref{tab:SM_contents} \\[1.5mm]
\cpp{SMHiggs}         & SM Higgs sector; pole mass and VEV& \ref{tab:SM_contents} \\[1.5mm]
\cpp{MSSM}            & 63 parameter general MSSM in SLHA2 conventions & \ref{tab:MSSM_contents} \\[1.5mm]
\cpp{SingletDM}       & Scalar singlet dark matter plus SM Higgs sector & \ref{tab:SingletDM_contents} \\[1.5mm]
\end{tabularx}
}
\caption{List of available \cpp{SubSpectrumContents} definitions.
\label{tab:list_of_subspectrum_contents}}
\end{table}

\section{List of \specbit~\cpp{SubSpectrum} wrappers}
\label{app:subspec_wrappers}

Here we provide a list of all the \cpp{SubSpectrum} wrapper classes which ship with the \gambitVer version of \specbit. Wrappers based on simple parameter containers with no RGE facilities are listed in Table \ref{tab:list_of_subspectrum_wrappers_simple}, while wrappers that connect to full spectrum calculators are listed in Table \ref{tab:list_of_subspectrum_wrappers_full}. The interface to the wrapped content of these classes is defined by their associated \cpp{SubSpectrumContents} classes, which are given in the ``Contents'' column of each table. The \cpp{SubSpectrumContents} classes are themselves described in Appendix \ref{app:subspec_contents}.

\begin{table}
\centering
\scriptsize{
\begin{tabularx}{\columnwidth}{l|X|l}
Name & Plain English Description  & Contents \\[2mm]
\hline
\cpp{MSSMSimpleSpec}           & General container for MSSM spectrum. Can be constructed from an SLHAea object & \cpp{MSSM} \\[1.5mm]
\cpp{SLHASimpleSpec}           & Base class for wrappers based on SLHAea. & n/a \\[1.5mm]
\multirow{2}{2cm}{\cpp{SMHiggs}\\\ \cpp{SimpleSpec}}
                         & Very simple wrapper for SM Higgs parameters & \cpp{SMHiggs} \\[1.5mm]
\cpp{SMSimpleSpec}             & General container for SM spectrum data. Can be constructed from an SLHAea object & \cpp{SM_slha} \\[1.5mm]
\multirow{2}{2cm}{\cpp{ScalarSingletDM}\\\ \cpp{SimpleSpec}}
                         & Simple wrapper for scalar singlet DM parameters, plus SM gauge and Yukawa couplings & \cpp{SingletDM} \\[1.5mm]
\end{tabularx}
}
\caption{List of available ``simple'' \cpp{SubSpectrum} wrappers, which wrap only parameter data and provide no RGE facilities, and are therefore not connected to any backend spectrum generators. Spectrum data is either entered directly as input parameters or retrieved from an external source, such as an SLHA file. The final column `Contents' gives the \cpp{SubSpectrumContents} definition (see Appendix \ref{app:subspec_contents}) to which the wrapper conforms.
\label{tab:list_of_subspectrum_wrappers_simple}}
\end{table}

\begin{table}
\centering
\scriptsize{
\begin{tabularx}{\columnwidth}{l|X|l}
Name & Plain English Description  & Contents \\[2mm]
\hline
\cpp{MSSMSpec}      & General templated wrapper for all \FlexibleSUSY MSSM models & \cpp{MSSM} \\[1.5mm]
\cpp{QedQcdWrapper} & Wraps QED$\times$QCD calculator from \SOFTSUSY & \cpp{SM} \\[1.5mm]
\cpp{SingletDMSpec} & Wraps \FlexibleSUSY spectrum for scalar singlet model  & \cpp{SingletDM} \\[1.5mm]
\end{tabularx}
}
\caption{List of available ``full'' \cpp{SubSpectrum} wrappers, which connect directly to backend spectrum generator code and provide RGE running facilities. The final column 'Contents' gives the \cpp{SubSpectrumContents} definition (see Appendix \ref{app:subspec_contents}) to which the wrapper conforms.
\label{tab:list_of_subspectrum_wrappers_full}}
\end{table}

\section{\cpp{SubSpectrum} wrapper \cpp{GetterMaps} and \cpp{SetterMaps} function signatures}
\label{app:gettermaps}

In this appendix we give details of the allowed function pointer signatures for the \cpp{GetterMaps} and \cpp{SetterMaps} objects which constitute the link between \cpp{SubSpectrum} interface functions, and backend spectrum data (see Sec. \ref{sec:parameter_box_wrapper}). In principle the system can be extended to allow any function signature. However, it cannot be done automatically, and thus at present only a limited set of options is available. If the object to which you want to interface does not conform to one of these signatures, you will need to write `helper' functions in your wrapper which alter the function signature. Some of the function signatures listed are designed to support such helper functions.

The list is given in Table \ref{tab:gettermap_sigs}. So, for example, to store a function pointer with the signature \cpp{double f(Model&)} in a \cpp{GetterMaps} collection, one needs to put it into the \cpp{map0} map, as it shown in the example in Sec. \ref{sec:model_member_functions}. The \cpp{map0W} (see Sec. \ref{sec:parameter_box_wrapper}), \cpp{map0_extraM} and \cpp{map0_extraI} maps work similarly, with function signatures as shown in the table.

Functions associated with indices require an extra step; they must be associated with lists of the allowed indices. This is made easy via the \cpp{FInfo} helper classes, which are listed in table \ref{tab:gettermap_sigs}. For example, say we wish to attached two-index functions with the signature \cpp{double Model::f(int,int)} to the \cpp{get} interface. The \cpp{fill_getter_maps} function for a wrapper \metavar{Wrap} would then be:
\begin{lstcpp}
static @\metavar{Wrap}@::GetterMaps
  @\metavar{Wrap}@::fill_getter_maps()
{
 static const std::set<int> i012 = initSet(0,1,2);
 GetterMaps map_collection;

 map_collection[Par::@\metavar{tag}@].map2[@\metavar{label}@]
   = MTget::FInfo2(&Model::f,i012,i012)

 return map_collection;
}
\end{lstcpp}
where \cpp{std::set<int> i012} is initialised with a helper function \cpp{initSet} (from the header \term{gambit/} \term{Utils/util_functions.hpp}). This set is then used to tell the \cpp{GetterMaps} that each index of the function can accept the values 0, 1 or 2. At present, it is not possible to restrict the allowed index values pair-wise (that is, \GB cannot automatically provide safeguards against users asking for specific invalid combinations of indices).

For more examples of how to add functions to \cpp{GetterMaps} and \cpp{SetterMaps} please see the source code for the various \cpp{SubSpectrum} wrappers that ship with \specbit. These are listed in Appendix \ref{app:subspec_wrappers}.

\begin{table*}
\centering
\begin{tabular}{l|l|l|l|l}
Data member name & Type & Helper class & ``Getter'' function signature & ``Setter'' function signature \\[2mm]
\hline
\hline
\cpp{map0}        & \cpp{fmap0}        & none          &\cpp{double Model::f()       } & \cpp{void Model::f(double)        } \\[1.5mm]
\cpp{map1}        & \cpp{fmap1}        & \cpp{FInfo1}  &\cpp{double Model::f(int)    } & \cpp{void Model::f(int,double)    } \\[1.5mm]
\cpp{map2}        & \cpp{fmap2}        & \cpp{FInfo2}  &\cpp{double Model::f(int,int)} & \cpp{void Model::f(int,int,double)} \\[1.5mm]
\cpp{map0W}       & \cpp{fmap0W}       & none          &\cpp{double Self::f()        } & \cpp{void Self::f(double)         } \\[1.5mm]
\cpp{map1W}       & \cpp{fmap1W}       & \cpp{FInfo1W} &\cpp{double Self::f(int)     } & \cpp{void Self::f(double,int)     } \\[1.5mm]
\cpp{map2W}       & \cpp{fmap2W}       & \cpp{FInfo2W} &\cpp{double Self::f(int,int) } & \cpp{void Self::f(double,int,int) } \\[1.5mm]
\cpp{map0_extraM} & \cpp{fmap0_extraM} & none          &\cpp{double f(Model&)        } & \cpp{void f(Model&,double)        } \\[1.5mm]
\cpp{map1_extraM} & \cpp{fmap1_extraM} & \cpp{FInfo1M} &\cpp{double f(Model&,int)     } & \cpp{void f(Model&,double,int)    } \\[1.5mm]
\cpp{map2_extraM} & \cpp{fmap2_extraM} & \cpp{FInfo2M} &\cpp{double f(Model&,int,int) } & \cpp{void f(Model&,double,int,int)} \\[1.5mm]
\cpp{map0_extraI} & \cpp{fmap0_extraI} & none          &\cpp{double f(Input&)        } & \cpp{void f(Input&,double)        } \\[1.5mm]
\cpp{map1_extraI} & \cpp{fmap1_extraI} & \cpp{FInfo1I} &\cpp{double f(Input&,int)     } & \cpp{void f(Input&,double,int)    } \\[1.5mm]
\cpp{map2_extraI} & \cpp{fmap2_extraI} & \cpp{FInfo2I} &\cpp{double f(Input&,int,int) } & \cpp{void f(Input&,double,int,int)} \\[1.5mm]
\end{tabular}
\caption{List of data members of the \cpp{GetterMaps} and \cpp{SetterMaps} classes (which are \cpp{std::map}s), and the corresponding signatures of function pointers which can be stored in those maps. Note, however, that functions requiring index input cannot be stored directly in the map; they must first be associated with sets of allowed values for those indices using the listed ``Helper class''; see text. In the table, the name \cpp{Self} refers to the \cpp{SubSpectrum} wrapper class itself (that is, these maps accept function pointers to members functions of the wrapper class).}
\label{tab:gettermap_sigs}
\end{table*}

\newcommand\strtagwidth{2.5cm}
\begin{table*}
\centering
\begin{tabular}{c|c|p{\strtagwidth}|c|c|c|c}
Quantity    &  Plain English Description   &   string &  tag  & \pbox{3cm}{number \\ \ctr of indices}  &  \pbox{3cm}{index 1 \\ \ctr range}  & \pbox{3cm}{index 2 \\ \ctr range} \\[2mm]
\hline
\hline
$m^{pole}_\gamma$ & Photon pole mass       &\cpp{"gamma"}   & \cpp{ Pole_Mass} & 0 & n/a & n/a \\[1.5mm]
$m^{pole}_g$ & Gluon pole mass        &\cpp{"g"    }   & \cpp{ Pole_Mass} & 0 & n/a & n/a \\[1.5mm]
$m^{pole}_Z$ & Z pole mass            &\cpp{"Z0"   }   & \cpp{ Pole_Mass} & 0 & n/a & n/a \\[1.5mm]
$m^{pole}_W$ & W pole mass            &\cpp{"W+"/"W-"} & \cpp{ Pole_Mass} & 0 & n/a & n/a \\[1.5mm]

$m^{pole}_t$ & Top quark pole mass    &\multirow{2}{*}{\parbox{\strtagwidth}{ \cpp{"u_3"/"ubar_3"/} \\\
                                       \cpp{"t"/"tbar"} }} & \cpp{ Pole_Mass} & 0 & n/a & n/a \\[1.5mm]
 & & & & & \\[1.5mm] 

$m^{pole}_b$ & Bottom quark pole mass &\multirow{2}{*}{\parbox{\strtagwidth}{ \cpp{"d_3"/"dbar_3"/} \\\
                                       \cpp{"b"/"bbar"} }} & \cpp{ Pole_Mass} & 0 & n/a & n/a \\[1.5mm]
 & & & & & \\[1.5mm] 

$m^{pole}_e$ & Electron pole mass     &\cpp{"e-"/"e+"}     & \cpp{ Pole_Mass} & 0 & n/a & n/a \\[1.5mm]
$m^{pole}_\mu$ & Muon pole mass         &\cpp{"mu-"/"mu+"}   & \cpp{ Pole_Mass} & 0 & n/a & n/a \\[1.5mm]
$m^{pole}_\tau$ & Tau pole mass          &\cpp{"tau-"/"tau+"} & \cpp{ Pole_Mass} & 0 & n/a & n/a \\[1.5mm]
$m^{pole}_{l_i}$ & Charged lepton pole masses &\cpp{"e-"/"e+"} & \cpp{ Pole_Mass} & 1 & {1,2,3} & n/a \\[1.5mm]
$m^{pole}_{\nu_i}$ & Neutrino pole masses       &\cpp{"nu"/"nubar"} & \cpp{ Pole_Mass} & 1 & {1,2,3} & n/a \\[1.5mm]
$\sin^2\theta^{\textrm{pole}}_W$ in (\ref{sthWpole}) & Weak mixing angle      &\cpp{"sinW2"} & \cpp{Pole_Mixing} & 0 & n/a & n/a \\[1.5mm]
$\alpha_e$ & Fine structure constant   &\cpp{"alpha"} & \cpp{dimensionless} & 0 & n/a & n/a \\[1.5mm]
$\alpha_s$ & Strong coupling constant  &\cpp{"alphaS"}& \cpp{dimensionless} & 0 & n/a & n/a \\[1.5mm]

$m_\gamma$ & Photon $\overline{MS}$ mass       &\cpp{"gamma"}    & \cpp{ mass1} & 0 & n/a & n/a \\[1.5mm]
$m_g$ & Gluon $\overline{MS}$ mass             &\cpp{"g"}        & \cpp{ mass1} & 0 & n/a & n/a \\[1.5mm]
$m_u$ & Up-type quark $\overline{MS}$ masses   &\cpp{"u"/"ubar"} & \cpp{ mass1} & 1 & {1,2,3}$^*$ & n/a \\[1.5mm]
$m_d$ & Down-type quark $\overline{MS}$ masses &\cpp{"d"/"dbar"} & \cpp{ mass1} & 1 & {1,2,3}$^*$ & n/a \\[1.5mm]
$m_l$ & Charged lepton $\overline{MS}$ masses  &\cpp{"e-"/"e+"}  & \cpp{ mass1} & 1 & {1,2,3}$^*$ & n/a \\[1.5mm]
\hline
$m_{h^0}$ &    Higgs boson pole mass &\cpp{"h0"}  & \cpp{Pole_Mass} & 0 & n/a & n/a \\[1.5mm]
$v$ in (\ref{Eq:vev}) & Higgs VEV             &\cpp{"vev"} & \cpp{mass1}     & 0 & n/a & n/a \\[1.5mm]
\end{tabular}

\caption{The spectrum contents available via the \cpp{get} function for the wrappers conforming to the \cpp{SM} and \cpp{SM_slha} (Standard Model) contents definitions (Sec. \ref{sec:subspectrum_contents}), except the Higgs pole mass and VEV, which are available from wrappers conforming to the \cpp{SMHiggs} definition and which contain {\em only} those two parameters. We found this split to be convenient since BSM models commonly alter the Higgs sector. The above quantities may all be obtained via \metavar{subspectrum}\cpp{.get(Par::}\metavar{tag},\metavar{label},\metavar{index 1},\metavar{index 2}) where \metavar{label} and \metavar{tag} are given in columns 3 and 4 respectively and \metavar{subspectrum} is defined as in Sec.~\ref{sec:subspectrum_structure}.  The final two arguments are optional and are filled when necessary by the indices from the allowed range given in columns 6 and 7. $^*$These indices only apply to \cpp{SM} contents; for \cpp{SM_slha} wrappers only \cpp{u_1}, \cpp{d_1}, and \cpp{d_2} are available.  Note that the weak mixing angle, provided in the form $\sin^2\theta^{\textrm{pole}}_W$, corresponds to the pole definition from Eq.~\ref{sthWpole}. The scale at which all $\overline{MS}$ parameters are defined is returned by \metavar{subspectrum}\cpp{.GetScale()} (see Sec.~\ref{sec:subspectrum_structure} and Fig.~\ref{fig:subspectrum_object}), unless the wrapper object does not support RGE running, in which case the scales should be in accordance with SLHA2 SMINPUTS definitions (this is the convention followed in the \cpp{SMSimpleSpec} wrapper class, for example. See Appendix~\ref{app:subspec_wrappers} for full lists of the currently available wrapper classes). Note that if definitions in accordance with SLHA2 SMINPUTS are desired, one can always access the parameters in the \cpp{SMInputs} structure of the parent \cpp{Spectrum} object rather than performing the required RGE running; for example \metavar{spectrum}\cpp{.get_SMInputs().mS} retrieves the strange quark $\overline{MS}$ mass at 2 GeV. See Fig.~\ref{fig:sminputs} for the definition of the \cpp{SMInputs} class, and see Fig.~\ref{fig:spectrum_structure} and \ref{fig:spectrum_object} for the structure of \cpp{Spectrum} objects.
\label{tab:SM_contents}}
\end{table*}

\begin{table*}
\centering
\renewcommand{\arraystretch}{0.7} 
\begin{tabular}{c|c|c|c|c|c}
Quantity    &  Plain English Description   &   string &  tag  &  \pbox{3cm}{index 1 \\ \ctr range}  & \pbox{3cm}{index 2 \\ \ctr range} \\[2mm]
\hline
\hline
$m^{pole}_{\tilde{g}}$ &   Gluino pole masses  &\cpp{ "~g"} &  \cpp{Pole_Mass} & n/a & n/a \\[1.5mm]
$m^{pole}_{W^\pm}$ &   W boson pole mass  &\cpp{"W+"} & \cpp{ Pole_Mass} & n/a & n/a \\[1.5mm]
$m^{pole}_{A^0}$  &   Psedoscalar Higgs pole mass  & \cpp{"A0"} & \cpp{ Pole_Mass} & n/a & n/a \\[1.5mm]
$m^{pole}_{H^+}$  &    Charged Higgs pole masses  &\cpp{"H+"} & \cpp{ Pole_Mass} & n/a & n/a \\[1.5mm]
$m^{pole}_{h^0}$  &   $CP$-even Higgs pole masses  & \cpp{"h0"} & \cpp{ Pole_Mass} & \cpp{\{1,2\}} & n/a \\[1.5mm]
$m^{pole}_{\tilde{u}}$ & up-type squark pole masses  & \cpp{"~u"} & \cpp{ Pole_Mass} & \cpp{\{1,2,3,4,5,6\}} & n/a \\[1.5mm]
$m^{pole}_{\tilde{d}}$ & down-type squark pole masses  & \cpp{"~d"} & \cpp{ Pole_Mass} & \cpp{\{1,2,3,4,5,6\}} & n/a \\[1.5mm]
$m^{pole}_{\tilde{e}}$ & slepton pole masses  & \cpp{"~e"} & \cpp{ Pole_Mass} & \cpp{\{1,2,3,4,5,6\}} & n/a \\[1.5mm]
$m^{pole}_{\tilde{\nu}}$ & sneutrino pole mass  & \cpp{"~nu"} & \cpp{ Pole_Mass} & \cpp{\{1,2,3\}} & n/a \\[1.5mm]
$m^{pole}_{\chi^\pm}$  &   chargino pole masses  & \cpp{"~chi+"} & \cpp{ Pole_Mass} & \cpp{\{1,2\}} & n/a \\[1.5mm]
$m^{pole}_{\chi^0}$ & neutralino pole masses  & \cpp{"~chi0"} & \cpp{ Pole_Mass} & \cpp{\{1,2,3,4\}} & n/a \\[1.5mm]
$Z_{\tilde{u}}$ in (\ref{Eq:sqrot}) & up-type squark mixing  & \cpp{"~u"} & \cpp{ Pole_Mixing} & \cpp{\{1,2,3,4,5,6\}} & \cpp{\{1,2,3,4,5,6\}} \\[1.5mm]
$Z_{\tilde{d}}$ in (\ref{Eq:sqrot}) & down-type squark mixing  & \cpp{"~d"} & \cpp{ Pole_Mixing} & \cpp{\{1,2,3,4,5,6\}} & \cpp{\{1,2,3,4,5,6\}} \\[1.5mm]
$Z_{\tilde{e}}$ in (\ref{Eq:slrot}) & slepton mixing  & \cpp{"~e"} & \cpp{ Pole_Mixing} & \cpp{\{1,2,3,4,5,6\}} & \cpp{\{1,2,3,4,5,6\}} \\[1.5mm]
$Z_{\tilde{\nu}}$ in (\ref{Eq:slrot}) & sneutrino mixing  & \cpp{"~nu"} & \cpp{ Pole_Mixing} & \cpp{\{1,2,3\}} & \cpp{\{1,2,3\}} \\[1.5mm]
$Z_{N}$ in (\ref{Eq:ChiRot}) & neutralino mixing  & \cpp{"~chi0"} & \cpp{ Pole_Mixing} & \cpp{\{1,2,3,4\}} & \cpp{\{1,2,3,4\}} \\[1.5mm]
$U_{+}$ in (\ref{Eq:ChaRot+}) & chargino mixing  & \cpp{"~chi+"} & \cpp{ Pole_Mixing} & \cpp{\{1,2\}} & \cpp{\{1,2\}} \\[1.5mm]
$U_{-}$ in (\ref{Eq:ChaRot-}) & chargino mixing  & \cpp{"~chi-"} & \cpp{ Pole_Mixing} & \cpp{\{1,2\}} & \cpp{\{1,2\}} \\[1.5mm]
$Z_{H}$ in (\ref{Eq:CPEvenRot}) & CP even Higgs mixing  & \cpp{"h0"} & \cpp{ Pole_Mixing} & \cpp{\{1,2\}} & \cpp{\{1,2\}} \\[1.5mm]
$Z_{A}$ in \ref{Eq:CPOddRot} & CP odd Higgs mixing  & \cpp{"A0"} & \cpp{ Pole_Mixing} & \cpp{\{1,2\}} & \cpp{\{1,2\}} \\[1.5mm]
$B\mu$ in (\ref{Higgses}) &  soft Higgs bilinear  &\cpp{"BMu"} & \cpp{ mass2} & n/a & n/a \\[1.5mm]
$m_{H_u}^2$ in (\ref{Higgses}) &  soft Higgs mass squared  &\cpp{"mHu2"} & \cpp{ mass2} & n/a & n/a \\[1.5mm]
$m_{H_d}^2$ in (\ref{Higgses}) &  soft Higgs mass squared  &\cpp{"mHd2"} & \cpp{ mass2} & n/a & n/a \\[1.5mm]
$m_{A}^2$ in (\ref{Eq:mA}) & $\overline{DR}$ pseudoscalar mass squared  &\cpp{"mA2"} & \cpp{ mass2} & n/a & n/a \\[1.5mm]
$\mathbf{m}^\mathbf{2}_Q$ in (\ref{sfermion masses}) &  soft squark mass squared  &\cpp{"mq2"} & \cpp{ mass2} & \cpp{\{1,2,3\}} & \cpp{\{1,2,3\}} \\[1.5mm]
$\mathbf{m}^\mathbf{2}_u$ in (\ref{sfermion masses}) &  soft up-squark mass squared  &\cpp{"mu2"} & \cpp{ mass2} & \cpp{\{1,2,3\}} & \cpp{\{1,2,3\}} \\[1.5mm]
$\mathbf{m}^\mathbf{2}_d$ in (\ref{sfermion masses}) &  soft down-squark mass squared  &\cpp{"md2"} & \cpp{ mass2} & \cpp{\{1,2,3\}} & \cpp{\{1,2,3\}} \\[1.5mm]
$\mathbf{m}^\mathbf{2}_L$ in (\ref{sfermion masses}) &  soft slepton mass squared  &\cpp{"ml2"} & \cpp{ mass2} & \cpp{\{1,2,3\}} & \cpp{\{1,2,3\}} \\[1.5mm]
$\mathbf{m}^\mathbf{2}_e$ in (\ref{sfermion masses}) &  soft slepton mass squared  &\cpp{"me2"} & \cpp{ mass2} & \cpp{\{1,2,3\}} & \cpp{\{1,2,3\}} \\[1.5mm]
$M_1$ in (\ref{gauginos}) &  Bino soft mass &\cpp{"M1"} & \cpp{ mass1} & n/a & n/a \\[1.5mm]
$M_2$ in (\ref{gauginos}) &  Wino soft mass &\cpp{"M2"} & \cpp{ mass1} & n/a & n/a \\[1.5mm]
$M_3$ in (\ref{gauginos}) &  Gluino soft mass &\cpp{"M3"} & \cpp{ mass1} & n/a & n/a \\[1.5mm]
$Mu$ in (\ref{eq:superpot}) &  Superpotential Higgs bilinear &\cpp{"Mu"} & \cpp{ mass1} & n/a & n/a \\[1.5mm]
$vu$ in (\ref{Eq:VEVS}) & Up-type Higgs VEV &\cpp{"vu"} & \cpp{ mass1} & n/a & n/a \\[1.5mm]
$vd$ in (\ref{Eq:VEVS}) & Down-type Higgs VEV &\cpp{"vd"} & \cpp{ mass1} & n/a & n/a \\[1.5mm]
$\mathbf{T}_u$ in (\ref{trilinear couplings}) & Trilinear &\cpp{"TYu"} & \cpp{ mass1} & \cpp{\{1,2,3\}} & \cpp{\{1,2,3\}} \\[1.5mm]
$\mathbf{T}_d$ in (\ref{trilinear couplings}) & Trilinear &\cpp{"TYd"} & \cpp{ mass1} & \cpp{\{1,2,3\}} & \cpp{\{1,2,3\}} \\[1.5mm]
$\mathbf{T}_e$ in (\ref{trilinear couplings}) & Trilinear &\cpp{"TYe"} & \cpp{ mass1} & \cpp{\{1,2,3\}} & \cpp{\{1,2,3\}} \\[1.5mm]
$g_1$ in (\ref{Eq:CovDer}) &  $U(1)_Y$ gauge coupling\footnote{GUT normalised} &\cpp{"g1"} & \cpp{ dimensionless} & n/a & n/a \\[1.5mm]
$g_2$ in (\ref{Eq:CovDer}) & $SU(2)_\mathrm{L}$ gauge coupling &\cpp{"g2"} & \cpp{ dimensionless} & n/a & n/a \\[1.5mm]
$g_3$ in (\ref{Eq:CovDer}) & $SU(3)_\mathrm{C}$ gauge coupling &\cpp{"g3"} & \cpp{ dimensionless} & n/a & n/a \\[1.5mm]
$\tan \beta$  & ratio of VEVs $v_u / v_d$ &\cpp{"tanbeta"} & \cpp{ dimensionless} & n/a & n/a \\[1.5mm]
$\sin^2 \theta_W$ in (\ref{Eq:sthW}) & weak mixing angle &\cpp{"sinW2"} & \cpp{ dimensionless} & n/a & n/a \\[1.5mm]
$\mathbf{Y}_u$ in (\ref{eq:superpot}) & Up-type quark Yukawa &\cpp{"Yu"} & \cpp{ dimensionless} & \cpp{\{1,2,3\}} & \cpp{\{1,2,3\}} \\[1.5mm]
$\mathbf{Y}_d$ in (\ref{eq:superpot}) & Down-type quark Yukawa &\cpp{"Yd"} & \cpp{ dimensionless} & \cpp{\{1,2,3\}} & \cpp{\{1,2,3\}} \\[1.5mm]
$\mathbf{Y}_e$ in (\ref{eq:superpot}) & lepton Yukawa &\cpp{"Ye"} & \cpp{ dimensionless} & \cpp{\{1,2,3\}} & \cpp{\{1,2,3\}}
\end{tabular}
\caption{The spectrum contents available via the \cpp{get} function for wrappers conforming to the \cpp{MSSM} spectrum contents definition (Sec. \ref{sec:subspectrum_contents}).  These quantities may be obtained via \metavar{subspectrum}.\cpp{get(Par::}\metavar{tag},\metavar{label},\metavar{index 1},\metavar{index 2}) where \metavar{label} and \metavar{tag} are given in columns 3 and 4 respectively and \metavar{subspectrum} is defined as in Sec.~\ref{sec:subspectrum_structure}.  The final two arguments are optional and are filled when necessary by the indices from the allowed range given in the last two columns. Note that weak mixing angle here is defined in the \DR scheme and is simply reconstructed from the gauge couplings using the expression given in Eq.~\ref{Eq:sthW}. \label{tab:MSSM_contents}}
\end{table*}

\begin{table*}
\centering
\begin{tabular}{c|c|c|c|c|c|c}
Quantity    &  Plain English Description   &   string &  tag  & \pbox{3cm}{number \\ \ctr of indices}  &  \pbox{3cm}{index 1 \\ \ctr range}  & \pbox{3cm}{index 2 \\ \ctr range} \\[2mm]
\hline
\hline
$m^{pole}_{S}$ &   Scalar pole mass  & \cpp{"s0"} & \cpp{ Pole_Mass} & 0 & n/a & n/a \\[1.5mm]
$m^{pole}_{h^0}$ &   Higgs pole mass  & \cpp{"h0"} & \cpp{ Pole_Mass} & 0 &  n/a & n/a \\[1.5mm]
$v$ in (\ref{Eq:vev}) & Higgs VEV &\cpp{"vev"} & \cpp{mass1} & 0 & n/a & n/a \\[1.5mm]
$\lambda$ in (\ref{Eq:SM_Lagrangian}) &  Higgs quartic coupling &\cpp{"lambda_h"} & \cpp{ dimensionless} & 0 & n/a & n/a \\[1.5mm]
$\lambda_{s}$ in (\ref{eqn:SingletDM_Lagrangian}) &  Scalar quartic coupling &\cpp{"lambda_S"} & \cpp{ dimensionless} & 0 & n/a & n/a \\[1.5mm]
$\lambda_{hs}$ in (\ref{eqn:SingletDM_Lagrangian}) &  Higgs quartic coupling &\cpp{"lambda_hS"} & \cpp{ dimensionless} & 0 & n/a & n/a \\[1.5mm]
$g_1$ in (\ref{Eq:CovDer}) &  $U(1)_Y$ gauge coupling &\cpp{"g1"} & \cpp{ dimensionless} & 0 & n/a & n/a \\[1.5mm]
$g_2$ in (\ref{Eq:CovDer}) & $SU(2)_W$ gauge coupling &\cpp{"g2"} & \cpp{ dimensionless} & 0 & n/a & n/a \\[1.5mm]
$g_3$ in (\ref{Eq:CovDer}) & $SU(3)_C$ gauge coupling &\cpp{"g3"} & \cpp{ dimensionless} & 0 & n/a & n/a \\[1.5mm]
$\sin^2 \theta_W$ in Eq. \ref{Eq:sthW} & weak mixing angle &\cpp{"sinW2"} & \cpp{ dimensionless} & 0 & n/a & n/a \\[1.5mm]
$\mathbf{Y}_u$ in (\ref{Eq:SM_Lagrangian})  & Up-type quark Yukawa &\cpp{"Yu"} & \cpp{ dimensionless} &2  & \cpp{\{1,2,3\}} & \cpp{\{1,2,3\}} \\[1.5mm]
$\mathbf{Y}_d$ in (\ref{Eq:SM_Lagrangian}) & Down-type quark Yukawa &\cpp{"Yd"} & \cpp{ dimensionless} &2  & \cpp{\{1,2,3\}} & \cpp{\{1,2,3\}} \\[1.5mm]
$\mathbf{Y}_e$ in (\ref{Eq:SM_Lagrangian})  & lepton Yukawa &\cpp{"Ye"} & \cpp{ dimensionless} &2  & \cpp{\{1,2,3\}} & \cpp{\{1,2,3\}}
\end{tabular}
\caption{The spectrum contents available via the \cpp{get} function for wrappers conforming to the \cpp{SingletDM} (scalar singlet) spectrum contents definition.  These quantities may be obtained via \metavar{subspectrum}.\cpp{get(Par::}\metavar{tag},\metavar{label},\metavar{index 1},\metavar{index 2}) where \metavar{label} and \metavar{tag} are given in columns 3 and 4 respectively and \metavar{subspectrum} is defined as in Sec.~\ref{sec:subspectrum_structure}.  The final two arguments are optional and are filled when necessary by the indices from the allowed range given in columns 6 and 7.}\label{tab:SingletDM_contents}
\end{table*}

\section{Adding support for new spectrum calculators}
\label{sec:adding_new_models}

While the \GB modules described in this paper are initially distributed with the MSSM and scalar
singlet models already implemented, the great advantage of this
framework is the manner in which new models can be systematically added. In
this section we describe how to do this.

The most simple way to extend \specbit is to just add a new \cpp{SubSpectrumContents} definition, which is described in Sec.\ \ref{sec:subspectrum_contents}. This creates a new standardised set of string label/tag/index sets to which new \cpp{SubSpectrum} wrappers may choose to conform. This is the first step that is required for adding a wrapper for an entirely new model, but it does not actually provide any real \capabilities{}, it only standardises the interface for wrappers which provide access to the same basic physics model.

The real work of interfacing to new spectrum information is done by writing a
\cpp{SubSpectrum} wrapper, which is the standardised \GB interface to any container
of spectrum information. This information may be supplied using a new spectrum
generator created with \FlexibleSUSY, a new model in \SARAH\ / \SPheno, or any available public or private spectrum generator for that model (that is written in C/C++ or Fortran; codes written in other languages can at present only be launched via calls to the operating system).

The new spectrum generator then needs to be interfaced with \specbit.
If the spectrum generator is written in \Cpp and has getters
and setters to access the spectrum information (and a run method) then
this task can be greatly simplified. It is also possible to use
the spectrum objects as a simple container for parameters, with
no RGE running facilities (i.e. no actual spectrum generator ``hooked up''),
which is the simplest use case.

There are therefore a number of different ways that the Spectrum
interface class can be connected to parameter data. We will tackle
these various cases in order of increasing complexity throughout the following
sections.

As well as designing the wrapper class, in order to actually use
the new spectrum object one must write \GB module functions that
construct the objects and return them as \GB \capabilities{} (running the
backend spectrum generator if required). This task
will be discussed in Sec. \ref{sec:wrapper_module_function}.

As a quick-reference for writers of new wrapper classes, in Sec. \ref{sec:specbit_adding_new_models}
we provide a checklist of tasks to be completed when writing a wrapper, with
references to the subsection of this guide in which the relevant detailed instructions
can be found.

\subsection{SubSpectrumContents definitions}
\label{sec:subspectrum_contents}

In this section we describe the mechanism which enforces the consistency of \cpp{SubSpectrum} objects wrapping different spectrum calculators. Recall that \cpp{SubSpectrum} objects are simply a virtual interface class; the real work of interfacing to external spectrum calculators is performed by wrapper classes which derive from \cpp{SubSpectrum}. In the process of defining this wrapper class, one must associate the wrapper with a \cpp{SubSpectrumContents} class. The \cpp{SubSpectrumContents} classes are simple \cpp{structs} that define which parameters are supposed to be contained by wrappers that interface to a calculator for a specific physics model. So, for example, two wrapper classes which wrap two different external spectrum calculators, but which calculate the spectrum for the same physics model, should both be associated with the same \cpp{SubSpectrumContents} class. The shared \cpp{SubSpectrumContents} class then acts as a promise that identical `get' calls made to each of the two wrappers should retrieve equivalent information. Furthermore, if a wrapper fails to make certain parameters accessible that are defined as part of the \cpp{SpectrumContents}, then a runtime error will occur when the spectrum is constructed. The error will explain that the wrapper does not conform to the declared contents and so cannot be used. This message should serve only to help wrapper writers not to forget any contents; it will not occur when using completed wrappers.

The first release of \specbit defines the set of \cpp{SubSpectrumContents} classes listed below:
\begin{lstcpp}
namespace SpectrumContents {
 struct SM              { SM(); };
 struct SM_slha         { SM_slha(); };
 struct SMHiggs         { SMHiggs(); };
 struct MSSM            { MSSM(); };
 struct SingletDM       { SingletDM(); };
}
\end{lstcpp}
and descriptions of each of them can be found in Appendix \ref{app:subspec_contents}. The above code shows the entire declaration for these classes, as can be found in \term{Models/include/gambit/Models/SpectrumContents/Regi}- \term{steredSpectra.hpp}, with the exception that each class is derived from a base class \cpp{SubSpectrumContents} (omitted here for brevity). The only definition required for these classes is their constructor, which for the above classes are defined in source files with matching names in the \term{Models/src/SpectrumContents/} directory. Defining the constructor, which defines the required parameters for that spectrum type, is very simple. It is easiest to explain with an example, so below we show the definition for the \cpp{ScalarSingletDM} class:
\begin{lstcpp}
SpectrumContents::
  ScalarSingletDM::ScalarSingletDM()
{
   using namespace Par;
   setName("ScalarSingletDM");
   addParameter(mass1,         "vev"      );
   addParameter(dimensionless, "lambda_hS");
   addParameter(Pole_Mass,     "h0");
   addParameter(Pole_Mass,     "S" );
}
\end{lstcpp}
First, a string name for the spectrum contents should be declared via the \cpp{setName} function. After that one simply declares all the parameters that should exist in the \cpp{SubSpectrum} wrapper by specifying the tag/string name/indices by which that parameter should be accessed. In the example the index definition argument is omitted, and so the parameter is assumed to be scalar-valued and require no index. To define an index requirement, one supplies as a final argument a vector of integers, specifying the dimension size of each required index. For example, to define a vector parameter with six entries, and a matrix parameter requiring two indices each of size three, one would write e.g.

\begin{lstcpp}
std::vector<int> v6; v6.push_back(6);
std::vector<int> m3x3; m3x3.push_back(3);
                       m3x3.push_back(3);
addParameter(Pole_Mass, "~d"  , v6);
addParameter(mass2    ,  "mq2", m3x3);
\end{lstcpp}
The \cpp{SubSpectrumContents} constructor definitions can always be consulted to check exactly what content is required to be retrievable from any wrapper that conforms to it. For reference purposes we provide tables in Appendix \ref{app:subspec_contents} that describe each of the pre-defined \cpp{SubSpectrumContents} sets, with cross-referencing to theoretical descriptions of each parameter. The descriptions of pre-defined \cpp{SubSpectrum} wrappers in Appendix \ref{app:subspec_wrappers} also refer to these tables to describe what they contain.

\subsection{Wrapping a simple parameter collection}
\label{sec:parameter_box_wrapper}

The most basic kind of object for which one may wish to construct an interface is a
simple class that contains parameter values. Such an object is far simpler
than the typical case that the \cpp{SubSpectrum} interface is designed to
wrap, and so many of its features are not needed. The functionality provided by wrapping
these simple objects in a \cpp{SubSpectrum} could be entirely
replaced by the standard \GB Model system (see Ref.\ \cite{gambit}), which already
provides a way to deal with simple parameter containers, however
it can be useful to interact with parameters via the Spectrum interface in cases
where they might alternatively be provided by a true spectrum generator. It is also
useful to examine this case simply to demonstrate the most basic
requirements that a wrapper class must fulfill.

So let us consider the following simple class, and construct a wrapper that
links it to the SubSpectrum interface. This wrapper is in fact implemented
in \GB and provides some scalar-singlet dark matter parameters and Standard
Model Higgs sector parameters for use by \darkbit; the full in-code
implementation can be found in
\begin{lstcpp}
Models/include/gambit/Models/SimpleSpectra/
ScalarSingletDMSimpleSpec.hpp.
\end{lstcpp}

\begin{lstcpplabel}[labelname={Example},caption={Simple ``Model'' class for the ScalarSingletDM model},label={ex:singlet_dm_model}]

namespace Gambit
{
  namespace Models
  {
    struct SingletDMModel
    {
      double HiggsPoleMass
      double HiggsVEV
      double SingletPoleMass
      double SingletLambda;
    };
  }
}
\end{lstcpplabel}
The namespace used here is not important, we specify it just to match
the actual code.

To begin wrapping this structure, one must define a specialisation
of the \cpp{SpecTraits} template class. The purpose of
this class is to communicate essential type information to the \cpp{SubSpectrum} base classes. Suppose that our wrapper
class is to be named \cpp{ScalarSingletDMSimpleSpec}; the associated
traits class should then be:
\begin{lstcpp}
namespace Gambit
{
 namespace Models
 {
   class ScalarSingletDMSimpleSpec;
 }
 template <>
 struct SpecTraits<
   Models::ScalarSingletDMSimpleSpec>
    : DefaultTraits
 {
  static std::string name()
  { return "ScalarSingletDMSimpleSpec"; }
  typedef
    SpectrumContents::ScalarSingletDM Contents;
 };
}
\end{lstcpp}
Here the namespace \emph{is} important; the original \cpp{SpecTraits} template is
declared in the \cpp{Gambit} namespace, so the specialisation must also
live in that namespace. Note that we forward declare the class \cpp{ScalarSingletDMSimpleSpec},
which will be our wrapper, because we need it as the template parameter for the
traits class.

The required members of this \cpp{SpecTraits} specialisation are as follows:
\begin{description}
\item[{\CPPidentifierstyle name()}] A function returning a \cpp{std::string} name for the wrapper. This is used in error messages so the class name is generally the logical choice.
\item[{\CPPidentifierstyle Contents}] A typedef that identifies the contents definition to which this wrapper will conform (see \cpp{SubSpectrumContents} in sec. \ref{sec:subspectrum_contents})
\end{description}
Other \cpp{SpecTraits} members can be defined, however for this simple example we do not need them (the defaults inherited from \cpp{DefaultTraits} will suffice). We will return to this in the more complicated examples of sections \ref{sec:model_member_functions} and \ref{sec:non_class_functions}.

We are now ready to define the wrapper itself. The declaration is as follows:
\begin{lstcpplabel}[labelname={Example},caption={SubSpectrum wrapper for the ScalarSingletDM model},label={ex:singlet_dm_wrapper}]
namespace Gambit
{
  namespace Models
  {
    class ScalarSingletDMSimpleSpec
      : public Spec<ScalarSingletDMSimpleSpec>
    {
     private:
      SingletDMModel params;

     public:
      ScalarSingletDMSimpleSpec
       (const SingletDMModel& p) : params(p) {}

      // Wrapper-side interface functions
      // to SingleDMModel
      double get_HiggsPoleMass()   const;
      double get_HiggsVEV()        const;
      double get_SingletPoleMass() const;
      double get_lambda_hS()       const;

      void set_HiggsPoleMass(double in);
      void set_HiggsVEV(double in);
      void set_SingletPoleMass(double in);
      void set_lambda_hS(double in);

      static GetterMaps fill_getter_maps();
      static SetterMaps fill_setter_maps();
    };
  }
}
\end{lstcpplabel}

Let us discuss what is going on here. First, the wrapper class must inherit from the \cpp{Spec} class, and provide its own type as the template parameter. This is because we employ the CRTP (curiously-recurring template pattern) for static polymorphism, to allow the base class access to the wrapper member functions.

Second, the wrapper contains an instance of the \cpp{SingletDMModel} object to which we want to interface. This is not necessary, but it is helpful for maintaining encapsulation.

Next is the constructor. The wrapper writer is quite free to do what they like with this; here we simply use it to initialise the member object.

Following the constructor are a series of ``getter'' and ``setter'' functions, which retrieve and set the parameter values we are interested in. These are the functions that we will ``hook up'' to the \cpp{SubSpectrum} ``get'' and ``set'' functions. These functions can access the hosted \cpp{SingletDMModel} in our example, and so could also perform extra tasks like calling functions of the \cpp{SingletDMModel} (if it had any) or performing unit conversions. Our later examples will demonstrate tasks like this. For now, it will suffice for these functions to have the following sort of definition:
\begin{lstcpp}
double get_lambda_hS() const
 { return params.SingletLambda; }

void set_lambda_hS(double in)
 { params.SingletLambda = in; }
\end{lstcpp}
We will refrain from listing the rest of the definitions in this example because they follow a similar pattern.

Finally, we get to the key part of the wrapper; the \cpp{fill\_getter\_maps} and \cpp{fill\_setter\_maps} functions. These are what define the actual interface to the hosted \cpp{Model} (and \cpp{Input}) objects. They are simply functions that fill a series of map (i.e. \cpp{std::map}) structures in the base class, where these maps are what define which functions are `hooked up' to the \cpp{get}, \cpp{set} and \cpp{has} functions of the \cpp{SubSpectrum} interface (see Sec. \ref{sec:subspectrum_structure}). The operation of these `filler' functions is most easily seen by looking at the definition of one, so let us examine the \cpp{fill\_getter\_maps} definition for our example:
\begin{lstcpp}
static ScalarSingletDMSimpleSpec::GetterMaps
  ScalarSingletDMSimpleSpec::fill_getter_maps()
{
   GetterMaps map_collection;
   typedef ScalarSingletDMSimpleSpec Self;

   map_collection[Par::mass1].map0W["vev"]
     = &Self::get_HiggsVEV;
   map_collection[Par::mass1].map0W["lambda_hS"]
     = &Self::get_lambda_hS;
   map_collection[Par::Pole_Mass].map0W["h0"]
     = &Self::get_HiggsPoleMass;
   map_collection[Par::Pole_Mass].map0W["S"]
     = &Self::get_SingletPoleMass;

   return map_collection;
}
\end{lstcpp}
Here \cpp{GetterMaps} is another type inherited from the base \cpp{Spec} class, and is the main container object that we use to associate a tag/string pair with a function pointer. So the first entry is what makes it possible for a call \cpp{subspectrum.get(Par::mass1,"vev")} to the interface class to in-turn call the \cpp{get\_HiggsVEV} member function of \cpp{Model}. The \cpp{fill\_getter\_maps()} function is used by the base \cpp{Spec} class to initialise a member variable of type \cpp{GetterMaps}, so here we are in fact defining how this member variable will be `filled'. The \cpp{fill\_setter\_maps} function works in direct analogy to the \cpp{fill\_getter\_maps} function, so we will skip discussion of it.

Note that neither of these filler-functions \emph{must} be defined, the wrapper will compile without them and the interface will simply not accept any tag/string pairs to the get/set functions. For example one may fill the `getter' maps, but not the `setter' maps, and it will then simply be impossible to change the parameters in the underlying \cpp{Model} object via the `set' functions, which is not necessarily problematic. However, if one fails to fill the `getter' maps in accordance with the declared \cpp{Contents} of the wrapper, then a runtime error will be thrown as soon as the wrapper object constructor is called, which will present a message telling the user that the wrapper is not correctly defined.

There is a subtlety here; note the \cpp{map0W} data member which is accessed from the \cpp{map\_collection} variable. This member is what defines the function signature of the function pointer to be stored in the \cpp{map\_collection}. This function signature must be known `in advance' in order for the base class to call the function pointer, so the function pointer needs to be placed in the correct location within the \cpp{map\_collection}. In our case the \cpp{map0W} data member expects functions with the signature
\begin{lstcpp}
  double Self::function(void);
\end{lstcpp}
that is, they should be a member function of the wrapper, take no arguments, and return a double. The equivalent \cpp{map0W} member within \cpp{SetterMaps} expects the signature
\begin{lstcpp}
  void Self::function(double);
\end{lstcpp}
that is, functions should be members of the wrapper, should accept a double, and return void.

Functions with a variety of other signatures can also be used, however they must be placed in the correct data member of \cpp{GetterMaps}/\cpp{SetterMaps} in order to work. A full list of the allowed function signatures, the way they should be stored in the \cpp{GetterMaps}/\cpp{SetterMaps}, and the way they can be accessed via the \cpp{SubSpectrum} get and set functions, is given in Appendix \ref{app:gettermaps}. We will see some of these other signatures in use in the more complicated wrapper examples.

\subsection{Interfacing directly with member functions of an external class}
\label{sec:model_member_functions}

When one wants to create a \cpp{SubSpectrum} wrapper for pre-existing classes, it may be the case that the pre-existing class already has ``getter'' and ``setter'' functions, such that one doesn't need to write new ones in the wrapper. Unfortunately, it is not possible to store functions of arbitrary signature in the \cpp{GetterMaps} and \cpp{SetterMaps}, and the pre-enabled set of permitted signatures is not very large. More can be added, however it is quite technically involved. If you have a special need to add more then please contact the authors for advice.

The practical use-cases of directly using external member functions is therefore very limited, however in \specbit we make use of this feature for our most complicated wrapper; the \cpp{MSSMSpec} wrapper which interfaces to \FlexibleSUSY model objects. Furthermore, new autogenerated \FlexibleSUSY spectrum generators will come with the required set of getters/setters with the correct function signature, so it is useful to describe this `shortcut' interface method just for the sake of the \FlexibleSUSY case.

Suppose that the external class to be wrapped has the following form:
\begin{lstcpp}
struct @\metavar{MyClass}@
{
 double get_par1() const;
 double get_par2(int i, int j) const;

 void set_par1(double in);
 void set_par2(int i, int j, double in);
};
\end{lstcpp}
It does not matter where the parameters themselves live at present, we just need the functions that set and retrieve them. In this example we will also show how to deal with functions requiring indices.

These functions can be inserted into the \cpp{GetterMaps} and \cpp{SetterMaps} of the wrapper via the filler functions as follows:
\begin{lstcpp}
static @\metavar{MyWrapper}@::GetterMaps
  @\metavar{MyWrapper}@::fill_getter_maps()
{
 GetterMaps map_collection;
 static const std::set<int> i012 = initSet(0,1,2);

 map_collection[@\metavar{tag1}@].map0["par1"]
   = &Model::get_par1;
 map_collection[@\metavar{tag2}@].map2["par2"]
   = &FInfo2( &Model::get_par2, i012, i012 );

 return map_collection;
}
\end{lstcpp}
and
\begin{lstcpp}
static @\metavar{MyWrapper}@::SetterMaps
  @\metavar{MyWrapper}@::fill_setter_maps()
{
 SetterMaps map_collection;
 static const std::set<int> i012 = initSet(0,1,2);

 map_collection[@\metavar{tag1}@].map0["par1"]
   = &Model::set_par1;
 map_collection[@\metavar{tag2}@].map2["par2"]
   = &FInfo2( &Model::set_par2, i012, i012 );

 return map_collection;
}
\end{lstcpp}
where these function signatures match the \cpp{map0} and \cpp{map2} members of the \cpp{GetterMaps} and \cpp{SetterMaps} classes, as shown in Appendix \ref{app:gettermaps}. This appendix also gives an explanation of the \cpp{FInfo2} helper class and how it is used to help specify the allowed indices to get and set functions.

In order to call these functions via function pointers, the \cpp{SubSpectrum} wrapper needs to know the \Cpp type of the hosted class. This can be supplied via a typedef in the \cpp{SpecTraits} struct;
\begin{lstcpp}
namespace Gambit
{
 namespace Models {class @\metavar{MyClass}@;}
 template <>
 struct SpecTraits<@\metavar{MyWrapper}@> : DefaultTraits
 {
  static std::string name()
  { return "@\metavar{MyWrapper}@"; }
  typedef SpectrumContents::@\metavar{MyContents}@ Contents;
  typedef @\metavar{MyClass}@ Model;
 };
}
\end{lstcpp}
The wrapper base classes will then be able to use the \metavar{MyClass} type via the \cpp{Model} typedef. In addition, the wrapper needs to have access to an instance of the \metavar{MyClass} type. This must be provided by overloading a special inherited member function \cpp{get_Model} in the wrapper, e.g.
\begin{lstcpp}
class @\metavar{MyWrapper}@
  : public Spec<@\metavar{MyWrapper}@>
{
 private:
  Model model;

 public:
  @\metavar{MyWrapper}@(Model& m) : model(m)

  Model& get_Model() { return model; }
  const Model& get_Model() const { return model; }

  static GetterMaps fill_getter_maps();
  static SetterMaps fill_setter_maps();
}
\end{lstcpp}
Note that here \cpp{Model} is the typedef for \metavar{MyClass}, which is learned via the \cpp{SpecTraits} struct. We assume in this example that an instance of \cpp{model} is carried as a data member of the wrapper class, which is a good idea for maintaining encapsulation, but it is not strictly required. All that is required is that \cpp{get_Model()} return an instance of the \cpp{Model} type. Note also that both \cpp{const} and non-\cpp{const} versions of \cpp{get_Model()} are required, because the wrapper base classes need both in order to maintain const-correctness when \cpp{const} wrapper objects are used.
\subsection{Interfacing with non-class functions}
\label{sec:non_class_functions}
In the previous \cpp{SubSpectrum} wrapper examples, we covered interfacing with member functions of wrapper objects, and with member functions of particular external classes. The former can be designed to access external functions or classes in arbitrary ways and so are the most generally useful, while the latter have limited use due to restrictions on the allowed function signature but are convenient for dealing with certain special cases such as \FlexibleSUSY classes. Now, we will deal with interfacing to plain functions that are not members of any class.

As with the other permitted kinds of functions, these are also restricted to specific function signatures. In fact, they do not provide any additional functionality over simply using member functions of the wrapper classes, and wrapper member functions are usually a better choice since they have easier access to data members of the wrapper class. However, they can be useful to avoid code repetition if the target functions are to be used in more than one wrapper, for example if they perform some common unit conversions or simple calculations.

The allowed non-class function signatures are described in Appendix \ref{app:gettermaps}. They may be connected to \cpp{GetterMaps} and \cpp{SetterMaps} as in the following example:
\begin{lstcpp}
double get_Pole_mElectron(const SMInputs& inputs)
 { return inputs.mE; }

static @\metavar{MyWrapper}@::GetterMaps
  @\metavar{MyWrapper}@::fill_getter_maps()
{
 GetterMaps map_collection;

 map_collection[Par::Pole_Mass].map0_extraI["e-_1"]
   = &get_Pole_mElectron;

 return map_collection;
}
\end{lstcpp}
This time, we have used the \cpp{map0_extraI} signature, which requires an \cpp{Input} object as an argument, and no indices. As with the \cpp{Model} object case, the wrapper needs to be informed of the type of the \cpp{Input} object, and provided an instance of it, via the \cpp{SpecTraits} class and \cpp{get_Input} functions, respectively:
\begin{lstcpp}
namespace Gambit
{
 template <>
 struct SpecTraits<@\metavar{MyWrapper}@> : DefaultTraits
 {
  static std::string name()
  { return "@\metavar{MyWrapper}@"; }
  typedef SpectrumContents::@\metavar{MyContents}@ Contents;
  typedef SMInputs Inputs;
 };
}
\end{lstcpp}
and
\begin{lstcpp}
class @\metavar{MyWrapper}@
  : public Spec<@\metavar{MyWrapper}@>
{
  ...
  Input& get_Input();
  const Input& get_Input() const;
  ...
}
\end{lstcpp}
Here, \cpp{Input} is imagined to be a class containing input information used to setup the wrapper, but which one may also wish to access via the \cpp{SubSpectrum} interface along with the rest of the spectrum data. But it is simply an arbitrary class which can be used to pass information to the interface functions.

\subsection{Index offsets}

In general, \cpp{SubSpectrum} wrapper interfaces use a one-based indexing system. For example when retrieving neutralino pole masses from the MSSM spectrum wrappers the lightest neutralino is retrieved with the index 1. However, not all external functions will follow this convention, so it is useful to have a system for converting between index systems. The wrapper system provides this functionality via the \cpp{index_offset} wrapper class member function. This function should be overridden in the wrapper in order to define an offset to be added to all index values input by users to the \cpp{get} and \cpp{set} functions of that wrapper. Its function signature is:
\begin{lstcpp}
static int index_offset() {return @\metavar{offset}@;}
\end{lstcpp}
So, for example, if \metavar{offset} is \cpp{-1}, then when a user calls e.g.\ \cpp{subspec.get(Pole_Mass,"~e",1)} the offset will be added to the input index of \cpp{1}, resulting in \cpp{0}, before being passed to the external function (which will therefore see the index as \cpp{0}). If the \cpp{index_offset} function is not overridden then a default null offset of \cpp{0} will be applied.

Note that this function should have \cpp{public} access.

\subsection{Interfacing with renormalisation group running functions}
\label{sec:interfacing_to_rge}
Now we have seen all the basic ways of connecting functions of various kinds to the \cpp{SubSpectrum} \cpp{get}/\cpp{set} interface. So far, however, the framework may seem like overkill for the tasks we have performed. Why not just call the external functions and store the return values in the appropriate place? Why store pointers to the functions themselves, or wrappers for those functions? The primary answer here is because we want to be able to call renormalisation group running functions, and have these affect all the values accessed by the \cpp{SubSpectrum} interface. If we simply copied the results of the spectrum calculation into an intermediate object, such as an \cpp{SLHAea} object, then this would not be possible. But with the framework we have set up, when the \cpp{get}/\cpp{set} functions are called it is possible for them to interact directly with the contents of external spectrum generator codes, and so the values they retrieve can be affected by \cpp{RGE} running.

If the external spectrum generator is a \Cpp code, then connecting the \cpp{SubSpectrum} wrapper to the \cpp{RGE} running facilities should be fairly easy. There are two main functions that need to be connected: \cpp{GetScale}, and \cpp{RunToScaleOverride}. These functions are not required by the \cpp{SubSpectrumContents} definitions, so whether or not a particular \cpp{SubSpectrum} can perform \cpp{RGE} running will need to be inferred from the \GB \capability{} it is given via whatever module function provides it.

The \cpp{GetScale} function is the simplest, and should return the scale at which all running parameters are defined. If there is an analogous function or parameter defined in the host \cpp{Model} object then the \cpp{GetScale} definition will simply be something like this:
\begin{lstcpp}
double @\metavar{MyWrapper}@::GetScale() const
{
  return model.get_scale();
}
\end{lstcpp}

The \cpp{RunToScaleOverride} function is wrapped by the user-side \cpp{RunToScale} function, which adds additional checks and behaviour modifications. But the \cpp{RunToScaleOverride} is what should be overridden in the wrapper and connected to any external \cpp{RGE} running functions. For example:
\begin{lstcpp}
void @\metavar{MyWrapper}@::RunToScaleOverride(double scale)
{
  model.run_to(scale);
}
\end{lstcpp}

Of course the details for hooking up these functions correctly are entirely dependent on how the external code works. The onus is therefore on the wrapper writer to understand the external code. However, the power of this framework is that once the wrapper is defined, writers of other module functions can interact with spectrum information in a totally generic way, even including \cpp{RGE} running.

\subsection{Constructing and returning \cpp{Spectrum} objects from module functions}
\label{sec:wrapper_module_function}
So far, we have described how to design a new wrapper class for external spectrum data. However, in \GB, a \cpp{Spectrum} object needs to be constructed from the wrapper and returned via a module function before the spectrum data can be used by other module functions. Here we discuss the general requirements of this process.

The details for constructing the wrapper depend on how the wrapper and its constructor are defined, which are up to the wrapper writer. We will therefore use the \cpp{SingletDM} example wrapper from Sec. \ref{sec:parameter_box_wrapper} to demonstrate the more general requirements for creating and returning the abstract interface. Let us first write out an example module function which does this, and then we will examine it line-by-line.

First, the module function rollcall declaration:
\begin{lstcppnum}
#define CAPABILITY SingletDM_spectrum
START_CAPABILITY
  // Create a Spectrum object from SMInputs,
  // SM Higgs parameters, and the SingletDM
  // parameters
  #define FUNCTION get_SingDM_spec
  START_FUNCTION(Spectrum)
  DEPENDENCY(SMINPUTS, SMInputs)
  ALLOW_MODEL_DEPENDENCE(StandardModel_Higgs,
                         SingletDM)
  MODEL_GROUP(higgs,   (StandardModel_Higgs))
  MODEL_GROUP(singlet, (SingletDM))
  ALLOW_MODEL_COMBINATION(higgs, singlet)
  #undef FUNCTION
#undef CAPABILITY
\end{lstcppnum}
Now the module function itself:
\begin{lstcppnum}
void get_SingDM_spec(Spectrum& result)
{
 using namespace Pipes::get_SingDM_spec;
 const SMInputs& sminputs =
  *myPipe::Dep::SMINPUTS;

 Models::SingletDMModel s; @\label{ex:singlet_begin}@
 s.HiggsPoleMass = *Param.at("mH");
 s.HiggsVEV = 1. / sqrt(sqrt(2.)*sminputs.GF);
 s.SingletPoleMass = *Param.at("mS");
 s.SingletLambda = *Param.at("lambda_hS"); @\label{ex:singlet_end}@

 Models::
   ScalarSingletDMSimpleSpec singletspec(s); @\label{ex:singlet_end2}@

 result = Spectrum(singletspec,sminputs,&Param); @\label{ex:full_spec}@
}
\end{lstcppnum}
Let us look first at the module function definition, and then afterwards we will examine how the rollcall declaration ensures that the necessary \dependencies{} are made available. The first line gives the function signature, which, as always, returns \cpp{void} and takes one argument by reference, which is the pre-allocated memory for the storage of the function result. Here we will create a \cpp{Spectrum} object so the `result' type is \cpp{Spectrum}. Next, a \dependency{} on an \cpp{SMInputs} object (as described in Sec. \ref{sec:sminputs}) is retrieved.

Lines \ref{ex:singlet_begin} to \ref{ex:singlet_end} show the construction of an instance of the \cpp{SingletDMModel} class, which is the object that we want to wrap. Here we set the parameters contained in the class by extracting them from the \cpp{Params} map supplied by \GB via the core hierarchical model system (see Sec. 5 of Ref. \cite{gambit}). On line \ref{ex:singlet_end2} we then use the \cpp{SingletDMModel} instance to construct our instance of the \cpp{ScalarSingletDMSimpleSpec} wrapper class.

Nearing the end of the module function, we construct on line \ref{ex:full_spec} the full \cpp{Spectrum} object out of the components we have prepared, namely the scalar singlet \cpp{SubSpectrum} object and an \cpp{SMInputs} object. More generally one can also supply the `low-energy' \cpp{SubSpectrum} component, however this shortened constructor will automatically generate an \cpp{SMSimpleSpec} (see Appendix \ref{app:subspec_wrappers}) wrapper based on \cpp{SMINPUTS} to fulfil this role. The final argument to the \cpp{Spectrum} constructor is the set of model parameters which were available to this \cpp{Spectrum} when it was constructed, which are retrieved via the \cpp{Params} map.

Finally, the newly-constructed object is moved into the \cpp{result} space of the module function, from where it can be distributed to other module functions by the \GB dependency resolver.

Now, let us return to the module function rollcall declaration and see how it requests the \dependencies{} that we require. The first two lines simply declare the name for the \capability{} that this function provides. The next two (define/start function) declare the \Cpp name of the function (which of course must match the function as defined elsewhere). Note here that we also declare the \cpp{result} type for this function as \cpp{Spectrum}.

Next, we declare the dependency we require. Here it is declared as \cpp{SMINPUTS}, with type \cpp{SMInputs}. This will then be made available to our module function by the \GB core.

The next few lines declare a joint dependence on model parameters from both \cpp{StandardModel_Higgs} and \cpp{SingletDM}, that is, they declare that this module function requires parameters from both of these models. For more details on this syntax please see Ref.\ \cite{gambit}.

With the rollcall declaration done we are finished, and the \cpp{Spectrum} object created by the module function \cpp{get_SingDM_spec} is made available to the rest of \GB under the \capability{} name \cpp{SingletDM_spectrum}.

\subsection{Scheme-dependence and other special dependency requirements}

All of the \capabilities{} associated with \cpp{Spectrum} types in \gambitVer are understood to provide running parameters in only one scheme. For example, spectra obtained via the \capability{} named \cpp{MSSM_spectrum} are understood to provide parameters in the \DR scheme, while spectra obtained via the \capability{} named \cpp{SM_spectrum} provide parameters in the \MSbar scheme. Thus, when adding new spectrum generators, users should check that they conform to these ``defaults'' if they write new module functions which promise to provide these existing capabilities (see tables \ref{tab:specbitsmcap}, \ref{tab:specbitmssmcap} and \ref{tab:specbitsingletdmcap}). If providing a spectrum in a different scheme then a different capability name should be used, for example \cpp{MSSM_spectrum_MSbar}. Of course this spectrum will then not be automatically usable by existing \GB module functions which expect the default \DR MSSM spectrum, because the new capability will not match with the dependencies of existing functions. Therefore further modification of the dependency hierarchy will generally be necessary to fully integrate such a spectrum with existing module functions. We recommend that users with special requirements of this kind contact the \GB authors for advice on how best to do this.

Along similar lines, some users may wish to receive certain spectrum-related information in a very specific format, for example requiring a running mass extracted at a specific loop-order or including specific corrections. Such specificity goes beyond the intended use cases for \cpp{Spectrum} objects, so in general this information needs to be recomputed at the point where it is needed from information available via the \cpp{Spectrum} wrappers. For example, one might perform some auxilliary calculations in the interface to a backend code that might have a special requirement for some input parameter. If the quantity is required repeatedly then users should consider creating a new \capability{}, for example \cpp{Higgs_mass_tree_level}, and adding a new module function which performs the necessary calculations to provide this capability.

\subsection{Controlling wrapper lifetimes}

That covers the construction of a simple wrapper, however there are a few more subtleties regarding the lifetime (in the sense of time between construction/destruction of the \Cpp objects as the code runs) of the various wrapper objects to be discussed. There are several constructors for the \cpp{Spectrum} object, and it is important to choose the correct one depending on how you want the member \cpp{SubSpectrum} objects to be treated. The options are listed below, with the argument accepting the \GB parameter container replaced with \cpp{<Param>} for brevity.

\subparagraph{{\CPPidentifierstyle Spectrum()}} --- Creates an empty object.
\subparagraph{{\CPPidentifierstyle Spectrum(const SubSpectrum\& he, const SMInputs\& smi, <Params>)}} --- Constructs a new object, automatically creating an \cpp{SMSimpleSpec} as the LE subspectrum, and cloning the ``HE'' \cpp{SubSpectrum} object supplied and taking possession of it.
\subparagraph{{\CPPidentifierstyle Spectrum(const SubSpectrum\& le, const SubSpectrum\& he, const SMInputs\& smi, <Params>)}} --- Construct new object, cloning both \cpp{SubSpectrum} objects supplied and taking possession of them.
\subparagraph{{\CPPidentifierstyle Spectrum(SubSpectrum\* const le, SubSpectrum\* const he, const SMInputs\& smi, const std)}} --- Construct new object, wrapping {\em existing} \cpp{SubSpectrum} objects. If the original objects are prematurely destructed then attempting to access them via the \cpp{Spectrum} interface will cause a segmentation fault.

\section{Worked example of writing a \cpp{SubSpectrum} wrapper}
\label{app:wrapper_worked_examples}

In this section we provide a stripped-down tutorial-style example of how to define a new \cpp{SubSpectrum} wrapper and return a \cpp{Spectrum} object from a module function. Comments and instructions will be kept to a bare minimum, and instead we will simply refer to the appropriate sections of the paper body in which further details can be found.

%
%
%
%

\subsection{FlexibleSUSY MSSM wrapper}
\label{app:FS_MSSM_wrapper}
The first step in adding a new \FlexibleSUSY spectrum generator to \specbit is to run \FlexibleSUSY and generate the \Cpp code for the new spectrum generator. It then needs to be added to the \GB cmake build system, and the appropriate headers included in certain key \specbit headers. As mentioned in Sec. \ref{Sec:MSSMspecgens} we provide instructions for these steps as separate documentation, which can be found in \term{gambit/doc/Adding_FlexibleSUSY_Models.txt}, and they will change when future technical improvements to the \GB \doublecrosssf{BOSS}{BOSS} \cite{gambit} system permit \FlexibleSUSY to be automatically backended. Thus we begin this example by assuming that these steps have been completed. From this point we can begin following the checklist given in Sec \ref{sec:specbit_adding_new_models}.

\noindent \ref{item:contents}) Choose or create a \cpp{SubSpectrumContents} definition (Sec. \ref{sec:subspectrum_contents}) --- The purpose of the \cpp{SubSpectrum} classes is to enforce consistency between wrappers which wrap the same essential physics, so one should always try to use an existing definition if possible. However, for the purposes of this example we will pretend that the definition for the MSSM did not previously exist. Thus, in \term{gambit/Models/include/gambit/Models/} \term{SpectrumContents/RegisteredSpectra.hpp} we add a line of code:
\begin{lstcpp}
struct MSSM: SubSpectrumContents { MSSM(); };
\end{lstcpp}
Next we create a new source file \term{gambit/Models/} \term{src/SpectrumContents/MSSM.cpp} with contents which define all the MSSM masses and parameters:
\begin{lstcpp}
#include "gambit/Models/SpectrumContents/         RegisteredSpectra.hpp"

namespace Gambit
{
  SpectrumContents::MSSM::MSSM()
  {
    setName("MSSM");
    using namespace Par;

    // useful index definitions
    std::vector<int> scalar = initVector(1);
    std::vector<int> v2     = initVector(2);
    std::vector<int> v3     = initVector(3);
    std::vector<int> v4     = initVector(4);
    std::vector<int> v6     = initVector(6);
    std::vector<int> m2x2   = initVector(2,2);
    std::vector<int> m3x3   = initVector(3,3);
    std::vector<int> m4x4   = initVector(4,4);
    std::vector<int> m6x6   = initVector(6,6);

    // Spectrum parameters
    //           tag,    name,  shape
    addParameter(mass2, "BMu" , scalar);
    addParameter(mass2, "mHd2", scalar);
    addParameter(mass2, "mHu2", scalar);

    addParameter(mass2, "mq2", m3x3);
    addParameter(mass2, "ml2", m3x3);
    addParameter(mass2, "md2", m3x3);
    addParameter(mass2, "mu2", m3x3);
    addParameter(mass2, "me2", m3x3);

    addParameter(mass1, "M1", scalar);
    addParameter(mass1, "M2", scalar);
    addParameter(mass1, "M3", scalar);
    addParameter(mass1, "Mu", scalar);
    addParameter(mass1, "vu", scalar);
    addParameter(mass1, "vd", scalar);

    addParameter(mass1, "TYd", m3x3);
    addParameter(mass1, "TYe", m3x3);
    addParameter(mass1, "TYu", m3x3);
    addParameter(mass1, "ad" , m3x3);
    addParameter(mass1, "ae" , m3x3);
    addParameter(mass1, "au" , m3x3);

    addParameter(dimensionless, "g1", scalar);
    addParameter(dimensionless, "g2", scalar);
    addParameter(dimensionless, "g3", scalar);

    addParameter(dimensionless, "sinW2", scalar);

    addParameter(dimensionless, "Yd", m3x3);
    addParameter(dimensionless, "Yu", m3x3);
    addParameter(dimensionless, "Ye", m3x3);

    addParameter(Pole_Mass, "~g", scalar);
    addParameter(Pole_Mass, "~d",    v6);
    addParameter(Pole_Mass, "~u",    v6);
    addParameter(Pole_Mass, "~e-",   v6);
    addParameter(Pole_Mass, "~nu",   v3);
    addParameter(Pole_Mass, "~chi+", v2);
    addParameter(Pole_Mass, "~chi0", v4);
    addParameter(Pole_Mass, "h0",    v2);
    addParameter(Pole_Mass, "A0", scalar);
    addParameter(Pole_Mass, "H+", scalar);

    addParameter(Pole_Mixing, "~d",    m6x6);
    addParameter(Pole_Mixing, "~u",    m6x6);
    addParameter(Pole_Mixing, "~e-",   m6x6);
    addParameter(Pole_Mixing, "~nu",   m3x3);
    addParameter(Pole_Mixing, "~chi0", m4x4);
    addParameter(Pole_Mixing, "~chi-", m2x2);
    addParameter(Pole_Mixing, "~chi+", m2x2);
    addParameter(Pole_Mixing, "h0",    m2x2);
    addParameter(Pole_Mixing, "A0",    m2x2);
    addParameter(Pole_Mixing, "H+",    m2x2);
  }
}
\end{lstcpp}
The \cpp{SpectrumContents} that we will use are now defined.

\noindent \ref{item:traits}) Write a \cpp{SpecTraits} specialization --- In this example, the \cpp{getter} and \cpp{setter} functions are defined in an external object, so the example in Sec. \ref{sec:non_class_functions} is relevant. It is best to define our \cpp{SpecTraits} class in the header file for the \cpp{SubSpectrum} wrapper class, because the definition needs to be available to the \Cpp compiler when when actual \cpp{SubSpectrum} class is defined. We thus begin a new header file \term{SpecBit/include/gambit/SpecBit/MSSM_FS.hpp} and add the following code (neglecting include guards):
\begin{lstcpp}
#include "gambit/Elements/spec.hpp"
#include "gambit/Models/SpectrumContents/         RegisteredSpectra.hpp"
#include "gambit/contrib/MassSpectra/        flexiblesusy/models/MSSM/MSSM_model_slha.hpp"

namespace Gambit
{
  namespace SpecBit
  {
    class MSSM_FS; //Forward declaration
  }

  template <>
  struct SpecTraits<SpecBit::MSSM_FS>
    : DefaultTraits
  {
    static std::string name() {return "MSSM_FS";}
    typedef SpectrumContents::MSSM Contents;
    using namespace flexiblesusy;
    typedef MSSM_slha<Two_scale> Model;
  };
}
\end{lstcpp}

\noindent \ref{item:wrapper}) Write the \cpp{SubSpectrum} wrapper class (Sec. \ref{sec:parameter_box_wrapper}) --- We can now declare the wrapper class in the same file:
\begin{lstcpp}
namespace Gambit
{
 namespace SpecBit
 {
  class MSSM_FS : public Spec<MSSM_FS>
  {
   private:
    Model fs_model;
    typedef MSSM_FS Self;

   public:
    // Interface function overrides
    static int index_offset() {return -1;}
    double GetScale() const;
    void SetScale(double scale);
    void RunToScaleOverride(double scale);

    //constructors / destructors
    MSSM_FS() {}
    MSSM_FS(Model& m) : fs_model(m) {}

    // Functions to allow the base `Spec' class
    // access to `Model'
    Model& get_Model() {return fs_model;}
    const Model& get_Model() const
     {return fs_model;}

    // Functions to fill SLHAea objects
    void add_to_SLHAea(int slha_version,SLHAstruct& slha) const;
    SLHAstruct getSLHAea(int getSLHAea) const;

    // Map filler overrides
    static GetterMaps fill_getter_maps();
    static SetterMaps fill_setter_maps();

    // Helper functions
    double get_tanbeta();
    double get_DRbar_mA2();
    double get_sinthW2_DRbar();
    double get_MAh1_pole_slha();
    double get_MHpm1_pole_slha();
    void set_MAh1_pole_slha(double);
    void set_MHpm1_pole_slha(double);
    void set_MGluino_pole_slha(double);
    void set_MZ_pole_slha(double);
    void set_MW_pole_slha(double);
    void set_MSu_pole_slha(double, int);
    void set_MSd_pole_slha(double, int);
    void set_MSe_pole_slha(double, int);
    void set_MSv_pole_slha(double, int);
    void set_MCha_pole_slha(double, int);
    void set_MChi_pole_slha(double, int);
    void set_Mhh_pole_slha(double, int);
    void set_ZD_pole_slha(double, int, int);
    void set_ZU_pole_slha(double, int, int);
    void set_ZE_pole_slha(double, int, int);
    void set_ZV_pole_slha(double, int, int);
    void set_ZH_pole_slha(double, int, int);
    void set_ZA_pole_slha(double, int, int);
    void set_ZP_pole_slha(double, int, int);
    void set_ZN_pole_slha(double, int, int);
    void set_UM_pole_slha(double, int, int);
    void set_UP_pole_slha(double, int, int);
    void set_ZH_pole_slha(double, int, int);
  };
 }
}
\end{lstcpp}
We now begin a new source file
\begin{itemize}[topsep=3pt]
\item[] \term{gambit/SpecBit/src/MSSM_FS.cpp}
\end{itemize}
to hold the definitions of the wrapper functions:
%
\begin{lstcppalt}
using namespace Gambit;
using namespace SpecBit;

// Fill an SLHAea object
void MSSM_FS::add_to_SLHAea(int slha_version,SLHAstruct& slha)
  const
{
  // Here one needs to extract all the SLHA
  // information from the spectrum object and
  // store it in the supplied SLHAea object.
  // This takes many lines of code, so we
  // exclude it from this example. For a
  // working MSSM example of this function see
  // "gambit/SpecBit/include/gambit/
  // SpecBit/MSSMSpec.hpp".
}

// Return a new (filled) SLHAea object
SLHAstruct MSSM_FS::get_SLHAea(SLHAstruct& slha)
  const
{
  SLHAstruct output;
  add_to_SLHAea(slha_version,output);
  return output;
}

void MSSM_FS::RunToScaleOverride(double scale)
{
  fs_model.run_to(scale);
}

double MSSM_FS::GetScale() const
{
  return fs_model.get_scale();
}

void MSSM_FS::SetScale(double scale)
{
  fs_model.set_scale(scale);
}

// `extra' function to compute TanBeta
double MSSM_FS::get_tanbeta()
{
  return fs_model.get_vu() / fs_model.get_vd();
}

// `extra' function to compute mA2
double MSSM_FS::get_DRbar_mA2()
{
  double tb = get_tanbeta();
  double cb = cos(atan(tb));
  double sb = sin(atan(tb));
  return fs_model.get_BMu() / (sb * cb);
}

// `extra' function to compute DRbar weak
// mixing angle
double MSSM_FS::get_sinthW2_DRbar()
{
  double sthW2 = Utils::sqr(fs_model.get_g1())
   * 0.6 / (0.6 * Utils::sqr(fs_model.get_g1())
   + Utils::sqr(fs_model.get_g2()));
  return sthW2;
}

// Wrapper functions for A0 and H+ getters, to
// retrieve only the non-Goldstone entries.
double MSSM_FS::get_MAh1_pole_slha()
{
  return fs_model.get_MAh_pole_slha(1);
}

double MSSM_FS::get_MHpm1_pole_slha()
{
  return fs_model.get_MHpm_pole_slha(1);
}

// Helper functions for manually setting pole
// masses. We can use a macro to automate this.
BeginLongMacro POLE_MASS_SETTER(NAME) \
void MSSM_FS::set_##NAME##_pole_slha(double mass,\
 int i)\
{\
  fs_model.get_physical_slha().NAME(i) = mass;\
}\ EndLongMacro

POLE_MASS_SETTER(MSu)
POLE_MASS_SETTER(MSd)
POLE_MASS_SETTER(MSe)
POLE_MASS_SETTER(MSv)
POLE_MASS_SETTER(MCha)
POLE_MASS_SETTER(MChi)
POLE_MASS_SETTER(Mhh)

// Similar for mixings
BeginLongMacro POLE_MIXING_SETTER(NAME) \
void MSSM_FS::set_##NAME##_pole_slha(double mass,\
 int i, int j)\
{\
  fs_model.get_physical_slha().NAME(i,j) = mass;\
}\ EndLongMacro

POLE_MIXING_SETTER(ZD)
POLE_MIXING_SETTER(ZU)
POLE_MIXING_SETTER(ZE)
POLE_MIXING_SETTER(ZV)
POLE_MIXING_SETTER(ZH)
POLE_MIXING_SETTER(ZA)
POLE_MIXING_SETTER(ZP)
POLE_MIXING_SETTER(ZN)
POLE_MIXING_SETTER(UM)
POLE_MIXING_SETTER(UP)
POLE_MIXING_SETTER(ZH)

// Higgs mass setters
void MSSM_FS::set_MAh1_pole_slha(double mass)
{
  fs_model.get_physical_slha().MAh(1) = mass;
}

void MSSM_FS::set_MHpm1_pole_slha(double mass)
{
  fs_model.get_physical_slha().MHpm(1) = mass;
}

// Pole masses with no indices
BeginLongMacro POLE_MASS_SETTER0(FNAME,VNAME)\
void MSSM_FS::set_##FNAME##_pole_slha(double\
 mass)\
{\
  fs_model.get_physical_slha().VNAME = mass;\
}\ EndLongMacro

POLE_MASS_SETTER0(MGluino,MGlu)
POLE_MASS_SETTER0(MZ,MVZ)
POLE_MASS_SETTER0(MW,MVWm)

// Filler function for getter function maps
MSSM_FS::GetterMaps MSSM_FS::fill_getter_maps()
{
  MSSM_FS::GetterMaps map_collection;

  static const std::set<int> i01 = initSet(0,1);
  static const std::set<int> i012 =
   initSet(0,1,2);
  static const std::set<int> i0123 =
   initSet(0,1,2,3);
  static const std::set<int> i012345 =
   initSet(0,1,2,3,4,5);

  /// mass2 - mass dimension 2 parameters

  // Functions from Model
  // (i.e. flexiblesusy::MSSM_slha<Two_scale>)
  {
    typename MTget::fmap0 tmp_map;
    tmp_map["BMu"]  = &Model::get_BMu;
    tmp_map["mHd2"] = &Model::get_mHd2;
    tmp_map["mHu2"] = &Model::get_mHu2;
    map_collection[Par::mass2].map0 = tmp_map;
  }

  // Extra functions from the wrapper
  map_collection[Par::mass2].map0W["mA2"] =
   &Self::get_DRbar_mA2;

  // Functions from Model with two indices
  {
    typename MTget::fmap2 tmp_map;
    tmp_map["mq2"]=FInfo2(&Model::get_mq2,i012,
     i012);
    tmp_map["ml2"]=FInfo2(&Model::get_ml2,i012,
     i012);
    tmp_map["md2"]=FInfo2(&Model::get_md2,i012,
     i012);
    tmp_map["mu2"]=FInfo2(&Model::get_mu2,i012,
     i012);
    tmp_map["me2"]=FInfo2(&Model::get_me2,i012,
     i012);
    map_collection[Par::mass2].map2 = tmp_map;
  }

  ///  mass1 - mass dimension 1 parameters

  // Zero-index Member functions of Model
  {
    typename MTget::fmap0 tmp_map;
    tmp_map["M1"] = &Model::get_MassB;
    tmp_map["M2"] = &Model::get_MassWB;
    tmp_map["M3"] = &Model::get_MassG;
    tmp_map["Mu"] = &Model::get_Mu;
    tmp_map["vu"] = &Model::get_vu;
    tmp_map["vd"] = &Model::get_vd;
    map_collection[Par::mass1].map0 = tmp_map;
  }

  // Two-index member functions of Model
  {
    typename MTget::fmap2 tmp_map;
    tmp_map["TYd"]=FInfo2(&Model::get_TYd,i012,
     i012);
    tmp_map["TYe"]=FInfo2(&Model::get_TYe,i012,
     i012);
    tmp_map["TYu"]=FInfo2(&Model::get_TYu,i012,
     i012);
    tmp_map["ad"] =FInfo2(&Model::get_TYd,i012,
     i012);
    tmp_map["ae"] =FInfo2(&Model::get_TYe,i012,
     i012);
    tmp_map["au"] =FInfo2(&Model::get_TYu,i012,
     i012);
    map_collection[Par::mass1].map2 = tmp_map;
  }

  ///  dimensionless - mass dimension 0 parameters

  // Zero index member functions of Model
  {
    typename MTget::fmap0 tmp_map;
    tmp_map["g1"]= &Model::get_g1;
    tmp_map["g2"]= &Model::get_g2;
    tmp_map["g3"]= &Model::get_g3;
    map_collection[Par::dimensionless].map0 =
     tmp_map;
  }

  // Zero-index `extra' functions from wrapper
  {
    typename MTget::fmap0W tmp_map;
    tmp_map["tanbeta"] = &Self::get_tanbeta;
    tmp_map["sinW2"]   = &Self::get_sinthW2_DRbar;
    map_collection[Par::dimensionless].map0W=tmp_map;
  }

  // Two-index member functions of Model
  {
    typename MTget::fmap2 tmp_map;
    tmp_map["Yd"]= FInfo2(&Model::get_Yd,i012,
     i012);
    tmp_map["Yu"]= FInfo2(&Model::get_Yu,i012,
     i012);
    tmp_map["Ye"]= FInfo2(&Model::get_Ye,i012,
     i012);
    map_collection[Par::dimensionless].map2 =
     tmp_map;
  }

  /// Pole_Mass - Pole mass parameters

  // Zero index member functions of Model
  {
    typename MTget::fmap0 tmp_map;
    tmp_map["W+"] = &Model::get_MVWm_pole_slha;
    tmp_map["~g"] = &Model::get_MGlu_pole_slha;
    map_collection[Par::Pole_Mass].map0 = tmp_map;
  }

  // Zero index `extra' functions from wrapper
  {
    typename MTget::fmap0W tmp_map;
    tmp_map["A0"] = &Self::get_MAh1_pole_slha;
    tmp_map["H+"] = &Self::get_MHpm1_pole_slha;
    map_collection[Par::Pole_Mass].map0W =
     tmp_map;
  }

  // One-index member functions of Model
  {
    typename MTget::fmap1 tmp_map;
    tmp_map["~d"] = FInfo1(
     &Model::get_MSd_pole_slha, i012345 );
    tmp_map["~u"] = FInfo1(
     &Model::get_MSu_pole_slha, i012345 );
    tmp_map["~e-"]= FInfo1(
     &Model::get_MSe_pole_slha, i012345 );
    tmp_map["~nu"]= FInfo1(
     &Model::get_MSv_pole_slha, i012 );
    tmp_map["h0"] = FInfo1(
     &Model::get_Mhh_pole_slha, i01 );
    tmp_map["~chi+"] = FInfo1(
     &Model::get_MCha_pole_slha, i01 );
    tmp_map["~chi0"] = FInfo1(
     &Model::get_MChi_pole_slha, i0123 );
    map_collection[Par::Pole_Mass].map1 = tmp_map;
  }

  /// Pole_Mixing - Pole mass parameters

  // Two-index member functions of Model
  {
    typename MTget::fmap2 tmp_map;
    tmp_map["~d"] = FInfo2(
     &Model::get_ZD_pole_slha, i012345, i012345);
    tmp_map["~nu"]= FInfo2(
     &Model::get_ZV_pole_slha, i012, i012);
    tmp_map["~u"] = FInfo2(
     &Model::get_ZU_pole_slha, i012345, i012345);
    tmp_map["~e-"]= FInfo2(
     &Model::get_ZE_pole_slha, i012345, i012345);
    tmp_map["h0"] = FInfo2(
     &Model::get_ZH_pole_slha, i01, i01);
    tmp_map["A0"] = FInfo2(
     &Model::get_ZA_pole_slha, i01, i01);
    tmp_map["H+"] = FInfo2(
     &Model::get_ZP_pole_slha, i01, i01);
    tmp_map["~chi0"] = FInfo2(
     &Model::get_ZN_pole_slha, i0123, i0123);
    tmp_map["~chi-"] = FInfo2(
     &Model::get_UM_pole_slha, i01, i01);
    tmp_map["~chi+"] = FInfo2(
     &Model::get_UP_pole_slha, i01, i01);
    map_collection[Par::Pole_Mixing].map2 =
     tmp_map;
  }

  return map_collection;
}

// Filler function for setter function maps
MSSM_FS::SetterMaps MSSMSpec_FS::
 fill_setter_maps()
{
  MSSM_FS::SetterMaps map_collection;

  static const std::set<int> i01 = initSet(0,1);
  static const std::set<int> i012 =
   initSet(0,1,2);
  static const std::set<int> i0123 =
   initSet(0,1,2,3);
  static const std::set<int> i012345 =
   initSet(0,1,2,3,4,5);

  /// mass2 - mass dimension 2 parameters

  // Zero index member functions of Model
  {
    typename MTset::fmap0 tmp_map;
    tmp_map["BMu"]  = &Model::set_BMu;
    tmp_map["mHd2"] = &Model::set_mHd2;
    tmp_map["mHu2"] = &Model::set_mHu2;
    map_collection[Par::mass2].map0 = tmp_map;
  }

  // Two-index member functions of Model
  {
    typename MTset::fmap2 tmp_map;
    tmp_map["mq2"]=FInfo2(&Model::set_mq2,i012,
     i012);
    tmp_map["ml2"]=FInfo2(&Model::set_ml2,i012,
     i012);
    tmp_map["md2"]=FInfo2(&Model::set_md2,i012,
     i012);
    tmp_map["mu2"]=FInfo2(&Model::set_mu2,i012,
     i012);
    tmp_map["me2"]=FInfo2(&Model::set_me2,i012,
     i012);
    map_collection[Par::mass2].map2 = tmp_map;
  }

  /// mass1 - mass dimension 1 parameters

  // Zero index member functions of Model
  {
    typename MTset::fmap0 tmp_map;
    tmp_map["M1"]= &Model::set_MassB;
    tmp_map["M2"]= &Model::set_MassWB;
    tmp_map["M3"]= &Model::set_MassG;
    tmp_map["Mu"]= &Model::set_Mu;
    tmp_map["vu"]= &Model::set_vu;
    tmp_map["vd"]= &Model::set_vd;
    map_collection[Par::mass1].map0 =
     tmp_map;
  }

  // Two-index member functions of model object
  {
    typename MTset::fmap2 tmp_map;
    tmp_map["TYd"]=FInfo2(&Model::set_TYd,i012,
     i012);
    tmp_map["TYe"]=FInfo2(&Model::set_TYe,i012,
     i012);
    tmp_map["TYu"]=FInfo2(&Model::set_TYu,i012,
     i012);
    tmp_map["ad"] =FInfo2(&Model::set_TYd,i012,
     i012);
    tmp_map["ae"] =FInfo2(&Model::set_TYe,i012,
     i012);
    tmp_map["au"] =FInfo2(&Model::set_TYu,i012,
     i012);
    map_collection[Par::mass1].map2 = tmp_map;
  }

  /// dimensionless - mass dimension 0 parameters
  //
  // Zero index member functions of Model
  {
    typename MTset::fmap0 tmp_map;
    tmp_map["g1"]= &Model::set_g1;
    tmp_map["g2"]= &Model::set_g2;
    tmp_map["g3"]= &Model::set_g3;
    map_collection[Par::dimensionless].map0 =
     tmp_map;
  }

  // Two-index member functions of Model
  {
    typename MTset::fmap2 tmp_map;
    tmp_map["Yd"]= FInfo2(&Model::set_Yd,i012,
     i012);
    tmp_map["Yu"]= FInfo2(&Model::set_Yu,i012,
     i012);
    tmp_map["Ye"]= FInfo2(&Model::set_Ye,i012,
     i012);
    map_collection[Par::dimensionless].map2 =
     tmp_map;
  }

  /// Pole_Mass parameters

  // Zero-index 'extra' functions from wrapper
  {
    typename MTset::fmap0W tmp_map;
    tmp_map["~g"] = &Self::set_MGluino_pole_slha;
    tmp_map["A0"] = &Self::set_MAh1_pole_slha;
    tmp_map["H+"] = &Self::set_MHpm1_pole_slha;
    tmp_map["W+"] = &Self::set_MW_pole_slha;
    map_collection[Par::Pole_Mass].map0W =
     tmp_map;
  }

  // One-index `extra' functions from wrapper
  {
    typename MTset::fmap1W tmp_map;
    tmp_map["~u"] = FInfo1W(
     &Self::set_MSu_pole_slha, i012345 );
    tmp_map["~d"] = FInfo1W(
     &Self::set_MSd_pole_slha, i012345 );
    tmp_map["~e-"]= FInfo1W(
     &Self::set_MSe_pole_slha, i012345 );
    tmp_map["~nu"]= FInfo1W(
     &Self::set_MSv_pole_slha, i012 );
    tmp_map["~chi+"] = FInfo1W(
     &Self::set_MCha_pole_slha, i01 );
    tmp_map["~chi0"] = FInfo1W(
     &Self::set_MChi_pole_slha, i0123 );
    tmp_map["h0"] = FInfo1W(
     &Self::set_Mhh_pole_slha, i01 );
    map_collection[Par::Pole_Mass].map1W =
     tmp_map;
  }

  /// Pole_Mixing parameters

  // Two-index `extra' functions from wrapper
  {
    typename MTset::fmap2_extraM tmp_map;
    tmp_map["~d"] = FInfo2W(
     &Self::set_ZD_pole_slha, i012345, i012345);
    tmp_map["~nu"]= FInfo2W(
     &Self::set_ZV_pole_slha, i012, i012);
    tmp_map["~u"] = FInfo2W(
     &Self::set_ZU_pole_slha, i012345, i012345);
    tmp_map["~e-"]= FInfo2W(
     &Self::set_ZE_pole_slha, i012345, i012345);
    tmp_map["h0"] = FInfo2W(
     &Self::set_ZH_pole_slha, i01, i01);
    tmp_map["A0"] = FInfo2W(
     &Self::set_ZA_pole_slha, i01, i01);
    tmp_map["H+"] = FInfo2W(
     &Self::set_ZP_pole_slha, i01, i01);
    tmp_map["~chi0"] = FInfo2W(
     &Self::set_ZN_pole_slha, i0123, i0123);
    tmp_map["~chi-"] = FInfo2W(
     &Self::set_UM_pole_slha, i01, i01);
    tmp_map["~chi+"] = FInfo2W(
     &Self::set_UP_pole_slha, i01, i01);
    map_collection[Par::Pole_Mixing].map2W =
     tmp_map;
  }

  return map_collection;
}
\end{lstcppalt}

\noindent \ref{item:modfunc}) Write a module function to construct and return a \cpp{Spectrum} interface object connected to the new wrapper (Sec. \ref{sec:wrapper_module_function}) --- Let us suppose that the new module function will be added to \specbit. Module functions that construct spectrum objects for the MSSM are declared in the header \term{SpecBit/include/gambit/SpecBit/SpecBit_MSSM_rollcall.hpp}. To this file, within the \capability{} block beginning \cpp{#define CAPABILITY unimproved\_MSSM\_spectrum}, we add the following declaration:
\begin{lstcpp}
#define FUNCTION get_MSSM_FS_spectrum
START_FUNCTION(Spectrum)
ALLOW_MODELS(MSSM63atQ)
DEPENDENCY(SMINPUTS, SMInputs)
#undef FUNCTION
\end{lstcpp}
We assume here that the new \FlexibleSUSY spectrum generator has been defined such that the soft masses are input at a user-specified scale $Q$, such that the matching \GB model parameters are \cpp{MSSM63atQ}. We therefore declare \cpp{ALLOW_MODELS(MSSM63atQ)}, so that we will be able to access these parameters in our module function. We will also need SM SLHA2 parameters, but it is more convenient to obtain them via an \cpp{SMInputs} object rather than directly from \GB model parameters, so we declare a \dependency{} on \cpp{SMINPUTS}, which can be provided by an existing \specbit module function.

Next we add the module function definition to an appropriate source file; \term{gambit/SpecBit/} \term{src/SpecBit_MSSM.cpp} is suitable for this example:
\begin{lstcpp}
// Runs MSSM spectrum generator with
// input at scale Q
void get_MSSM_FS_spectrum (Spectrum& result)
{
 namespace myPipe = Pipes::get_MSSM_FS_spectrum;
 namespace ss = softsusy;
 const SMInputs& sminputs =
  *myPipe::Dep::SMINPUTS;
 flexiblesusy::MSSM_input_parameters input;

 // Transfer input parameters to FlexibleSUSY
 input.Qin = *myPipe::Param.at("Qin");

 //double valued parameters
 input.TanBeta     = *Param.at("TanBeta");
 input.SignMu      = *Param.at("SignMu");
 input.mHu2IN      = *Param.at("mHu2");
 input.mHd2IN      = *Param.at("mHd2");
 input.MassBInput  = *Param.at("M1");
 input.MassWBInput = *Param.at("M2");
 input.MassGInput  = *Param.at("M3");

 //3x3 matrices; filled with the help of
 // a convenience function
 #define FILL_3X3_SYM \
  fill_3x3_symmetric_parameter_matrix
 #define FILL_3X3 \
  fill_3x3_parameter_matrix
 input.mq2Input = FILL_3X3_SYM("mq2", Param);
 input.ml2Input = FILL_3X3_SYM("ml2", Param);
 input.md2Input = FILL_3X3_SYM("md2", Param);
 input.mu2Input = FILL_3X3_SYM("mu2", Param);
 input.me2Input = FILL_3X3_SYM("me2", Param);
 input.Aeij = FILL_3X3("Ae", Param);
 input.Adij = FILL_3X3("Ad", Param);
 input.Auij = FILL_3X3("Au", Param);

 // Construct spectrum generator input objects

 // SoftSUSY object used to set quark and lepton
 // masses and gauge couplings in QEDxQCD
 // effective theory.
 softsusy::QedQcd oneset;

 // Fill QedQcd object with SMInputs values
 oneset.setPoleMt(sminputs.mT);
 oneset.setPoleMtau(sminputs.mTau);
 oneset.setMbMb(sminputs.mBmB);
 oneset.setMass(ss::mDown, sminputs.mD);
 oneset.setMass(ss::mUp,   sminputs.mU);
 oneset.setMass(ss::mStrange, sminputs.mS);
 oneset.setMass(ss::mCharm,   sminputs.mCmC);
 oneset.setAlpha(ss::ALPHA, 1./sminputs.alphainv);
 oneset.setAlpha(ss::ALPHAS, sminputs.alphaS);
 oneset.setMass(ss::mElectron, sminputs.mE);
 oneset.setMass(ss::mMuon,     sminputs.mMu);
 oneset.setPoleMZ(sminputs.mZ);

 // Run everything to Mz
 oneset.toMz();

 // Create spectrum generator object
 flexiblesusy::MSSM_spectrum_generator<Two_scale>
   spectrum_generator;

 // Generate spectrum
 spectrum_generator.run(oneset, input);

 // Create SubSpectrum wrapper objects
 MSSM_FS mssmspec(spectrum_generator.get_model())
 QedQcdWrapper qedqcdspec(oneset,sminputs);

 // Check for problems during spectrum generation
 if( spectrum_generator.get_problems()
      .have_problem() )
 {
   std::ostringstream msg;
   problems.print_problems(msg);
   invalid_point().raise(msg.str());
 }

 // Package QedQcd SubSpectrum object, MSSM
 // SubSpectrum object, and SMInputs struct
 // into a 'full' Spectrum object
 result = Spectrum(qedqcdspec,mssmspec,
                   sminputs,myPipe::Param);
}
\end{lstcpp}

The example is now complete. Various aspects related to error checking, code re-use between module functions, and altering the behaviour of the spectrum generator via \YAML options, have been removed in order to keep the example simpler. It is also useful to write several `auxiliary' module functions to perform helper tasks on the new spectrum, for instance transforming the \capability{}, extracting \cpp{SLHAea} objects, and similar tasks. See the module functions in \term{"gambit/SpecBit/src/SpecBit_MSSM.cpp"} for examples of these.

\startglossary

\gitem{backend}\input{"glossary/backend.glossentry"}
\gitem{backend function}\input{"glossary/backend_function.glossentry"}
\gitem{backend requirement}\input{"glossary/backend_requirement.glossentry"}
\gitem{backend variable}\input{"glossary/backend_variable.glossentry"}
\gsfitem{BOSS}\input{"glossary/BOSS.glossentry"}
\newcommand{\seecompdatabase}{see Sec.\ 10.7 of Ref.\ \cite{gambit}}
\gitem{capability}\input{"glossary/capability.glossentry"}
\gitem{dependency}\input{"glossary/dependency.glossentry"}
\gitem{dependency resolver}\input{"glossary/dependency_resolver.glossentry"}
\gitem{dependency resolution}\input{"glossary/dependency_resolution.glossentry"}
\newcommand{\deptreefig}{Fig.\ 5 of Ref.\ \cite{gambit}}
\gitem{dependency tree}\input{"glossary/dependency_tree.glossentry"}
\gitem{frontend}\input{"glossary/frontend.glossentry"}
\gitem{frontend header}\input{"glossary/frontend_header.glossentry"}
\gitem{module}\input{"glossary/module.glossentry"}
\gitem{module function}\input{"glossary/module_function.glossentry"}
\gitem{physics module}\input{"glossary/physics_module.glossentry"}
\gitem{rollcall header}\input{"glossary/rollcall_header.glossentry"}
\gitem{type}\input{"glossary/type.glossentry"}

\finishglossary

\bibliography{R1}

\end{document}